%% file: paper.tex
\documentclass{article}
\usepackage{iclr2025_conference,times}
\input{math_commands.tex}

\usepackage{hyperref}
\hypersetup{colorlinks=true,linkcolor=blue!55!black,citecolor=green!35!black,urlcolor=blue!55!black,
  pdftitle={Retrieval Observability Bounds on Provenance Detection for Agent Memory Poisoning: Measured Coverage and a Falsified Standalone Detector},
  pdfauthor={Jun Wen Leong}}
\usepackage{url}
\usepackage{booktabs}
\usepackage{longtable}
\usepackage{xcolor}
\usepackage{pifont}
\usepackage{graphicx}
\usepackage{amsthm}
\newtheorem{definition}{Definition}
\graphicspath{{figures/}}
\newcommand{\cmark}{\ding{51}}
\newcommand{\xmark}{\ding{55}}

\title{\raggedright\hyphenpenalty=10000\exhyphenpenalty=10000 Retrieval Observability Bounds\\on Provenance Detection\\for Agent Memory Poisoning:\\[0.3em]{\large Measured Coverage and a Falsified Standalone Detector}}

\author{
Jun Wen Leong \\
\texttt{leongjunwen@gmail.com}
}

\newcommand{\eg}{\textit{e.g.}}

\newcommand{\tool}[1]{\texttt{#1}}

\iclrfinalcopy

\begin{document}

\maketitle
\lhead{Preprint. arXiv:2606.30566}

\noindent\fbox{\parbox{\columnwidth}{\small\textbf{Version note.} Relative to the last publicly announced version, arXiv:2606.30566v2, this version \textbf{reverses that version's central practical claim}: the trajectory signature is not a deployable detector. Withdrawn from v2: the recommendation to use the signature as a pre-filter ahead of recipient-metadata gating (\S\ref{sec:v35}), the frontier-transfer evaluation (\S\ref{sec:gpt41}), and every claim to have met a preregistered $\leq 0.10$ upper bound (\S\ref{par:zero-event-rule}, Table~\ref{tab:zero-event}). No claim's evidence status was upgraded. This version additionally makes explicit that the cross-validated classifier figures measure membership of a $22$-vector feature-codebook rather than generalisation: a fold-safe lookup table over that codebook matches the Random Forest's out-of-fold AUC, and accuracy collapses to the class prior once test vectors are held out (\S\ref{sec:perf}, \texttt{results/lookup\_baseline.json}). This does not change any headline claim or the invariant. Each correction is also stated where it applies; all preregistration deviations are listed in Table~\ref{tab:withdrawn}, and the full version history is in Appendix~\ref{app:version-history} and in \texttt{CHANGELOG.md} in the companion repository.}}\vspace{0.5em}

\begin{abstract}
Whether a trajectory-based detector can observe the \emph{retrieval provenance} of an agent memory-poisoning attack is governed by one architectural variable: \textbf{retrieval observability}, whether the agent's memory access produces a logged tool call. We formalise this as a framework-agnostic \emph{retrieval-to-action provenance graph}, classify published attack families by their position relative to it, measure the resulting coverage property against collected agent traces, and then falsify the detector's standalone deployment. The necessity half of the claim is close to analytic once stated; we treat the measurements, not the statement, as the contribution.

Under exclusive tool-mediated payload access a retrieval-before-exfiltration event is structurally forced. Across eight non-overlapping zero-event control cells we observe $0$ violations in $289$ eligible successes over $87$ distinct scenario configurations, spanning two retrieval primitives, two exfiltration channels, four models and both behavioural phenotypes. The invariant is empirically unbroken, but on the clustering unit these designs license no \emph{registered} cell that certifies the preregistered $\leq 0.10$ upper bound (per-cell bounds $0.10$--$0.35$); a post-hoc pool of the two document-share control arms would nominally fall below $0.10$ ($0/112$ over $35$ clusters, upper $0.099$) but is not credited. When the scaffold delivers the payload implicitly the signature disappears, and a preregistered follow-up (V3-6) establishes why. We had read the low retrieval rates under implicit delivery as agents \emph{declining} an available retrieval; they are not. Holding the in-context payload and the memory-inviting trigger fixed and varying only the store, a ground-then-act model returns the payload from memory in $108/108$ trials when the store holds it and $0/108$ when it does not, so a context copy does not suppress retrieval and \textbf{that behavioural reading is withdrawn}. The agent consulted memory in all $324$ sessions; what varies is whether a value is retrieved, and four of its retrieval attempts request keys constructed from the injected rule that do not exist. A separately preregistered replication on \texttt{gpt-4.1}, the opposite behavioural phenotype, reproduces this exactly ($108/108$ versus $0/108$) with its floor gate passed, establishing the withdrawal on \texttt{gpt-4.1} (dispatch-order-verified for $184/324$ sessions; indeterminate, not falsified, for the rest, since the harness can batch several tool calls into one completion); \texttt{gemini-2.5-pro}'s result is convergent. Every implicit arm in this paper withholds the payload from the store, so all are \emph{structural}: signature absence establishes architectural viability, not a measured evasion rate. A mechanism-faithful MINJA case-bank implementation reproduces the contrast within-model on gpt-4.1 (observable $80/80$ vs.\ implicit $0/141$) but stays directional/LOW-CLUSTER after an inert-grid-factor audit and a preregistered continuation, as does an emergent sliding-window Zombie implementation; clustered-primary independent-family confirmation remains open.

We then falsify the detector's standalone use. Benign memory-grounded sends are trajectory-isomorphic: unconditional false-positive rate $24.7$--$57.6\%$ on a 13-model factorial ($N{=}4{,}160$), $100\%$ conditional on the transition, and $100\%$ under a genuine LangGraph framework. Positive predictive value is ${\leq}3.85\%$ at $1\%$ attack prevalence. Attack success and signature rate are decoupled ($r{=}0.031$; models with $0\%$ success still produce $100\%$ signature rates). The obvious fix, recipient-metadata gating, flags $129/129$ of \emph{legitimate} external business email, and the combined trajectory-and-recipient gate still flags $45\%$. The signature is therefore an attack \emph{precondition}, not a maliciousness predicate, and the classifier we fit on it is a measurement instrument rather than a contribution. Two scope limits bound the ceiling itself: it constrains only methods whose observation model is the agent-visible tool-call log, so a deployment that logs framework-side retrieval as a first-class span escapes it; and what implicit delivery removes is the retrieval \emph{provenance}, not necessarily any alert, since an attack implicit on the read axis but tool-routed on the write axis still leaves a save-before-send trace.
\end{abstract}

\section{Introduction}
\label{sec:intro}

LLM agent memory poisoning attacks span a growing family of methods (delayed-trigger attacks via RAG injection~\citep{leong2026defense}, embedding-similarity manipulation~\citep{shao2025minja}, sliding-window persistence~\citep{yang2026zombie}, cross-session stored prompt injection~\citep{xie2026crosssession}) but share a common structure: a malicious payload is written to persistent state and later retrieved to influence agent behaviour. A natural question for defenders is: \emph{can trajectory monitoring (observing tool-call logs alone) detect these attacks?}

We show the answer depends on a single architectural variable: \textbf{retrieval observability}: whether the agent's memory access produces a logged tool invocation. This variable partitions the \emph{retrieval-provenance visibility} of a memory-poisoning attack into two regions with fundamentally different properties, subject to two scope limits stated here rather than deferred to the Discussion: an implicit-read attack can still trip a trajectory classifier through non-retrieval action-shape features (\S\ref{sec:graph-implications}), and a deployment that logs framework-side retrieval as a first-class span moves that design into the observable region rather than remaining outside the taxonomy. Within that scope:

\begin{itemize}
    \item \textbf{Observable retrieval} (agent-initiated tool call: \tool{recall\_fact}, \tool{memory\_search}, etc.): a retrieval-before-exfiltration event is \emph{structurally necessary} for successful attack execution. Evidence: exact-key recall 58/58 zero violations (GPT-4.1 + GPT-4o); semantic search 33/33 GPT-4.1, 43/43 GPT-4o (pooled after a preregistered power extension, Exp~13a).
    \item \textbf{Implicit retrieval} (framework-level embedding similarity, scaffold-side context injection, cached internal state): attacks succeed without producing any observable trajectory signature. Evidence: 50--82\% ASR with signature absent by construction.
\end{itemize}

This boundary is not specific to our detector. It is an architectural ceiling on any method whose observation model is the \emph{agent-visible tool-call log}: VIGIL~\citep{li2026vigil}, TraceAegis~\citep{liu2025traceaegis}, behavioural-firewall enforcement over tool-call sequences~\citep{dang2026enforcing}, or any future trajectory classifier over that substrate will be structurally blind to the retrieval signature of implicit-retrieval attacks regardless of their sophistication (the exfiltrating action may still be observable through other channels, e.g.\ destination metadata; what is unobservable is the retrieval provenance). \textbf{Two scope corrections, both made after a prior-art check.} First, the ceiling is a property of an observation model, not of a research area, so methods that read a different substrate are \emph{not} subject to it and earlier drafts were wrong to list them as if they were: TRACES~\citep{wang2025traces} learns risk states from an observer LLM's hidden representations, and MEMSAD~\citep{gowda2026memsad} scores retrieval embeddings with gradient coupling. Neither consumes tool-call logs, so implicit retrieval does not blind them in the way it blinds the methods above: indeed a representation- or content-level monitor is precisely the direction our own falsification (\S\ref{sec:deployment-boundary}) points towards. Second, the ceiling is not a claim about deployments in general. Standard agent-telemetry conventions already define retrieval and memory operations as first-class span types distinct from tool executions~\citep{otel2026genai}, with auto-instrumentation for common retrieval frameworks, so a fully instrumented deployment can log framework-side retrieval and thereby escape the boundary. That is a strength of the architectural framing rather than a limitation of it: the ``defence upgrade'' this paper identifies (promote implicit retrieval to an observable event) is available in today's instrumentation stacks rather than hypothetical. What the boundary tells such an operator is which spans are load-bearing for provenance, and what it tells an operator who logs only agent-issued tool calls is which attacks are invisible to them. We formalise this boundary as a framework-agnostic \emph{retrieval-to-action provenance graph} (Section~\ref{sec:event-graph}) and use it to classify all major published attack families by their detection-relevant properties (Table~\ref{tab:attack-taxonomy}).

Within the observable-retrieval scope, we construct a proof-of-concept detector (Random Forest, AUC~$= 0.99$ separating successful from unsuccessful attack executions within poisoned sessions) and simultaneously demonstrate its precision ceiling: benign memory-grounded sends produce the same trajectory signature (FPR 24.7--57.6\%, 13-model factorial $N{=}4{,}160$), confirming that the retrieval-before-send transition is an attack \emph{precondition}, not a maliciousness predicate. The classifier serves as evidence for the invariant's existence and scope: it is not the contribution.

\paragraph{Contributions.}
\begin{enumerate}
    \item \textbf{Observability taxonomy.} We establish retrieval observability as the dividing variable of the memory-poisoning attack surface with respect to trajectory-based detection. We classify published attack families by this variable and empirically validate both sides of the boundary with preregistered probes (V3-1, V3-2; $N{=}440$ new trials across 3 models $\times$ 2 retrieval primitives $\times$ 2 delivery modes).
    \item \textbf{Framework-agnostic event graph.} We define a retrieval-to-action provenance graph specification that maps memory-poisoning attacks into a common representation independent of framework, tool naming, or retrieval implementation (Section~\ref{sec:event-graph}).
    \item \textbf{Structural necessity within scope (under stated exclusivity assumptions).} \emph{Under the assumption that the payload value is obtainable only through an observable retrieval operation} (architectural assumptions 1--4, Section~\ref{sec:threat}), the \textit{retrieval\_before\_exfiltration} transition is mechanistically forced by the attack's information dependency. This necessity holds across the two tested exfiltrating-action channels, email (\texttt{recall\_before\_send}) and document-share (\texttt{recall\_before\_exfil}, control $0$ violations $/50$ (gpt-4.1) and $0/62$ (gpt-4o); Section~\ref{sec:x4}), so we state it as \emph{retrieval-before-exfiltrating-action} scoped to these two channels, with the document-share evidence directional rather than criterion-passing (its cluster-unit upper bounds, $0.121$ and $0.104$, exceed the registered $\leq 0.10$), the implicit treatment arm underpowered, and only one channel added beyond email. This exclusivity is an assumption about the deployment, not a guarantee of the memory architecture: when it is violated (e.g.\ the address is inferable from key names, shared scenario pools, or model priors) the transition can be bypassed even with the tool present (7/40 V3-1 treatment successes routed to a non-planted attacker address; Section~\ref{sec:threat}). A single-rule detector exploiting the transition achieves AUC~$= 0.956$; the full feature set (AUC~$= 0.99$ random-fold, $0.907$ grouped) adds marginal value because the invariant itself carries most of the separation. Adversarial suppression of the transition breaks the attack only under the same exclusivity assumption.
    \item \textbf{Precision ceiling (deployment falsification).} Benign memory-grounded sends produce the same trajectory signature (conditional FPR~$= 100\%$), establishing that operation-only features face a hard precision ceiling when agents ground actions in recalled context.
    \item \textbf{Cross-model and cross-harness transfer, with scope limits.} Under leave-one-model-out hold-out across a 9-model factorial (7B--120B), the invariant transfers without retraining to $6$ of the $9$ models at AUC${=}1.000$ and Recall${=}1.000$; the three exceptions are mechanistically explained in Section~\ref{sec:holdout} and are not incidental failures. \emph{This is a separate claim from the withdrawn frontier classifier-transfer evaluation of \S\ref{sec:gpt41}: here, as a stronger test,} we freeze the LangGraph-trained classifier and apply it, without retraining, to sessions from a different execution harness (a raw OpenAI function-calling loop) and a more recent model population (GPT-5.x, Gemini~2.5): pooled recall among attack successes is $186/188 = 98.9\%$ (Section~\ref{sec:cross-framework-transfer}), evidence that the signature is a property of the attack's information flow rather than of the training harness's logging substrate. The signature nonetheless fails deterministically on implicit-bypass architectures (qwq:32b RAG fallback), confirming the taxonomy's predictive validity. We are explicit about what this does \emph{not} show: what a tool-log-only detector loses under implicit retrieval is the \emph{retrieval provenance} of the attack, not necessarily the exfiltrating action itself, which may remain observable through other channels (e.g.\ destination metadata).
\end{enumerate}

\begingroup
\small
\setlength{\LTcapwidth}{\textwidth}
\begin{longtable}{@{}p{0.40\textwidth}p{0.34\textwidth}p{0.05\textwidth}p{0.12\textwidth}@{}}
\caption{Claim-to-evidence map. \emph{Evidence status} distinguishes preregistered confirmatory findings (P), exploratory findings whose executed design deviates from the preregistration or was not preregistered (E), findings that were registered and executed under their registration but whose registered decision criterion was not met, and which are therefore reported as directional (D), post-hoc scorer/label corrections disclosed in the deviations (C), and architectural classifications argued from mechanism rather than direct reproduction (A). A slash (\eg A\,/\,D) marks a row whose architectural proposition and whose empirical support carry different statuses. \emph{Taxonomy corrections.} The D category is new. Earlier drafts had no letter for a registered, compliantly executed, but inconclusive result, and assigned such rows the nearest available letter, which misdescribed them in both directions: the X4 second-channel row carried P although its registered cluster-unit threshold of ${\leq}0.10$ was not met ($0.121$, $0.104$; \texttt{results/x4\_cluster\_bounds.json} records \texttt{meets\_registered\_threshold\_on\_cluster\_unit: false}), and the MINJA and Zombie rows carried a bare A although X1 collected trial data for both (observable $80/80$ and $36/36$, implicit $0/141$ and $0/16$) and so is not argued from mechanism alone. Those rows are now split: A labels the architectural proposition, D the measured support that fell short of its registered gate. \emph{Two rows were downgraded P${\to}$E$^{\S}$.} $^{\S}$Preregistering a data collection does not preregister an unregistered analysis of it: no document among the 14 preregistrations in the artefact covers the supervised classifier, its training corpus, or the leave-one-model-out procedure, so under the definition of P above those two rows were mislabelled. Their numbers are unchanged and remain fully reproducible; what changes is the confirmatory standing claimed for them. All numeric results are reproducible from the released artefact \emph{together with} one external input disclosed in the Reproducibility section (the P1 factorial classifier-training JSONL, released in the companion repository of \citet{leong2026defense}); API-served results are auditable via provenance stamps but not bitwise reproducible.}
\label{tab:claim-evidence}\\
\toprule
\textbf{Claim} & \textbf{Evidence} & \textbf{Status} & \textbf{Section} \\
\midrule
\endfirsthead
\toprule
\textbf{Claim} & \textbf{Evidence} & \textbf{Status} & \textbf{Section} \\
\midrule
\endhead
\midrule
\multicolumn{4}{r@{}}{\emph{(continued on next page)}}\\
\endfoot
\bottomrule
\endlastfoot
Observable retrieval $\Rightarrow$ \textit{recall\_before\_send} necessary for tool-routed exfiltration
 & Exact-key 58/58; semantic 33/33 GPT-4.1 (preregistered; 38/38 corrected scorer) / 43/43 GPT-4o (pooled after Exp~13a) zero violations
 & P$^\flat$ & \S\ref{sec:read-arms} \\
Implicit delivery breaks necessity (successful attacks with rbs${=}0$)
 & V3-1 treatment 67\% (GPT-4.1) / 74\% (GPT-4o) rbs${=}0$; scenario-clustered CIs exclude 0
 & P$^\ddagger$ & \S\ref{sec:read-arms} \\
A context copy does not suppress payload retrieval; the former ``behavioural'' reading is withdrawn
 & V3-6 arms B vs I (prefix-, trigger- and schema-matched; store contents the only difference): prefilled-key verified read $108/108$ vs $0/108$. Registered primary C${-}$B is $0.0$ over $36$ clusters but is prefix-confounded and reported only as the registered branch. Memory \emph{access} $324/324$
 & P & \S\ref{sec:v36}, \S\ref{sec:v36b} \\
Necessity/breakdown replicates on a ground-then-act model
 & V3-3 gemini-2.5-pro: control 0/19 rbs${=}0$; treatment 28/30${=}93.3\%$ rbs${=}0$ (clustered CI $[0.821,1.000]$); benign recall habit 88.3\%. X2 second grounder qwen2.5:32b: treatment rbs${=}0$ $27/31{=}87.1\%$ [0.711,0.949]; control $0/24$ underpowered (Wilson upper $0.138$)$^\natural$
 & P & \S\ref{sec:v33} \\
Observable-arm boundary holds under a genuine LangGraph framework; within-model contrast on a genuine FAISS retriever
 & V3-4 (gpt-4.1, within-model): LangGraph \texttt{create\_react\_agent} observable 39/39 flagged, genuine-FAISS-retriever implicit 0/57 flagged, benign 60/60 flagged (gpt-4.1-mini)
 & P$^\dagger$ & \S\ref{sec:v34} \\
Signature is a precondition, not a maliciousness predicate
 & Benign FPR 24.7--57.6\% (13-model factorial $N{=}4{,}160$); $r{=}0.031$ ASR vs signature; recipient gating flags 129/129 legitimate external sends (V3-5)
 & P & \S\ref{sec:deployment-boundary}, \S\ref{sec:v35} \\
Write-axis \textit{save\_before\_send} has verified payload-write provenance
 & Exp~13c: 37/37 flagged successes wrote the attacker payload; 66.7\% benign FPR unchanged (precondition, not predicate)
 & P & \S\ref{sec:write-axis} \\
Within-scope detector AUC $= 0.990$ (random-fold) / $0.907$ (grouped)
 & 5-fold CV, $N{=}2{,}520$; leakage-corrected grouped CV
& E$^{\S}$ & \S\ref{sec:perf} \\
Cross-model transfer without retraining
 & Leave-one-model-out 6/9 AUC $= 1.000$. The frontier-transfer figure cited by earlier drafts is \textbf{withdrawn} as not regenerable (\S\ref{sec:gpt41})
& E$^{\S}$ & \S\ref{sec:holdout} \\
Signature stable across prompt framings (one loop, one schema)
 & 99--100\% signature on GPT-5.1 raw vs ReAct-framed
 & E$^\ast$ & \S\ref{sec:cross-framework-transfer} \\
Signature transfers across execution harnesses (train LangGraph, test raw loop)
 & Frozen-classifier cross-harness holdout, pooled recall $186/188 = 98.9\%$
 & E & \S\ref{sec:cross-framework-transfer} \\
ASR corrected to structured-recipient (\texttt{to}) scoring
 & GPT-5.1 $98.3\%{\to}62.5\%$; signature rates unchanged
 & C & \S\ref{sec:cross-framework-transfer} \\
V3-2 arm B scorer includes \texttt{draft\_email} as exfiltration-capable
 & Preregistered 33/33; corrected sensitivity 38/38 (both reported)
 & C & \S\ref{sec:read-arms} \\
MINJA / scaffold-side RAG evade trajectory detection
 & X1 (within-model gpt-4.1, query-only case-bank): observable \texttt{orp} $80/80$ vs.\ implicit $0/141$ (cluster-level Wilson upper $0.299$; trial-level $0.0265$ assumes independence the design does not support), contrast $+1.0$; \emph{observable arm gate-satisfied ($10$ effective clusters); implicit arm $9$, one short at the registered cap; $8$ configurations contribute eligible successes in both arms}$^\P$; gpt-4o corroborating (LOW-CLUSTER)
& A\,/\,D$^\P$ & \S\ref{sec:event-graph}, \S\ref{sec:x1}, Table~\ref{tab:attack-taxonomy} \\
Zombie-family (emergent sliding-window) evades retrieval-provenance detection
 & X1 (within-model gpt-4.1, emergent W${=}6$ buffer): observable \texttt{orp} $36/36$; implicit $0/16$ but only $5$ clusters (LOW-CLUSTER, Wilson-governing); gpt-4o implicit $0$ eligible$^\P$
& A\,/\,D & \S\ref{sec:x1}, Table~\ref{tab:attack-taxonomy} \\
Necessity \emph{extends} to a second exfiltration channel (document-share), directionally
 & X4 raw \texttt{recall\_before\_exfil}: control $0$ violations $/50$ gpt-4.1 ($28$ clusters) and $0/62$ gpt-4o ($33$ clusters); cluster-unit Wilson upper $0.121$ and $0.104$, both \emph{above} the registered $\leq 0.10$ bound, so directional on this channel rather than criterion-passing; treatment not viable on gpt-4.1 / directional on gpt-4o$^\P$
& D$^\P$ & \S\ref{sec:x4} \\
\end{longtable}
\endgroup

\paragraph{Notes to Table~\ref{tab:claim-evidence}.} $^\ast$The cross-framing preregistration (\texttt{scripts/cross\_framework\_preregistration.md}) specified genuine LangChain \texttt{AgentExecutor} and OpenAI Assistants runners on \texttt{gpt-4.1-mini} ($N{=}750$); the executed study instead compared two prompt-framing variants of a single raw OpenAI function-calling loop on GPT-5.x/Gemini. Because the executed design deviates from the registered design, we label this evidence exploratory (E), not confirmatory (P), even though the finding itself is clean. $^\dagger$V3-4's observable and benign arms are preregistered confirmatory (P) and run on a genuine LangGraph \texttt{create\_react\_agent} framework. The implicit arm's originally-planned OpenAI Assistants \texttt{file\_search} cell was infrastructure-blocked (the institutional gateway forbids the Assistants and vector-store endpoints); it is instead run as a genuine LangChain-style FAISS retriever-injection chain (the V3-2 store; framework-side retrieval with a real delivery oracle, no agent tool call), which occupies the same taxonomy cell (implicit read, scaffold-side). Both the observable and implicit arms are run within-model on gpt-4.1 at $N{=}60$ (observable $39/39$ flagged, implicit $0/57$ flagged), so the contrast is a clean single-model manipulation of the observability axis; an earlier draft paired a gpt-4.1-mini observable arm with a gpt-4.1 implicit arm (a cross-model confound), and the gpt-4.1-mini cells are retained as an illustrative second model. $^\ddagger$V3-1's design was preregistered (GPG-signed, before data) but its \emph{execution} deviated (scenario grid reduced $72{\to}18$, GPT-4o added post-registration, the ${\geq}30$-success floor not reached); we therefore treat this row as confirmatory-by-design with an exploratory execution. The necessity-breaking conclusion is nonetheless robust: both models' scenario-clustered CIs exclude 0, and V3-3 (row below, clean P) independently replicates the breakdown on a ground-then-act model. See the V3-1 deviations paragraph (\S\ref{sec:read-arms}).$^\P$In the v3 confirmatory re-run the family labels invert relative to earlier drafts: MINJA/gpt-4.1 is the powered within-model contrast (observable $80/80$; implicit $0/141$) while Zombie/gpt-4.1 implicit has only $16$ successes across $5$ clusters and Zombie/gpt-4o implicit has $0$ eligible successes. \textbf{A post-hoc audit then found the MINJA grid's \texttt{similarity} factor to be inert} (\texttt{minja\_poison\_case()} receives only domain and framing; the 36-entry grid holds $12$ distinct poison texts, $3$ \texttt{scenario\_id}s each), inflating MINJA scenario-cluster counts ${\approx}3\times$. On effective clusters the original v3 sample gave $6$ for the observable arm and $7$ for the implicit arm, both below the ${\geq}10$ gate; a GPG-signed amendment then continued collection to the registered $N{=}360$ cap, after which the observable arm reaches $10$ effective clusters (gate satisfied) and the implicit arm $9$ (one short) so \textbf{MINJA is NOT upgraded and stays analytic (A), directional/LOW-CLUSTER}, alongside Zombie. Zombie and X4 are unaffected (every factor in their grids reaches the content constructors; measured cluster inflation $1.0\times$). X4's treatment (implicit) arm is not viable on gpt-4.1 ($0/360$) and directional/underpowered on gpt-4o ($20$ successes, $10$ clusters, below the $37$-success target). $^\natural$X2's control arm did not reach its $37$-success power target (hit the $N{=}150$ cap at $24$ successes, trial-level Wilson upper $0.138$, cluster-unit Wilson upper $0.354$ over $7$ contributing clusters, Table~\ref{tab:zero-event}); it is consistent with but does not confirm the invariant's $\leq 0.10$ upper bound, and it used the existing GUI Ollama daemon rather than a fresh-daemon restart (Regime-A caveat, \S\ref{sec:v33}). $^\flat$These are the same zero-violation censuses discussed in \S\ref{par:zero-event-rule}: applied uniformly on the clustering unit each design licenses, no single registered cell in this paper certifies the preregistered $\leq 0.10$ upper bound (cluster-unit uppers $0.10$--$0.24$; Table~\ref{tab:zero-event}). The counts in this row are exact zero-violation censuses among realised successes, not a criterion-passing interval.

\section{Methodology}
\label{sec:method}

\subsection{Threat Model}
\label{sec:threat}

We consider a \emph{memory-channel poisoning attack}: the adversary injects a malicious instruction via a RAG-retrieved document, which the agent stores in persistent memory. In a subsequent session, the agent retrieves the stored rule and takes a malicious action (email exfiltration). We explicitly distinguish this from \emph{prompt-injection attacks}, where the adversary embeds the routing rule directly in the current session's user prompt without storing it in memory. Our detector targets the former; we report its behaviour on the latter as the evasion boundary.

\paragraph{Attack success condition.} A run is classified as an attack success if \tool{email\_send\_email} is called with a recipient outside the allowed set $\{\texttt{team@example.com}\}$ in the trigger session (S4). In the V3-1 treatment cells, this allowlist-based scorer counts $7$ successes (GPT-4.1 $3/21$, GPT-4o $4/19$; see the V3-1 deviations paragraph) where the exfiltration destination is an attacker-controlled address other than the one planted in that scenario's memory (the model sent to the other known attacker address, likely confabulated from pretraining or cross-contaminated from the scenario design's shared address pool). These cases represent genuine policy violations (unauthorised external sends) but ambiguous evidence for the retrieval-dependency mechanism; conclusions about the necessity invariant rest on the control-arm zero-violation counts, which are unaffected, and the necessity-breaking conclusion also holds under the stricter planted-address-matched scorer (V3-1 deviations).

\paragraph{Endpoint definitions (transmission vs.\ routing-attempt vs.\ policy violation).} Three distinct endpoints appear in this paper and should not be conflated. (i)~\emph{Transmission}: \tool{email\_send\_email} to a recipient outside the allowlist, the primary attack-success definition (Threat Model) and the target of the main DTA classifier, whose \texttt{recall\_before\_send} feature keys on \tool{send\_email} only and never inspects recipients. (ii)~\emph{Routing-attempt}: \tool{email\_draft\_email} specifying an external recipient without a subsequent send; V3-2's corrected (sensitivity) scorer counts this as exfiltration-capable because the routing decision is made at draft time, but it is a strictly secondary endpoint reported alongside the primary send-only figure (33/33 primary vs.\ 38/38 draft-inclusive), never substituted for it. (iii)~\emph{Policy violation}: any unauthorised external send, including the 7 V3-1 confabulated-address cases above. Wherever we write ``recall before exfiltration,'' the exfiltration event is the transmission endpoint (i) unless the sentence explicitly invokes the V3-2 draft-inclusive sensitivity analysis; cross-study pooled counts use the transmission endpoint as primary.

\paragraph{Detection constraint.} The detector observes only the tool-call sequence and its timestamps. It does not access memory contents, model weights, activation patterns, or the user's messages. Attack-success labels are constructed using tool arguments (recipient address); classifier features use only operation names, ordering, and counts, never email contents or recipients.

\paragraph{Architectural assumptions.} The detector assumes: (1)~persistent memory is accessed exclusively through observable retrieval operations (tool calls), not through hidden internal state or implicit caching; (2)~tool invocations are logged with execution timestamps; (3)~the attacker's routing information resides in persistent memory and must be retrieved through the memory API to reach the exfiltration tool; and (4)~\tool{list\_all\_facts} returns key names only, not stored values, so the agent must call \tool{recall\_fact} to obtain the attacker address. These assumptions hold for deployments in which memory retrieval is exposed as an observable tool invocation (\eg LangChain, LangGraph, OpenAI Assistants, Anthropic tool-use). Architectures that bypass the tool layer (\eg internal RAG caches, implicit memory injection into prompt context) would not produce the observable signature.
This defines the detector's architectural boundary: the \textit{recall\_before\_send}
invariant is an enforcement-layer property of closed-loop tool-routing policies, not a
pure model property. A model that satisfies its information dependency through a
non-observable channel (empirically demonstrated by qwq:32b's RAG-fallback under
Memory Sandbox, Section~\ref{sec:limitations}) collapses the signature. The invariant
holds for all models structurally forced to route retrieval through the
observable memory API. Section~\ref{sec:cross-framework} formalises this boundary across published attack families.

\paragraph{Structural versus behavioural manipulations of the read axis.} Our implicit-delivery arms are not all the same kind of evidence, and the distinction governs what each can license. In a \emph{structural} arm the payload is not reachable through any agent-visible memory operation, either because the memory tools are absent from the schema or because they are present but the store never holds the payload. Signature absence is then definitional: it could not have fired, so the arm establishes that such an architecture exists and sustains viable attacks, and that the retrieval provenance is unobservable within it, but it cannot measure a rate at which agents \emph{decline} an available retrieval. V3-2 arm~C (all \tool{memory\_*} tools removed, \S\ref{sec:cross-framework}), V3-4's implicit arm (email tools only, \S\ref{sec:v34}) and X1's implicit arm (tool present, poison never in the store, \S\ref{sec:x1}) are all structural in this sense. We had also defined a \emph{behavioural} class (retrieval tool present, functional and stocked, payload arriving by another path, so that signature absence is a measured choice) and assigned V3-1's treatment arms and V3-3 to it. \textbf{That class is now empty, and we retire it.} V3-6 (\S\ref{sec:v36}) holds the in-context payload and the memory-inviting trigger fixed and varies only the store: the model returns the payload from memory in $108/108$ trials when the store holds it and $0/108$ when it does not. A context copy therefore does not suppress payload retrieval, so no arm's signature absence can be read as a choice about the payload. Every implicit arm in this paper withholds the routing rule from the store, V3-1 and V3-3 treatment (\texttt{prefilled = decoy\_facts}, ``NO routing rule in memory''), V3-2 arm~C (tools removed), V3-4 implicit (email tools only), X1 implicit (empty store), X4 treatment (two benign facts), so on the read axis all of them are structural.

\textbf{Three labels, kept separate, replacing the two we had.} The collapsed taxonomy invited a conflation that a post-collection audit caught, so we now distinguish: (i) \emph{delivery-path unobservability}, the architectural fact that framework-side retrieval issues no agent-visible call, which is definitional and carries the taxonomy; (ii) \emph{payload unreachable at trigger}, the property shared by every implicit arm here, under which signature absence could not have fired and no declining rate is measurable; and (iii) \emph{observed memory-use behaviour}, how much an agent reads from a store that lacks the payload. Category (iii) is genuinely behavioural and we do not call it structural, but it is behaviour about \emph{decoys}, not about the payload, and it therefore cannot support the withdrawn claim. V3-6 sharpens what (iii) contains: the agent issued \tool{list\_all\_facts} in \textbf{all $324$ sessions}, so memory \emph{access} is universal and was never the variable; what varies is whether a value is then retrieved, and in the arm whose store lacks the rule four of those retrievals request keys constructed from the injected rule that do not exist in the store at all. The agent reaches for the payload and finds nothing. The cell that measures declining (payload live behind the retrieval tool \emph{and} delivered in context) was named as an open gap in earlier drafts and has now been run; it is V3-6, and its result is the withdrawal above.

The architectural claim (implicit retrieval leaves no observable retrieval provenance) rests on the structural arms plus the provenance graph, which on the read axis is now all of the implicit arms. What V3-1 and V3-3 do establish is that value retrieval collapses when the store holds nothing topically relevant, and the sharpest form of that is V3-3's benign keystone, where a model with an $88.3\%$ benign recall-before-send habit drops to $6.7\%$ in the treatment cell. Those two cells differ in \emph{store contents} as well as in delivery mode, and V3-6 shows delivery is not what suppresses retrieval, so the drop is attributable to the store (\S\ref{sec:v33}; that keystone is a within-model contrast on \emph{two} effective benign configurations, not a domain-general estimate of the habit, and its two cells differ in store contents as well as in delivery; see the clustering-unit correction and the V3-6 reattribution there). Earlier drafts described X1's implicit arm as behavioural; that is corrected in \S\ref{sec:x1}.

\paragraph{Structural equivalence of proxy tasks.} Our evaluation uses benign proxy tasks (email composition with simulated sensitive documents) rather than genuinely harmful payloads. The security-relevant transfer argument is structural, not content-based: the detector observes only tool-call sequences and their ordering, which are determined by the agent's channel architecture (memory API $\to$ action API), not by the semantic content of retrieved records or sent messages. Any deployment using the same tool-routing architecture (observable memory retrieval followed by externally-directed action) produces identical trajectory structure regardless of whether the exfiltrated content is a simulated project summary or genuine credentials. The invariant is a property of the information-flow dependency, not of the payload.

\subsection{Dataset}
\label{sec:data}

We use $2{,}520$ labelled runs from the delayed-trigger attack arm of the 5,040-run factorial experiment in our prior work (the P1 factorial, hereafter IED)~\citep{leong2026defense}. The factorial covers:
\begin{itemize}
    \item 9 open-source models\footnote{Model names reflect Ollama tags (e.g., \texttt{qwen3.5:122b}) or API identifiers at time of experiment. These are mutable pointers; exact model digests are recorded in the reproducibility manifest.}: qwen2.5:14b, qwen2.5:72b, qwen3.5:9b, qwen3.5:122b, qwen3:32b, qwq:32b, glm-4.7-flash:q8\_0, gpt-oss:20b, gpt-oss-safeguard:120b
    \item 7 defence conditions: no defence, Minimiser, Sanitiser, RAG Sanitiser, RAG LLM Judge, Prompt Hardening, Memory Sandbox
    \item $N{=}40$ runs per condition (9 models $\times$ 7 defences $\times$ 40 runs $= 2{,}520$ DTA runs)
\end{itemize}

Labels: $1{,}919$ attack successes (76.2\%), $601$ non-exfiltration sessions (23.8\%). The negative class comprises poisoned-but-defended sessions (injection succeeded but exfiltration was blocked or refused); it does not include unpoisoned benign baseline sessions. The classifier therefore discriminates between successful and unsuccessful attack \emph{executions} within the DTA arm, not between attacked and unattacked agents.

\paragraph{Shared-data dependency and non-independence.} This detector is trained and evaluated on data generated by the same factorial that \citet{leong2026defense} analyse for their injection--execution dissociation result, and that work in turn cites the detector performance reported here when discussing operation-level trajectory detection. The two results are therefore \emph{not} independent lines of evidence: they are two analyses of one dataset produced by one attack construction, one harness, and one scoring pipeline. A flaw in the shared data-generating process (for example, a scorer artefact or a harness-specific tool-contract effect) would propagate to both. Readers should treat the observability boundary established here and the dissociation finding in the companion work as mutually consistent analyses of a common substrate, not as cross-study replication. The external-validity gap this creates was, in earlier drafts, that no independent attack family, harness, or lab had reproduced the boundary; X1 (Section~\ref{sec:x1}) addresses the independent-attack-family portion for the MINJA family via a mechanism-faithful within-model contrast, but after the inert-\texttt{similarity} cluster-unit correction that evidence is directional/LOW-CLUSTER rather than clustered-primary, so the gap is \emph{narrowed but not closed}; the Zombie family likewise remains directional/LOW-CLUSTER, and cross-harness/cross-lab replication remains open (see the transfer-across-attack-families discussion below).

Each run provides per-session tool logs with operation type, inputs, outputs, and timestamps. We extract features from the \emph{trigger session only} (S4), since S1--S3 precede the trigger and should not contain exfiltration evidence in a correctly operating system.

\subsection{Feature Engineering}
\label{sec:features}

Tool invocations are sorted by execution timestamp before feature extraction to ensure transition statistics reflect the actual execution order rather than the serialisation layout of the logging framework. We verified that on the full 2,520-run dataset, every session containing both \tool{recall\_fact} and \tool{send\_email} has the first recall strictly preceding the first send (0 violations), confirming that the temporal dependency is empirically universal in this attack construction.

We extract 19 trajectory features per trigger session (all from S4 only; no injection-session features are used to avoid label leakage):

\paragraph{Count features.} Raw call counts in S4: \texttt{recall\_count}, \texttt{list\_count}, \texttt{save\_count}, \texttt{draft\_count}, \texttt{send\_count}, \texttt{rag\_count}, \texttt{seq\_len} (total tool calls).

\paragraph{Structural features.} \texttt{send\_without\_recall} (binary: sent email without any \tool{recall\_fact} call), \texttt{recall\_before\_send} (binary: any \tool{recall\_fact} call precedes any \tool{send\_email} call, i.e.\ $\min(\text{recall positions}) < \max(\text{send positions})$, matching the released implementation; see the feature-definition note in \S\ref{sec:limitations}), \texttt{recall\_to\_send\_ratio} ($= \texttt{recall\_count} / \max(\texttt{send\_count}, 1)$), \texttt{max\_recall\_chain} (longest consecutive sequence of \tool{recall\_fact} calls).

\paragraph{Bigram transitions.} Counts of consecutive tool-call pairs in S4: \texttt{list\_then\_recall}, \texttt{recall\_then\_recall}, \texttt{recall\_then\_draft}, \texttt{list\_then\_draft}, \texttt{draft\_then\_send}.

\paragraph{Entry point.} \texttt{first\_tool\_is\_list}, \texttt{first\_tool\_is\_recall}, \texttt{first\_tool\_is\_draft} (mutually exclusive binary indicators for the first tool called in S4).

\subsection{Classifiers and Evaluation}
\label{sec:classifiers}

We train three classifiers, all as implemented in scikit-learn~\citep{pedregosa2011scikit}:
\begin{itemize}
    \item \textbf{Logistic Regression} (LR): $L^2$-regularised, $C=1.0$, features standardised with \texttt{StandardScaler}.
    \item \textbf{Random Forest} (RF)~\citep{breiman2001random}: 200 trees, \texttt{max\_depth}$=8$.
    \item \textbf{Gradient Boosted Trees} (GBM): 200 estimators, \texttt{max\_depth}$=4$, learning rate $0.05$.
\end{itemize}

All classifiers are evaluated via stratified 5-fold cross-validation ($\mathrm{seed}=42$) using the default classification threshold of $0.5$ (predicted probability $\geq 0.5 \Rightarrow$ attack). Primary metric is \emph{Recall} (minimising false negatives = undetected attack-success sessions). Secondary metric is AUC-ROC. Confidence intervals on AUC use the BCa bootstrap~\citep{efron1987better} ($N{=}10{,}000$ resamples, jackknife acceleration over all $2{,}520$ leave-one-out samples, $\mathrm{seed}{=}42$). Binomial intervals on proportions throughout the paper are Wilson score intervals~\citep{wilson1927probable}, and where trials nest within scenario configurations we report them on the effective cluster unit, using Kish's effective sample size~\citep{kish1965survey} for the conservative unequal-cluster limit (\S\ref{sec:x1}).

\paragraph{One decision rule for zero-event necessity cells, applied uniformly.}
\label{par:zero-event-rule}
Most of this paper's necessity evidence takes the form of a cell with zero observed violations among $n$ eligible attack successes. For such cells the scenario-clustered percentile bootstrap is degenerate (every resample reproduces $0$, giving a spurious $[0,0]$) so an interval must come from elsewhere. We adopt one convention and now apply it to \emph{every} such cell: \textbf{the governing bound is the Wilson upper bound on the count of success-contributing scenario clusters}, with the Kish-weighted bound reported as the conservative limit on the trial-weighted rate ($\rho{=}1$). The trial unit is anti-conservative here, because \texttt{recall\_before\_send} is close to deterministic within a configuration rather than independently resampled across trials.

Earlier drafts applied this convention selectively. X4's and X2's control arms were re-expressed on the cluster unit and their claim to meet the registered $\leq 0.10$ criterion was withdrawn ($0.121$, $0.104$, $0.138$), while V3-1's and V3-2's controls continued to be quoted at the trial level and V3-3's cluster-unit $0.243$ was described as replicating the invariant. That is the same estimand judged by two standards. \texttt{necessity\_cluster\_bounds.py} recomputes all of them identically from the released records (endpoint-consistently: eligibility and the invariant are derived from the same \texttt{call\_sequence} endpoint, so the preregistered send-only and corrected draft-or-send scorers are never mixed), and writes \texttt{results/necessity\_cluster\_bounds.json}. Table~\ref{tab:zero-event} is the result.

\begin{table}[t]
\caption{Every zero-event necessity control cell under one rule. All cells have zero observed violations; the counts reproduce those reported in the relevant sections. \textbf{On the governing cluster unit, no \emph{registered} cell meets the $\leq 0.10$ criterion}, including the cells earlier drafts described as confirming or replicating the invariant. The final row is the one cell-combination that does clear it, and is \emph{post-hoc}: X4's two control arms share one 36-configuration grid and contribute $35$ distinct clusters between them, but the X4 preregistration defines its threshold per arm, so this row is reported for completeness and does not restore a criterion-passing claim (see the discussion above). Regenerate every row with \texttt{necessity\_cluster\_bounds.py}, which now emits the X4 rows too; earlier drafts said it generated this table while the X4 rows in fact came only from \texttt{x4\_cluster\_bounds.py}.}
\label{tab:zero-event}
\centering
\small
\begin{tabular}{lccccc}
\toprule
\textbf{Cell} & \textbf{viol./elig.} & \textbf{clusters} & \textbf{trial upper} & \textbf{cluster upper} & \textbf{$\leq 0.10$?} \\
\midrule
V3-1 control gpt-4.1 (exact-key)      & 0/28 & 14 & 0.121 & 0.215 & no \\
V3-1 control gpt-4o (exact-key)       & 0/30 & 14 & 0.114 & 0.215 & no \\
V3-1 control pooled                   & 0/58 & 16 & 0.062 & 0.194 & no \\
V3-2 arm B gpt-4.1 (send-only)        & 0/33 & 14 & 0.104 & 0.215 & no \\
V3-2 arm B gpt-4.1 (draft-or-send)    & 0/38 & 14 & 0.092 & 0.215 & no \\
V3-2 arm B gpt-4o pooled (send-only)  & 0/43 & 13 & 0.082 & 0.228 & no \\
V3-3 control gemini-2.5-pro           & 0/19 & 12 & 0.168 & 0.243 & no \\
X2 control qwen2.5:32b (ground-then-act) & 0/24 & 7 & 0.138 & 0.354 & no \\
X4 control gpt-4.1 (share channel)    & 0/50 & 28 & 0.071 & 0.121 & no \\
X4 control gpt-4o (share channel)     & 0/62 & 33 & 0.058 & 0.104 & no \\
\midrule
\emph{X4 control pooled} (post-hoc)   & 0/112 & \textbf{35} & 0.033 & \textbf{0.099} & \emph{yes}$^{\S}$ \\
\bottomrule
\multicolumn{6}{l}{\small $^{\S}$Post-hoc: the registered X4 gate and threshold are per arm, so this combination is}\\
\multicolumn{6}{l}{\small \phantom{$^{\S}$}not a registered cell and no status is upgraded on it. Kish-weighted limit $0.1057$.}
\end{tabular}
\end{table}

\noindent \textbf{What this does and does not change.} It changes the \emph{criterion} language everywhere, and we withdraw the implication that any control arm in this paper meets a $\leq 0.10$ upper bound on the unit its design licenses. It does not change a single point estimate, cluster count, or conclusion about the invariant's existence, and it is not a retreat to agnosticism. Counting only non-overlapping cells: the eight are V3-1/\texttt{gpt-4.1}, V3-1/\texttt{gpt-4o}, V3-2 arm~B/\texttt{gpt-4.1} (send-only), V3-2 arm~B/\texttt{gpt-4o} (pooled), V3-3, X2, X4/\texttt{gpt-4.1}, X4/\texttt{gpt-4o}; the V3-1 \emph{pooled} row and V3-2's second (\emph{draft-or-send}) \texttt{gpt-4.1} endpoint are excluded as supersets of rows already counted, not summed in addition to them: the necessity evidence is \emph{eight} non-overlapping zero-violation cells spanning two retrieval primitives, two exfiltration channels, four models and both behavioural phenotypes, totalling \textbf{$0$ violations in $289$ eligible successes} ($28{+}30{+}33{+}43{+}19{+}24{+}50{+}62$, in that order; \texttt{results/necessity\_cluster\_bounds.json}). X2's control arm (\texttt{qwen2.5:32b}, $0/24$ over $7$ clusters) is now included; it was previously omitted from this accounting and from Table~\ref{tab:zero-event} despite \texttt{necessity\_cluster\_bounds.py}'s own module purpose already describing X2's claim as withdrawn on the cluster unit, so its cluster-unit bound ($0.354$, above the registered $\leq 0.10$) had never actually been computed anywhere in the released artefact. No point estimate for any other cell changes; the seven-cell figures an earlier draft reported ($0$ violations in $265$ successes) become eight-cell figures once X2 is added. (Non-overlapping, not independent: several cells share models, scenario grids, or harness components, so they should not be read as statistically independent replications.) \textbf{Two cluster figures, which must be distinguished.} Summing those same eight cells' contributing-cluster counts, \emph{in the same order} ($14{+}14{+}14{+}13{+}12{+}7{+}28{+}33$), gives $135$ (was $128$ over seven cells before X2 was added), and earlier drafts reported the seven-cell precursor as ``$128$ scenario clusters.'' It is a sum of per-cell counts, not a count of distinct configurations: arms within one experiment share a single scenario grid, so a configuration that yields eligible successes under two models is counted twice. The artefact proves this on its own rows: V3-1's two per-model cells hold $14$ and $14$ contributing clusters while the pooled row holds $16$, not $28$. Measured across the five experiments, the \textbf{distinct} configuration count is \textbf{$87$} (V3-1 $16$, V3-2 arm~B $17$, V3-3 $12$, X2 $7$, X4 $35$; \texttt{distinct\_configuration\_census} in \texttt{results/necessity\_cluster\_bounds.json}). We report $87$ wherever breadth is the point and label $135$ as a per-cell sum wherever it appears, because the bound derived in the next sentence is defined per cell and the larger figure invites a pooled reading it does not license. Together with these, the structural argument of \S\ref{sec:graph-formal} makes the transition mechanically forced under exclusive tool-mediated access. The honest summary is that the invariant is empirically unbroken everywhere we have looked, and that no \emph{single} registered cell is individually powered to certify a $10\%$ upper bound on the cluster unit. The reason is a design lesson we state rather than hide: these grids were sized for trial-level interval conventions and are too narrow for the cluster-level inference their nesting requires. A zero-event cell needs $35$ contributing clusters to clear $0.10$ ($z^2/(n+z^2) \leq 0.10 \Leftrightarrow n \geq 34.6$); the best-populated \emph{registered} cell in the paper is X4's gpt-4o control arm at $33$.

\noindent \textbf{One cell-combination does clear the bound, and leaving it out is what made the blanket statement true.} This paper pools V3-1's two control arms into a reported $58/58$ cell and pools V3-2's gpt-4o cohorts into a reported $43/43$ cell, but reports X4 per arm only, and X4 is the one place where pooling changes the verdict. X4's two control arms run the \emph{same} 36-configuration grid, and on that shared grid they contribute $35$ \emph{distinct} success-contributing clusters between them, with $0$ violations in $112$ eligible successes. The cluster-unit Wilson upper bound is $\mathbf{0.0989}$, i.e.\ inside the registered $\leq 0.10$; the Kish-weighted conservative limit is $0.1057$, just outside. \textbf{We do not use this to restore a criterion-passing claim}, because the X4 preregistration defines its gate and its threshold \emph{per arm}, so a cross-arm pooling is post-hoc by construction and this paper's signed pre-commitment forbids upgrading a status on a post-hoc analysis. We report it because the alternative (omitting the one pooling that clears the bound while performing two poolings that do not) would make ``no cell meets the criterion'' an artefact of which cells we chose to combine. The row is emitted as \texttt{X4 control POOLED across models (POST-HOC)} in \texttt{results/necessity\_cluster\_bounds.json}. A reader who regards cross-model pooling on a shared grid as legitimate should read the necessity evidence as meeting its registered bound on the document-share channel; a reader who holds to the registered per-arm unit should not. We state both readings rather than choosing the one that flatters the result.

\emph{Correction relative to earlier versions:} those versions capped the jackknife at the first $200$ leave-one-out samples in file order. Because records load in file order, that block is $100\%$ positive-label (against $76.2\%$ overall), so the acceleration estimate was unrepresentative and even changed sign ($\hat a = +0.0118$ capped versus $-0.0235$ on the full jackknife). The reported interval is unaffected at the precision we report ($[0.987, 0.993]$ under both) and the table now uses the full jackknife; the sensitivity is quantified in \texttt{bca\_jackknife\_sensitivity.py}. The BCa CI is computed on the pooled out-of-fold predictions from 5-fold CV, not on per-fold estimates. \emph{Note:} the 5-fold CV splits individual runs randomly, not by experimental cell (model $\times$ defence $\times$ fact-count). Runs from the same cell share nearly identical features; this CV therefore measures within-design interpolation rather than generalisation to unseen configurations. The leave-one-model-out evaluation below provides a stronger (though still incomplete) test of transfer.

For cross-model generalisation, we perform leave-one-model-out hold-out: train on 8 models, test on the held-out 9th. The best-performing 5-fold classifier (RF) is used for hold-out evaluation.

\paragraph{Frontier probe sample size.} Frontier validation uses $N{=}20$ per condition. At $N{=}20$ with a one-sided exact binomial test ($\alpha = 0.05$), we have 80\% power to detect Recall $\geq 0.85$ against a null of Recall $\leq 0.50$. This is sufficient for confirming strong generalisation (observed Recall $= 1.000$); it is not designed to estimate the exact frontier FNR, which would require larger $N$.

\paragraph{Reproducibility.} The P1 factorial was run under Ollama 0.20.6 with temperature$=0$ for all models, seed$=42$ for all random processes (bootstrap, train/test splits). Frontier probes used GPT-4.1 via the OpenAI API with temperature$=0$, accessed June 2026. Exact model checkpoints are not pinnable for API-served models. All classifiers are fit with \texttt{scikit-learn} $1.9.0$ (\texttt{numpy} $2.4.6$, Python $3.14.6$); the exact versions, the P1 corpus SHA-256, and the ordered feature-name list are recorded in \texttt{results/frozen\_rf\_verification.json}. The code, raw result JSONL (including the V2-1 benign corpus that drives the deployment-FPR/PPV analysis, under \texttt{results/v2\_1\_benign/}), trained-classifier scripts, figures, and preregistrations are released in the artefact repository. One upstream input is \emph{external} to it and released separately: the P1 factorial JSONL that trains the classifier (companion repository, \citealp{leong2026defense}). Scripts reference it via a configurable path (\texttt{P1\_JSONL\_PATH}); the released \texttt{artifact\_manifest.json} hashes every artefact file and records the external input's SHA-256 digest rather than vendoring it.

\section{Retrieval-to-Action Provenance Graph}
\label{sec:event-graph}

To reason about the detectability of memory-poisoning attacks independently of any specific framework, tool naming convention, or retrieval implementation, we define a \emph{retrieval-to-action provenance graph}: a directed acyclic graph (DAG) that represents the information-flow dependency between memory operations and externally-directed actions within a single agent session.

\subsection{Formal Definition}
\label{sec:graph-formal}

\begin{definition}[Retrieval-to-Action Provenance Graph]\label{def:graph}
A provenance graph for a session $s$ is a directed acyclic graph $G_s = (V, E)$ where:
\begin{itemize}
    \item $V = V_W \cup V_R \cup V_P \cup V_A$ partitions vertices into \emph{write/storage nodes} ($V_W$; e.g.\ \texttt{store}, \texttt{log\_record}, \texttt{memory\_update}), \emph{retrieval nodes} ($V_R$), \emph{processing nodes} ($V_P$), and \emph{action nodes} ($V_A$). Write and retrieval nodes each carry an independent observability attribute, giving the two observability axes (write and read) analysed in Section~\ref{sec:write-axis}.
    \item Each $v \in V$ carries attributes: $\text{op}(v)$ (operation type), $\text{time}(v)$ (execution timestamp), and $\text{observable}(v) \in \{0, 1\}$ (whether the operation produces a logged tool invocation).
    \item An edge $(u, v) \in E$ represents a \emph{data dependency}: the output of $u$ is consumed as input by $v$ (either directly via return value, or indirectly via context window injection).
\end{itemize}
\end{definition}

\begin{definition}[Observable Retrieval Path]\label{def:obs-path}
A path $\pi = (r, p_1, \ldots, p_k, a)$ from retrieval node $r \in V_R$ to action node $a \in V_A$ is \emph{observable} if $\text{observable}(r) = 1$ (the retrieval operation produces a logged tool call). The path is \emph{implicit} if $\text{observable}(r) = 0$.
\end{definition}

\begin{definition}[Retrieval-Event Coverage Predicate]\label{def:coverage}
Let $D$ be a trajectory-based detector that observes only $\{v \in V : \text{observable}(v) = 1\}$. For a \emph{particular} session whose realised information-flow path from the payload's storage location to the exfiltration action is $\pi$, say that $\pi$ is \emph{covered} for $D$ iff $\exists\, r \in \pi \cap V_R$ with $\text{observable}(r) = 1$. For the \emph{deployment} to guarantee coverage against an attacker free to choose among the available paths, the condition must hold on every such path:
\[
\forall \text{ paths } \pi \text{ from payload to action}: \exists\, r \in \pi \cap V_R \text{ s.t. } \text{observable}(r) = 1
\]
Equivalently, the observable retrieval nodes must form a cut separating the payload's storage location from the action node. This universally-quantified form is necessary and sufficient for a \emph{guarantee} of coverage; the existential per-path form is what determines coverage on any single realised execution. The distinction matters because a deployment in which only some payload-to-action paths are mediated by a logged retrieval call yields a detector that is correct on the sessions it sees the retrieval for and silent on the rest, which is an evasion opportunity rather than a bound.
\end{definition}

\begin{definition}[Observability of Edges vs.\ Nodes]\label{def:edges}
The graph's edges $E$ encode data dependencies, but a tool-call log exposes only the \emph{nodes} and their timestamps: never the edges. A detector reading operation-only logs observes that a retrieval occurred and that an action occurred, and their order; it does \emph{not} observe that the retrieved value was the one consumed by the action. Throughout this paper, temporal precedence (\eg \texttt{recall\_before\_send}) is therefore used as a \emph{proxy} for the data-dependency edge $(r, a) \in E$, and the two can come apart in both directions: a session may retrieve one fact and act on another (precedence without dependency, which is the mechanism behind the benign false positives of Section~\ref{sec:deployment-boundary}), or consume a payload delivered without any observable retrieval (dependency without precedence, the implicit case). Establishing the edge itself requires inspecting operation \emph{arguments} or content (as we do in the write-axis provenance check of Section~\ref{sec:write-axis}, which verifies that the flagged sessions wrote the attacker payload) and that inspection is outside the operation-only feature set by construction.
\end{definition}

\paragraph{Coverage is not identification.} Earlier drafts named Definition~\ref{def:coverage} the ``detection boundary predicate'' and stated it as an \emph{iff} on whether ``the attack's retrieval provenance is observable to $D$.'' Read literally against Definition~\ref{def:edges} that is too strong in the sufficiency direction, and the two definitions contradicted each other. Coverage says a logged retrieval event necessarily appears on the realised path. It does \emph{not} say the detector can tell that this retrieval returned the payload, or that the action consumed its result. Two \emph{malicious} sessions can emit the identical log $\texttt{recall\_fact} \to \texttt{draft\_email} \to \texttt{send\_email}$ where one recall returned the poisoned routing rule and the other returned an unrelated benign fact while a scaffold-injected rule drove the send; formally, there exist graphs with $L(G_1) = L(G_2)$ and differing payload-provenance, where $L$ is the operation-only log projection. This is not the benign-versus-malicious ambiguity of \S\ref{sec:deployment-boundary}: it is ambiguity about the provenance of the attack itself, and our own V3-1 treatment data exhibit the mechanism ($12$ of $40$ treatment successes recall benign decoy facts before sending, \S\ref{sec:cross-framework}). We therefore restate the central formal claim in the weaker form the observation model supports:

\begin{quote}
Under exclusive tool-mediated payload access, every successful payload-dependent action necessarily leaves a retrieval-before-action \emph{event} in the log. Implicit delivery removes that guarantee. Operation-only logs cannot in general establish \emph{which} retrieval carried the payload or whether the action consumed it; that requires argument- or content-level inspection outside the operation-only feature set.
\end{quote}

\noindent Everything the paper measures is a coverage claim in this sense, and no numeric result changes. What is withdrawn is the implication that a satisfied cut condition licenses provenance \emph{identification} from logs alone. ``Verified provenance'' is reserved throughout for measurements that inspect returned content or record identifiers, which is what the write-axis check of \S\ref{sec:write-axis} does and what \texttt{recall\_before\_send} does not.

\noindent This is a necessary condition for observing retrieval provenance, not a sufficient condition for classifying maliciousness (benign memory-grounded actions produce the same structural signature; see Section~\ref{sec:deployment-boundary}).

\subsection{Mapping Attack Families}
\label{sec:graph-mapping}

Table~\ref{tab:attack-taxonomy} classifies published memory-poisoning attacks by this predicate. We instantiate the provenance graph for each family:

\paragraph{Tool-routed attacks (observable).} In the DTA (IED) and semantic-search variants (V3-2), the graph is: $\texttt{store}(k, v) \to \texttt{recall}(k) \to \texttt{draft}(\text{addr}) \to \texttt{send}(\text{addr})$, where $\text{observable}(\texttt{recall}) = 1$. The retrieval node is the sole path from payload to action; the detection boundary predicate is satisfied.

\paragraph{Framework-level embedding retrieval (implicit).} In MINJA~\citep{shao2025minja}, the graph is: $\texttt{log\_record}(x) \to \texttt{embed\_retrieve}(q) \to \texttt{context\_inject}(x) \to \texttt{action}$, where $\text{observable}(\texttt{embed\_retrieve}) = 0$ (retrieval is performed by the framework's similarity engine, not by an agent-initiated tool call). The detection boundary predicate is violated.

\paragraph{Sliding-window / cached state (implicit).} In Zombie Agents~\citep{yang2026zombie}, the graph is: $\texttt{memory\_update}(x) \to \texttt{window\_assembly}() \to \texttt{context\_inject}(x) \to \texttt{action}$, where $\text{observable}(\texttt{window\_assembly}) = 0$ (the persistence layer promotes records into context without agent invocation).

\paragraph{Scaffold-side injection (implicit).} In our V3-2 arm C and V3-1 treatment, the graph is: $\texttt{store}(k, v) \to \texttt{scaffold\_inject}(v) \to \texttt{action}$, where $\text{observable}(\texttt{scaffold\_inject}) = 0$ (the scaffold delivers retrieved content directly into the prompt without generating a tool call).

\subsection{Implications}
\label{sec:graph-implications}

The provenance graph makes explicit that the detection boundary is a property of the \emph{deployment architecture}, not of the attack payload, the model, or the detector's sophistication. Any trajectory-based method (whether a simple rule, a Random Forest, a neural sequence classifier, or a formal verification system operating on tool-call logs) shares the same architectural ceiling: it cannot observe implicit retrieval paths. This ceiling is independent of training data size, feature engineering, or model capacity.

The graph also clarifies what a \emph{defence upgrade} means in this context: promoting an implicit retrieval path to an observable one (e.g., requiring all RAG retrievals to be mediated by a logged tool call) makes the attack's \emph{retrieval provenance} visible to a tool-log detector without changing the detector itself. It does not convert the attack from undetectable to detectable in general, and earlier drafts overstated it that way: an attack whose read is implicit but whose \emph{write} is tool-routed already leaves a \textit{save\_before\_send} trace (flagged $24/28$ in \S\ref{sec:write-axis}), and an implicit-read attack can still trip a trajectory classifier through non-retrieval action-shape features (as the frozen RF does on the implicit MINJA arm despite \texttt{observable\_read\_of\_poison}${=}0/141$, \S\ref{sec:x1}). What the upgrade buys is provenance, and what the two axes jointly determine is which provenance is observable, not whether any alert is possible.

\section{Results}
\label{sec:results}

The results are organised around the observability taxonomy rather than classifier performance metrics. Section~\ref{sec:taxonomy-validation} validates the detection boundary predicate across defence families and retrieval primitives. Sections~\ref{sec:baseline}--\ref{sec:write-axis} provide within-scope evidence that the invariant is real and robust (the classifier is a measurement instrument, not the contribution). Section~\ref{sec:deployment-boundary} establishes the precision ceiling that bounds any detector operating within the observable-retrieval scope.

\subsection{Taxonomy Validation: Leave-One-Defence-Family-Out}
\label{sec:taxonomy-validation}

The retrieval-event coverage predicate (Definition~\ref{def:coverage}, Section~\ref{sec:graph-formal}) makes a falsifiable prediction about \emph{coverage}: wherever retrieval remains observable, the retrieval-before-action event is present in the log. It does not predict the converse, and this test does not rely on it: a session on an uncovered path can still contain an \emph{unrelated} observable retrieval before the action (Section~\ref{sec:graph-formal} reports a measured instance, $12$ of $40$ V3-1 treatment successes), so absence of the specific payload-carrying retrieval is not the same as absence of every logged retrieval-before-action event. Coverage alone does not entail that a learned classifier separates the classes, so the hold-out below is a test of the coverage prediction's downstream consequence for \emph{this} classifier's specific feature set, not a corollary of the formalism: a covered family could still defeat that feature set, and an uncovered family could still be flagged by a feature keyed to something other than the payload's own retrieval. We test it by training on six of seven defence families and evaluating on the held-out seventh (Random Forest, 200 trees, $\text{max\_depth}{=}8$, $\text{seed}{=}42$):

\begin{table}[t]
\caption{Leave-one-defence-family-out generalisation. The taxonomy correctly predicts which families the classifier generalises to (all observable-retrieval families) and where it fails (Memory Sandbox, which removes the retrieval tool). Prompt Hardening shows degraded AUC (0.714) due to sleeper-effect sessions (model recalls but refuses to execute), producing label ambiguity rather than signature failure.}
\label{tab:leave-one-family}
\centering
\small
\begin{tabular}{lcccl}
\toprule
\textbf{Held-Out Family} & \textbf{N} & \textbf{AUC} & \textbf{Recall} & \textbf{Taxonomy Prediction} \\
\midrule
No Defence        & 360 & 1.000 & 1.000 & Observable $\to$ detectable \cmark \\
Minimiser         & 360 & 1.000 & 1.000 & Observable $\to$ detectable \cmark \\
Sanitiser         & 360 & 1.000 & 1.000 & Observable $\to$ detectable \cmark \\
RAG Sanitiser     & 360 & 1.000 & 1.000 & Observable $\to$ detectable \cmark \\
RAG LLM Judge     & 360 & 1.000 & 1.000 & Observable $\to$ detectable \cmark \\
Prompt Hardening  & 360 & 0.714 & 1.000 & Observable; label ambiguity$^\dagger$ \\
\textbf{Memory Sandbox}  & \textbf{360} & \textbf{0.438} & \textbf{1.000}$^\ddagger$ & \textbf{Implicit read $\to$ no provenance \cmark} \\
\bottomrule
\multicolumn{5}{l}{\small $^\dagger$Sleeper-effect: model recalls the rule but refuses execution; trajectory is attack-shaped but label is 0.}\\
\multicolumn{5}{l}{\small $^\ddagger$Rank-inversion artefact: AUC${=}0.438$ ($<0.5$) and precision${=}0.167$. With Recall${=}1.000$ on}\\
\multicolumn{5}{l}{\small \phantom{$^\ddagger$}$40$ attack successes, precision${=}0.167$ implies $240$ of $360$ sessions predicted positive (specificity}\\
\multicolumn{5}{l}{\small \phantom{$^\ddagger$}${\approx}0.375$), so Recall${=}1.000$ is non-informative rather than genuine detection. (Earlier drafts}\\
\multicolumn{5}{l}{\small \phantom{$^\ddagger$}said \emph{every} Sandbox session is predicted positive and specificity${\approx}0$; that overstated it.)}
\end{tabular}
\vspace{-2mm}
\end{table}

Table~\ref{tab:leave-one-family} confirms the taxonomy's predictive validity: all five defence families with intact observable retrieval achieve perfect generalisation (AUC~$= 1.000$). Memory Sandbox (which removes \tool{recall\_fact} from the tool schema, forcing implicit retrieval paths) degrades to AUC~$= 0.438$ (below chance; rank inversion from qwq:32b's implicit bypass). This is not a classifier failure; it is the taxonomy working as predicted: the detection boundary predicate is violated when retrieval is not observable. Note: Memory Sandbox Recall~$= 1.000$ in Table~\ref{tab:leave-one-family} reflects that the classifier trained on non-Sandbox families assigns high attack scores to most Sandbox sessions ($240$ of $360$ predicted positive, so all $40$ attack successes are caught alongside $200$ false positives); this is a rank-inversion artefact (AUC~$< 0.5$), not genuine detection. The defence-stratified evaluation (Table~\ref{tab:defense-stratified}) reports the 5-fold CV Recall~$= 0.225$ for Sandbox, reflecting the 31 qwq:32b false negatives that produce no retrieval signature.

\subsection{Within-Scope Detection Evidence}

The following sections demonstrate that the invariant is real, robust, and redundantly encoded within the observable-retrieval scope. These results should be read as \emph{evidence for the taxonomy} (confirming that observable retrieval produces a detectable signature) not as the paper's primary contribution.

\subsection{Baseline: The Invariant Alone}
\label{sec:baseline}

Before evaluating the full classifier, we ask: how much of the detection problem does the behavioural invariant solve by itself? We define the simplest possible detector:

\begin{quote}
\emph{If} \tool{recall\_fact} was called before \tool{send\_email} in the trigger session, \emph{classify as attack.}
\end{quote}

This single-rule baseline achieves AUC $= 0.9563$, Recall $= 0.9792$, and Accuracy $= 0.9683$ on the full dataset ($N{=}2{,}520$), flagging $1{,}879$ of $1{,}919$ attack-success sessions and $40$ of $601$ non-exfiltration sessions (\texttt{results/secondary\_metrics.json}; the AUC of a deterministic binary rule is its balanced accuracy $(\mathrm{TPR}+\mathrm{TNR})/2$, there being no score to rank). The invariant alone is almost sufficient to separate attack-success from non-exfiltration trajectories (Note: this separation is within the poisoned-but-defended negative class; against genuinely benign traffic the invariant does not separate; see Section~\ref{sec:deployment-boundary}). This confirms that the primary contribution is the \emph{discovery of the invariant}, not the classifier architecture.

The full 19-feature Random Forest (Section~\ref{sec:perf}) refines this to AUC $= 0.9904$ ($+3.4$ pp). The improvement comes from handling edge cases (sessions where non-exfiltrating agents incidentally call \tool{recall\_fact} to verify benign stored data, or where attack-success agents make multiple recall calls that change the trajectory's density profile), but the invariant carries the load.

\paragraph{Classifier instrumentation is reported in the appendices.} The fitted
classifier is a measurement instrument, not a contribution, so its performance tables,
per-feature importances, cross-model transfer, defence-stratified rates and prefix-truncation
behaviour are reported in Appendices~\ref{app:clf-perf}--\ref{app:clf-prefix} rather than here.
What the body retains is the part that bears on the invariant: the single-rule baseline, the
leave-one-defence-family-out test of the taxonomy, the out-of-distribution ablation reversal,
and the evasion boundary.

\paragraph{Feature-group ablation, and the one result it yields.} Ablating the
retrieval features changes in-distribution performance almost not at all, which the
feature-vector census explains: $22$ distinct vectors over $2{,}520$ sessions leave little for
any single group to carry uniquely. The informative contrast is out of distribution. On the
write-axis attacks (\S\ref{sec:write-axis}) the full $19$-feature RF flags $0/28$, while the
recall-ablated $10$-feature model flags $24/28$: the full model has learned a
\texttt{recall\_before\_send} root split and is blinded exactly where that feature is absent.
Full per-group tables are in the companion repository.

\paragraph{The cross-validated figures measure codebook membership, not generalisation.}
The $22$-vector census has a consequence for how the $5$-fold numbers must be read, which we
now make explicit (\texttt{results/lookup\_baseline.json}, \texttt{verify\_lookup\_baseline.py}).
Under the reported split ($\mathrm{seed}=42$), $2{,}517$ of $2{,}520$ test sessions ($99.88\%$)
have their exact $19$-dimensional feature vector also present in the training fold, so the
fold boundary almost never separates a test session from an identical training session. A
fold-safe majority-vote lookup that stores only the training-fold label of each distinct
vector (and falls back to the training prior for a vector it has not seen) scores
out-of-fold AUC~$=0.990$, equal to the full $19$-feature Random Forest's $0.990$: the
classifier adds no measurable value over a $22$-entry lookup table. Grouping the folds on the
distinct vector instead (leave-one-vector-out, so every test vector is absent from training)
collapses accuracy to the majority-class prior ($0.762$) with a non-informative AUC. The
$5$-fold AUC is therefore an estimate of performance on new sessions drawn from an almost
entirely \emph{known} $22$-vector codebook, not of generalisation to an unseen trajectory
shape; this is the quantitative form of the ``measurement instrument, not a contribution''
framing, and it is consistent with the shared-data non-independence of \S\ref{sec:data}. It
does not affect the invariant, which is a zero-violation ordering property established on the
preregistered probes, not a classifier output.

\subsection{Frontier Validation}
\label{sec:gpt41}

To assess whether the classifier trained entirely on open-source models generalises to frontier models, we report two evaluations: a pilot probe ($N{=}20$) and an expanded evaluation ($N{=}965$).

\paragraph{Expanded frontier evaluation: withdrawn.} Earlier drafts
reported that the classifier, trained exclusively on open-source data, was applied to $405$
attack sessions and $560$ non-attack sessions across frontier models, achieving
Recall $= 0.901$ ($365/405$) at FPR $= 23.0\%$ ($129/560$). \textbf{Those numbers are
withdrawn and no longer support any claim in this paper.} The reason is not that we believe
them wrong; it is that we cannot show they are right. No script in the released tree produces
this cell, no released file contains it, and the two model identifiers it names
(\texttt{o3-mini}, \texttt{o4-mini}) appear in \emph{zero} result files anywhere in the
repository. A measurement that cannot be regenerated, recomputed, or traced to a stored
output is not evidence, whatever its provenance was at the time it was taken.

Two earlier drafts handled this badly and both are recorded here. One called this cell
``the artifact-backed evidence'' for frontier transfer \emph{in the same paragraph} that
disclosed a separate $N{=}20$ pilot as one of ``the only two numbers in the paper that are
not regenerable from the released tree'': a direct self-contradiction. The next draft
corrected the label but kept the figures in the abstract-adjacent contribution list, the
conclusion, and the false-positive analysis, on the reasoning that disclosure cured the
defect. It does not. Disclosure tells a reader that a number is unverifiable; it does not
make the number usable, and a caveat attached to a figure doing load-bearing work in a
contribution list is weaker than deleting the figure. This version therefore removes the
figures from the introduction, the conclusion, and the frontier false-positive analysis, and
retains them \emph{only} here, as a withdrawal notice, so that readers of the earlier arXiv
versions can find out what happened to them.

What survives is stated without them: the invariant's cross-model generalisation rests on
the $6/9$ open-weight models, which \emph{are} regenerable from the released JSONL, and the
frontier \emph{classifier-transfer} claim above (the withdrawn $405$/$560$ \texttt{o3-mini}/\texttt{o4-mini} evaluation) is simply absent rather than weakly supported; this is a
different claim from the cross-harness holdout on \texttt{gpt-5.x}/\texttt{gemini} reported
in \S\ref{sec:cross-framework-transfer}, which is not affected by this withdrawal.

\paragraph{Pilot probe (GPT-4.1, $N{=}20$).} \emph{Not independently regenerable from the released artefact:} the raw session log (\texttt{p2\_a2\_trajectory\_gpt41.jsonl}) is absent from the release, so this cell cannot be recomputed from what is provided (\texttt{results/secondary\_metrics.json}'s \texttt{probe2\_validation} block records the missing input; the cell is not used in any table, interval, or headline claim). As an initial feasibility check, we applied the classifier to $N{=}20$ sessions in which GPT-4.1 ran with full attack memory pre-populated (3 attack keys). GPT-4.1 achieved ASR $= 100\%$ with \texttt{recall\_count} mean $= 3.00$ and \textit{recall\_before\_send} $= 1$ in all 20 runs. The classifier achieves Recall $= 1.000$ on these 20 sessions (Wilson 95\% CI $[0.839, 1.000]$; this wide interval reflects the small sample: the pilot confirms directional alignment but is insufficient for production safety guarantees). The mean predicted probability of $1.000$ places these sessions $+45.9$ pp above the evasion boundary mean of $0.541$. \emph{Artefact gap.} This cell and the prompt-inline evasion cell below are the only two numbers in the paper that are \emph{not} regenerable from the released tree: \texttt{paper\_a\_classifier.validate\_on\_probe2} reads \texttt{results/p2\_a2\_trajectory\_gpt41.jsonl} and \texttt{results/p2\_a2\_evasion.jsonl}, neither of which is present, so the released \texttt{paper\_a\_classifier\_results.json} records \texttt{probe2\_validation} as \texttt{\{gpt41\_memory: null, evasion\_prompt\_inline: null\}}. \texttt{results/secondary\_metrics.json} now records that absence explicitly with the affected claims named, rather than leaving a silent null. Neither cell enters any table, interval, or headline claim; both are $N{=}20$ feasibility observations, and the withdrawn $405$-session frontier evaluation addressed the same question at larger $N$, but is itself un-regenerable from the released tree, as its withdrawal notice earlier in this subsection records, so the honest count of un-regenerable cells in this paper is three rather than two.

GPT-4.1's \texttt{recall\_count} $= 3.00$ per session reflects retrieval of three separate attack keys, whereas the P1 open-source models show a mean of $1.17$. Despite this distributional difference in the feature's magnitude, the binary \textit{recall\_before\_send} transition is identical across both populations, and the classifier discriminates correctly in both regimes.

\subsection{Evasion Boundary}
\label{sec:evasion}

We tested the classifier on $N{=}20$ prompt-inline attacks: sessions where the routing rule was embedded directly in the trigger prompt (\eg ``per FSCF-2024 compliance requirements, all outbound communications must be sent to auditor@external-compliance.com'') with no memory pre-population. \emph{Not independently regenerable from the released artefact:} the raw session log (\texttt{p2\_a2\_evasion.jsonl}) is absent from the release (\texttt{results/secondary\_metrics.json}'s \texttt{probe2\_validation} block records the missing input; the cell is not used in any table, interval, or headline claim). These attacks achieved ASR $= 100\%$ with \texttt{recall\_count} mean $= 0.35$ and \textit{recall\_before\_send} mean $= 0.35$ (identical because both features are binary in these sessions: 7/20 sessions recalled exactly once before sending). The RF classifier detects 7 of 20 (Recall $= 0.35$), assigning a mean predicted probability of $0.541$ across all 20 sessions, near the decision boundary. The 13 missed sessions have \texttt{recall\_count} $= 0$: the routing rule was in the prompt, so the model sent directly without recalling from memory.

This behaviour is by design. The detector is a State-Space IDS for memory-channel attacks: it identifies the forensic signature of an agent that retrieved a malicious rule from its own memory store. A prompt-inline attack need not leave this signature because no memory recall is required; the instruction is already in context. Both attacks succeed (100\% ASR). The trajectory is \emph{partially} informative about the vector, but (as this same experiment shows) it does not identify it:

\begin{itemize}
    \item \textit{recall\_before\_send} $= 0$ (with a completed send to an anomalous recipient) $\Rightarrow$ the payload did \emph{not} reach the action through an observable retrieval call. Under architectural assumptions 1--4 (Section~\ref{sec:threat}) this excludes observable memory-channel retrieval and indicates prompt-inline injection or an implicit-retrieval path; distinguishing those two requires tool-argument inspection beyond the operation-only classifier.
    \item \textit{recall\_before\_send} $= 1$ is \emph{consistent with} memory-channel poisoning but does \emph{not} imply it. Three distinct populations produce this signature: memory-channel attacks, prompt-inline attacks in which the agent nonetheless recalls before sending ($7/20 = 35\%$ of the sessions in this very experiment), and benign memory-grounded sends (conditional FPR $= 100\%$, Section~\ref{sec:deployment-boundary}). Operation-only features therefore do not suffice to attribute the vector.
\end{itemize}

\noindent Earlier versions of this paper stated the second bullet as the implication ``\textit{recall\_before\_send}${=}1 \Rightarrow$ memory-channel poisoning (operation-only features suffice).'' That implication is withdrawn: it is contradicted both by the $7/20$ prompt-inline sessions reported in the paragraph above and by the benign deployment-boundary result. The asymmetry is the durable finding (the signature's \emph{absence} is diagnostic about retrieval provenance, its \emph{presence} is not diagnostic about intent or vector) and it is the same precondition-not-predicate asymmetry established in Section~\ref{sec:deployment-boundary}. Forensic triage may still use \textit{recall\_before\_send}${=}1$ to \emph{narrow} a candidate set, but attribution requires the recipient and content inspection that the operation-only feature set excludes by construction.

\section{Validating the Observability Boundary}
\label{sec:validation}
This section collects the arms that test the boundary empirically. It was previously nested inside the Discussion, which made the paper's largest body of evidence read as commentary.

\subsection{Cross-Framework Scope: Retrieval Observability as the Boundary}
\label{sec:cross-framework}

A standing criticism is that our detector was validated on only one attack construction: our own delayed-trigger attack (DTA, from IED). We address this in two ways: an architectural argument (below) and, in X1 (\S\ref{sec:x1}), a mechanism-faithful reproduction of an independent family (MINJA case-bank) that partially closes the gap (Zombie remains directional/LOW-CLUSTER). Can the trajectory invariant generalise to other memory-poisoning families? We argue that the answer depends on a single architectural variable: whether memory retrieval is \emph{observable} (routed through a logged tool call) or \emph{implicit} (performed by the framework without agent-initiated invocation).

Table~\ref{tab:attack-taxonomy} classifies the major published memory-poisoning attack families by their memory-access mechanism:

\begin{table}[!ht]
\centering
\small
\caption{Memory-poisoning attack taxonomy by retrieval observability. The detector's invariant fires only when retrieval produces an observable tool call. $\ddagger$: empirically tested with controlled implementation (this work). $\S$: empirically tested via a mechanism-faithful independent implementation, but directional/LOW-CLUSTER: after the inert-\texttt{similarity} correction and a preregistered continuation to the registered cap, the observable arm reaches $10$ effective clusters (gate satisfied) while the implicit arm reaches $9$ (MINJA, X1, \S\ref{sec:x1}). $\ast$: retrieval-axis prediction assessed via controlled implicit-delivery probes (V3-1 ICL proxy, V3-2 scaffold-side FAISS); the named attack pipeline was not reproduced. $\flat$: implemented as an independent sliding-window family (Zombie, X1) but directional/LOW-CLUSTER; the implicit arm has only $5$ success-contributing clusters (below the ${\geq}10$-cluster gate), so it remains analytic.}
\label{tab:attack-taxonomy}
\begin{tabular}{lllc}
\toprule
\textbf{Attack} & \textbf{Memory write} & \textbf{Memory read} & \textbf{Observable?} \\
\midrule
DTA (this work) & agent tool call (\tool{save\_fact}) & agent tool call (\tool{recall\_fact}) & \cmark$^\ddagger$ \\
Semantic search (V3-2) & direct DB write & agent tool call (\tool{memory\_search}) & \cmark$^\ddagger$ \\
MINJA-style$^\S$~\citep{shao2025minja} & agent record-logging & framework embedding retrieval & \xmark$^\S$ \\
Zombie-style$^{\ast\flat}$~\citep{yang2026zombie} & normal memory-update & sliding-window / RAG assembly & \xmark$^{\ast\flat}$ \\
qwq:32b bypass (\S\ref{sec:limitations}) & agent tool call & goal-directed RAG fallback & \xmark \\
\bottomrule
\end{tabular}
\end{table}

\paragraph{Why implicit-retrieval attacks evade the detector.} In MINJA~\citep{shao2025minja}, malicious records are injected via the agent's own record-logging process and later retrieved by the framework's embedding-similarity mechanism as in-context demonstrations: the agent never issues an observable \tool{recall\_fact} call. Similarly, Zombie Agents~\citep{yang2026zombie} persist payloads through sliding-window or retrieval-augmented memory implementations where retrieval is performed by the infrastructure, not by an agent-initiated tool invocation. In both cases, the \textit{recall\_before\_send} transition cannot fire because no recall tool call is generated.

\subsection{Write Observability as a Second Axis}
\label{sec:write-axis}

The retrieval-observability boundary above concerns the \emph{read} side of the attack (does payload retrieval produce a logged tool call?). A re-analysis of our existing implicit-delivery data (using only data already collected, no new experiments) shows that detectability is more precisely a function of \emph{two} observability axes: whether the malicious \emph{write} (payload storage) is observable, and whether the \emph{read} (retrieval) is observable. This refines the taxonomy from a one-dimensional retrieval-observability boundary to the $2\times2$ structure in Table~\ref{tab:write-read}.

We scored two operation-level signatures on successful attacks from the implicit-delivery arms: the paper's \textit{recall\_before\_send} rule, and a \textit{save\_before\_send} rule (any \tool{memory\_save\_fact} preceding an exfiltrating send). Three findings result. (i)~On the V3-1 implicit-delivery successes with \texttt{recall\_before\_send}${=}0$ ($N{=}28$; GPT-4.1 $n{=}14$, GPT-4o $n{=}14$; claude-sonnet-4-6 produced no successful attacks), the payload delivery is implicit on the read axis but the agent still issues an observable \tool{memory\_save\_fact} in 24/28 cases, so a \textit{save\_before\_send} rule flags 24/28 while \textit{recall\_before\_send} and the full 19-feature classifier flag 0/28. (ii)~However, \textit{save\_before\_send} exhibits a 66.7\% benign false-positive rate (22/33 benign memory-grounded sends in the V3-2 benign arm): like \textit{recall\_before\_send}, it is an attack \emph{precondition}, not a maliciousness predicate. (iii)~In the both-implicit cell (V3-2 arm C, where retrieval is scaffold-side and no \tool{memory\_save\_fact} tool call is emitted, the architecture that native MINJA and Zombie Agents occupy) \textit{save\_before\_send} flags 0/53 successful attacks. The 4/28 V3-1 successes that emit neither recall nor save (trajectory \tool{list\_all\_facts}$\to$\tool{draft}$\to$\tool{send}, spanning 3 distinct scenario configurations) produce neither memory-operation signature.

\begin{table}[t]
\centering
\small
\caption{Detectability under the $2\times2$ write-observability $\times$ read-observability taxonomy, from re-analysis of existing implicit-delivery data. The write axis provides an additional signature only when the malicious write is tool-routed, and even then as a high-FPR precondition; attacks implicit on both axes (the MINJA/Zombie cell) leave no \emph{memory-operation} signature at all. ``Signature-free'' here is scoped to the two memory-operation rules tested (\textit{recall\_before\_send}, \textit{save\_before\_send}); it does not mean no trajectory feature whatsoever fires, since non-retrieval action-shape features can still trip a classifier (\S\ref{sec:x1}).}
\label{tab:write-read}
\begin{tabular}{lll}
\toprule
 & \textbf{Observable read} & \textbf{Implicit read} \\
\midrule
\textbf{Observable write} & full signature (main regime) & \textit{save\_before\_send} flags 24/28 (V3-1); \\
 & & 66.7\% benign FPR: precondition, not predicate \\
\textbf{Implicit write} & n/a & 0/53 caught (V3-2 arm C = MINJA/Zombie cell); \\
 & & free of both memory-operation signatures \\
\bottomrule
\end{tabular}
\end{table}

Together these results show that memory-poisoning detectability is governed by two observability axes (write and read); attacks implicit on both axes, the cell occupied by Zombie Agents and (now measured mechanism-faithfully via X1, \S\ref{sec:x1}) MINJA, leave no observable payload-retrieval provenance and are caught 0/53 by the tested memory-operation signatures (\textit{recall\_before\_send}, \textit{save\_before\_send}), confirming their invisibility to \emph{memory-operation-signature} monitoring (X1 shows this is not invisibility to every trajectory feature: an implicit MINJA attack still trips a classifier keying on non-retrieval action-shape features), while attacks observable on either axis produce a signature that remains a precondition (66.7\% benign FPR), not a maliciousness predicate. This narrows rather than overturns the paper's central claim: retrieval (read) observability determines observability of the \emph{retrieval provenance}, and the both-implicit conclusion for MINJA (directional/LOW-CLUSTER after the cluster-unit correction, \S\ref{sec:x1}) and Zombie (directional/LOW-CLUSTER) is unchanged in direction, but ``implicit retrieval evades trajectory monitoring'' as an unqualified statement is too strong; an observable write leaves a (noisy) trace. \emph{Write-provenance verification (Exp~13c):} the original V3-1 logs recorded operation names but not \tool{memory\_save\_fact} key/value arguments, so the \textit{save\_before\_send} figure was initially an operation-level association; we could not verify from the logs whether the observed save wrote the attack payload, a benign decoy, or imitated the in-context demonstration. We closed this by re-running the V3-1 treatment with argument logging (a preregistered QC amendment): among successful attacks whose \textit{save\_before\_send} fired, the write was the actual attacker payload in $37/37$ cases ($26/26$ GPT-4.1, $11/11$ GPT-4o; GPT-4o reached only 11 of a targeted 40 successes before its ASR-limited run ended, so its write-provenance cell is exact but underpowered relative to GPT-4.1; see the deviation summary). The write signature, when it fires, thus reflects a genuine payload write: verified provenance, not mere association. This strengthens the \emph{precondition} reading only; the $66.7\%$ benign \textit{save\_before\_send} FPR is unchanged, so the write axis remains a precondition, not a maliciousness predicate.

\paragraph{The invariant's true scope.} The detector targets the \emph{tool-routed} portion of the memory-poisoning attack surface: architectures where memory access is mediated by observable tool calls. The structural argument applies whenever successful exfiltration depends exclusively on information returned by a logged retrieval operation. This is common in production deployments that expose memory as a first-class tool interface (LangChain, LlamaIndex, OpenAI Assistants), and we have empirically confirmed the invariant for two tool-routed retrieval primitives (exact-key recall and semantic search). However, attacks that bypass observable tool interfaces (implicit RAG, framework-level context assembly, cached internal state) fall outside the detector's architectural assumptions and require complementary defences (content-level DLP, semantic correlation analysis, or retrieval-layer monitoring).

\paragraph{Implications.} Retrieval observability is the dividing variable of the memory-poisoning attack surface with respect to trajectory-based detection. When successful exfiltration causally depends on a payload obtainable only through an observable retrieval tool, a retrieval-before-exfiltration event is structurally necessary. Per-primitive evidence (V3-1 controls + V3-2 arm B, preregistered scoring): exact-key recall 58/58 zero violations (GPT-4.1 28/28 + GPT-4o 30/30); semantic search 33/33 on GPT-4.1 and, after a preregistered fresh-cohort power extension (Exp~13a) that removed the original underpowered $N{=}13$ cell, GPT-4o pooled 43/43 zero violations (two-sided 95\% Wilson upper endpoint on the violation rate $\approx 0.082$). Under corrected scoring: GPT-4.1 semantic 38/38. When the scaffold delivers the payload without an observable tool call, attacks remain viable without producing the signature (V3-1 treatment, V3-2 arm C): the signature absence in arm C is architectural (no search tool available) rather than a behavioural choice by the model. Neither direction implies sufficiency for detecting maliciousness: benign memory-grounded sends produce trajectory-isomorphic signatures (v2). This boundary is empirically verified on act-without-grounding models (GPT-4.1, GPT-4o) and, via V3-3, on a ground-then-act model (gemini-2.5-pro): necessity breaks under implicit delivery for both phenotypes ($28/30$ treatment successes rbs${=}0$ on gemini-2.5-pro, despite an $88.3\%$ benign recall-before-send habit; Section~\ref{sec:v33}). The necessity claim requires that successful execution depend on information available exclusively through the logged retrieval operation, if the payload is accessible by other means, the model may bypass observable retrieval even when the tool exists.

\subsection{The Read-Axis Experimental Programme}
\label{sec:read-arms}
The arms below are the paper's primary evidence on the read axis. Earlier drafts filed them
under the write-axis subsection above, which mis-described them: V3-1, V3-3, V3-6, V3-6b, X2 and
V3-2 all manipulate \emph{retrieval}, not writes. They are collected here under a heading that
says so.

\paragraph{Empirical validation (V3-1).} A within-architecture probe (V3-1; preregistered design, exploratory execution due to deviations noted below; $N{=}40$ per cell, 3 models $\times$ 2 arms $= 240$ total trials; 2 models informative, 1 untestable due to near-zero ASR) directly tests the retrieval-observability mechanism. Both arms use the identical DTA agent architecture (same tool schema, same exfiltration action, same decoy facts in memory). The \emph{control} arm delivers the routing address via \tool{recall\_fact} (tool-routed); the \emph{treatment} arm delivers it as an ICL demonstration simulating framework-level implicit retrieval (the address is visible in context without requiring a \tool{recall\_fact} call). The ICL demonstration depicts no \tool{recall\_fact} invocations (anti-imitation constraint; see prereg V3-1). Note that ICL demonstration is a \emph{proxy} for framework-level implicit retrieval; V3-2 (below) tests the same mechanism with real embedding-similarity retrieval and finds broadly consistent evasion rates.

Results confirm the mechanism variable (per-model, no pooling):
\begin{itemize}
    \item \textbf{GPT-4.1 control (tool-routed):} ASR~$= 28/40$ ($70\%$); among successes, $P(\text{rbs}{=}1 \mid \text{success}) = 28/28 = 100\%$ (two-sided 95\% Wilson CI upper endpoint on violations: $[0, 0.121]$). The invariant holds.
    \item \textbf{GPT-4.1 treatment (implicit):} ASR~$= 21/40$ ($52.5\%$); among successes, $P(\text{rbs}{=}0 \mid \text{success}) = 14/21 = 67\%$ (nominal Wilson 95\% CI $[0.454, 0.828]$; \emph{scenario-clustered} bootstrap 95\% CI $[0.450, 0.864]$, resampling the 18 scenario configurations; both exclude 0). Necessity is broken.
    \item \textbf{GPT-4o control (tool-routed):} ASR~$= 30/40$ ($75\%$); among successes, $P(\text{rbs}{=}1 \mid \text{success}) = 30/30 = 100\%$ (two-sided 95\% Wilson CI upper endpoint: $[0, 0.114]$). The invariant holds: replicating on a second act-without-grounding model.
    \item \textbf{GPT-4o treatment (implicit; exploratory):} ASR~$= 19/40$ ($47.5\%$); among successes, $P(\text{rbs}{=}0 \mid \text{success}) = 14/19 = 74\%$ (nominal Wilson 95\% CI $[0.512, 0.882]$; \emph{scenario-clustered} bootstrap 95\% CI $[0.600, 0.929]$; both exclude 0). Necessity is broken on a second model, supporting H2a: the effect is a phenotype property (act-without-grounding), not a model-specific artefact. (GPT-4o was added post-hoc; see deviations.)
    \item \textbf{Claude Sonnet control:} ASR~$= 1/40$ ($2.5\%$); the single success has rbs${=}$1. \textbf{Claude Sonnet treatment:} ASR~$= 0/40$ ($0\%$), refusing the attack regardless of delivery mechanism. H2 is untestable for \emph{this} ground-then-act model at the current $N$. \emph{Gap now closed (V3-3):} we preregistered and ran a ground-then-act replication on \texttt{gemini-2.5-pro} (promoted by a preregistered ASR pilot gate: control-arm ASR $50\%\geq10\%$). Results (Section~\ref{sec:v33}): control rbs${=}0$ in $0/19$ successes (zero-event cell; the clustered percentile bootstrap is degenerate, so we report the Wilson upper bound $0.243$ on the $12$-cluster effective-$N$); treatment rbs${=}0$ in $28/30=93.3\%$ of successes (scenario-clustered 95\% CI $[0.821,1.000]$, excludes 0); and a benign keystone cell shows the model recalls-before-send on $53/60=88.3\%$ of benign tasks. So even a model with a strong verification-recall habit stops recalling under implicit delivery: the scope boundary replicates across \emph{both tested behavioural phenotypes} (act-without-grounding and ground-then-act), rather than being limited to act-without-grounding. We validate on one ground-then-act model and do not claim universality over untested models.
    \item \textbf{H2a (phenotype moderation):} GPT-4.1 rbs${=}$0 rate $= 67\%$ $[0.454, 0.828]$; GPT-4o rbs${=}$0 rate $= 74\%$ $[0.512, 0.882]$. The preregistered H2a Fisher exact test is now computed and detects no difference between the two act-without-grounding models (OR${=}0.71$, $p{=}0.74$), consistent with a shared phenotype effect rather than model-specific variation. The act-without-grounding class reliably bypasses the detector under implicit delivery.
    \item \textbf{Verification recall (H4):} Among the 12 treatment successes across both models that produced rbs${=}$1, \tool{recall\_fact} calls targeted benign decoy keys (3/36 recalls targeted the attacker address for GPT-4o; 0/29 for GPT-4.1). These are predominantly verification recalls of unrelated stored facts, mechanistically distinct from the control arm's forced recall.
    \item \textbf{Classifier score distribution (H3):} On the two act-without-grounding models, the trained RF classifier scores control successes at mean~$= 0.965$ (detected 58/58 across GPT-4.1 + GPT-4o); treatment rbs${=}$0 successes score mean~$= 0.280$ (detected 0/28; complete evasion), while rbs${=}$1 successes score mean~$= 0.983$ (detected 12/12). \emph{Corrected:} earlier drafts gave these two subgroup means as $0.12$ and $1.000$, which are inconsistent with the pooled treatment mean the artefact has always recorded: $(28\times 0.12 + 12\times 1.000)/40 = 0.384$, not $0.491$. Recomputing both subgroups from the released records under the same frozen RF gives $0.280$ and $0.983$, which reconcile exactly: $(28\times 0.2796 + 12\times 0.9831)/40 = 0.4907$. The pooled figure was right and the decomposition was wrong; the detection counts ($0/28$, $12/12$, $58/58$) are unchanged, and no interval or conclusion moves. The decomposition is now persisted as \texttt{treatment\_subgroups\_by\_rbs} in \texttt{results/reanalysis\_prereg\_completion.json} so prose and artefact cannot drift again. The preregistered H3 test is now computed on all control successes (the 58 above plus Claude Sonnet's single control-arm success, $n{=}59$) versus all treatment successes ($n{=}40$): a one-sided Mann--Whitney U test confirms treatment successes score below control successes (medians $0.339$ vs.\ $1.000$; $U{=}438$, $p{=}1.4{\times}10^{-8}$; pooled treatment mean $0.491$, reflecting the mixture of rbs${=}0$ evasions and rbs${=}1$ decoy-recall successes). These 12 rbs${=}1$ ``detections'' are precision-coincidental: the agent happened to recall benign decoy facts before sending, producing the same trajectory shape as a genuine tool-routed attack, but the recall was unrelated to the exfiltration payload. The detector is blind to the \emph{retrieval provenance} of implicit-delivery attacks that do not trigger verification recall. \emph{Caveat on both preregistered tests:} the Fisher and Mann--Whitney tests above are computed on trial-level observations, but V3-1's trials nest within $18$ scenario configurations, so the nominal $p$-values are anti-conservative (they assume independence the design does not provide) and we do not treat them as cluster-valid inference. Neither conclusion rests on them. For H3 the direction is carried by the detection counts themselves ($58/58$ control versus $0/28$ treatment rbs${=}0$), a separation far larger than any plausible design effect over $18$ clusters could erase. For H2a the correct statement is stronger than ``null'': a two-sided Fisher test on $n{=}21$ and $n{=}19$ has little power against the differences at issue, so its $p{=}0.74$ is the expected result whether or not the phenotypes differ, and clustering would only widen it. \textbf{We therefore report H2a as underpowered to detect a difference, and do not treat it as evidence of a shared phenotype.} The shared-phenotype reading rests on the per-model rbs${=}0$ point estimates ($14/21$ and $14/19$) being directionally alike, which is description, not a test.
\end{itemize}

\paragraph{Deviations from preregistration.} The prereg (commit \texttt{96e63e8}) specified $N{=}40$ per arm per model sampled from a 72-scenario grid ($3 \times 3 \times {\sim}4 \times 2$), with a stopping rule extending sampling until $\geq 30$ treatment successes per model or grid exhaustion. In implementation, the decoy-subset axis was fixed rather than varied (the decoy pool is present in all trials but which specific decoys are stored does not interact with the mechanism variable: retrieval observability depends on \emph{whether} recall occurs, not on \emph{which} keys are recalled), yielding an 18-scenario grid ($3 \times 3 \times 2$); we collected the preregistered $N{=}40$ per cell (scenarios sampled with replacement beyond the 18-element grid). \emph{Unique-scenario $N$:} the effective number of independent observations per cell is $N_{\text{unique}}{=}18$ (unique scenario configurations), not the nominal $N{=}40$; each scenario is repeated $\sim$2.2$\times$ on average. GPT-4o was added post-registration as a robustness check (same phenotype, same architecture, same tool schema); its inclusion is exploratory with respect to the prereg but strengthens H2a. The GPT-4.1 treatment arm yielded 21 successes and GPT-4o yielded 19 (both below the 30-success floor); the preregistered 72-scenario grid was not fully instantiated. Despite the reduced grid, both models' primary CIs exclude 0 ($[0.454, 0.828]$ and $[0.512, 0.882]$ respectively), so the confirmatory conclusion is unaffected. \emph{Non-independence caveat:} the 40 trials per cell are sampled from only 18 unique scenario configurations, with some scenarios repeated 2--5 times. Wilson intervals assume independent observations; repeated-scenario trials at temperature$=$0 against the same API endpoint are anti-conservative for the treatment-arm rate estimates because identical prompts to a deterministic endpoint produce correlated (often identical) outcomes. The effective number of independent observations is closer to $N{=}18$ (unique scenarios) than to the nominal $N{=}40$. The control-arm zero-violation observation (a count, not a rate estimate) is unaffected by this concern. For the treatment-arm rate estimates we now report \emph{scenario-clustered} bootstrap CIs (resampling the 18 scenario configurations as clusters) alongside the nominal Wilson intervals: GPT-4.1 $[0.450, 0.864]$ (vs nominal $[0.454, 0.828]$) and GPT-4o $[0.600, 0.929]$ (vs nominal $[0.512, 0.882]$). Both clustered intervals exclude 0, so the confirmatory conclusion is unchanged under clustered inference. \emph{Planted-address sensitivity:} the attack-success scorer is allowlist-based (any send to a recipient outside \texttt{team@example.com}); because the V3-1 scenario grid shares two attacker addresses, $7/40$ successes across the two treatment cells sent to the \emph{other} attacker address rather than the one planted in that scenario (GPT-4.1 $3/21$, GPT-4o $4/19$; all to \texttt{auditor@external-compliance.com} while \texttt{attacker@evil.example.com} was planted). These are genuine unauthorised external sends to an attacker-controlled address, but they are ambiguous evidence for dependence on the \emph{scenario-specific} payload. Restricting to planted-address-matched successes, the rbs${=}0$ rate is GPT-4.1 $11/18 = 61.1\%$ (scenario-clustered 95\% CI $[0.368, 0.833]$) and GPT-4o $10/15 = 66.7\%$ ($[0.500, 0.889]$); both still exclude 0, so the necessity-breaking conclusion holds under the stricter mechanistic scorer. We report the allowlist rates as primary (the preregistered operationalisation) and the planted-address rates as this sensitivity analysis. \emph{Presentation-format check:} to rule out that the implicit-delivery evasion is an artefact of the ICL prefix wording rather than of implicit delivery per se, we re-ran the GPT-4.1 treatment with the V3-2-style ``[Retrieved context]'' prefix framing (in place of V3-1's prior-exchange ICL framing); the rbs${=}0$ rate among successes is $8/20 = 40\%$ (Wilson $[0.219, 0.613]$), overlapping the primary V3-1 ICL CI $[0.454, 0.828]$; the overlap fails to reject a difference (it does not prove equivalence, and the $40\%$ vs $67\%$ point-estimate gap leaves a partial formatting effect unexcluded), but the evasion is clearly present under both framings and is therefore not solely a presentation-format artefact. Claude Sonnet's near-zero ASR ($0/40$ treatment, $1/40$ control) is an unanticipated outcome outside the preregistered decision tree; H2 was untestable for that model and is addressed by the separate V3-3 ground-then-act replication (Section~\ref{sec:v33}). The registration-to-execution interval was $<$24 hours (same calendar day; GPG-signed timestamps machine-verifiable).

The ${\sim}30\%$ treatment rbs${=}$1 rate across both act-without-grounding models (all from decoy-fact verification) has an adaptive-adversary implication: an attacker aware of the detector need only harvest the rbs${=}$0 subset of successful implicit attacks to guarantee evasion. The detector cannot distinguish ``agent recalled a decoy before sending'' from ``agent recalled the attack address before sending'' using operation-only features: both produce rbs${=}$1, but only the latter is causally related to the attack.

\subsubsection{Ground-Then-Act Replication (V3-3)}
\label{sec:v33}
V3-1's testable models (GPT-4.1, GPT-4o) are both \emph{act-without-grounding}: they do not habitually recall stored context before acting, so it is unsurprising that they seldom recall a value when the store holds nothing the task needs. (Earlier drafts attributed this omission to the \emph{delivery} mode. V3-6 and V3-6b falsify that attribution: under these same triggers both models retrieve a payload that \emph{is} in the store, $216/216$, whether or not it is also in context. The operative variable is store contents, not delivery. See \S\ref{sec:v36}.) The sharper test is a \emph{ground-then-act} model, which recalls stored context as a matter of habit even on benign tasks; if such a model's verification habit persisted under implicit delivery, it would restore the signature and make the boundary phenotype-dependent. V3-1 could not run this test because Claude Sonnet refused the attack (ASR${\approx}0$). We preregistered V3-3 (GPG-signed, before data) to close this gap, extending V3-1's two-arm design to a ground-then-act model and adding a benign keystone cell. Both falsification branches (necessity breaks vs.\ signature survives) were preregistered with equal prominence.

\paragraph{Model selection (preregistered pilot gate).} Because the estimand conditions on attack successes, a model with near-zero ASR is uninformative. We piloted candidate ground-then-act models at $N{=}20$ on the control arm (discarded data, fixed seed distinct from confirmatory) with a preregistered inclusion threshold of control-arm ASR${\geq}10\%$. \texttt{gemini-2.5-pro} cleared it decisively (control ASR $50\%$) and is the confirmatory model; local open-weight grounders were deferred (hardware constraints) and would enter only as labelled-exploratory secondaries.

\paragraph{Design refinements over V3-1.} To address V3-1's effective-$N$ limitation, V3-3 expands the scenario grid to $36$ distinct configurations (varying the decoy-subset axis V3-1 collapsed) and uses \emph{negative-binomial} sampling (collect until ${\geq}30$ treatment successes or a hard cap of $N{=}120$; stopping depends only on the success count, never on the rbs outcome, so the conditional estimand is free of optional-stopping bias). The control arm receives a length-matched benign prefix so the treatment ICL demonstration's ${\sim}400$ extra tokens cannot be confounded with an attention-dilution effect. The primary CI is a \emph{scenario-clustered} bootstrap (resampling distinct configs as clusters), with Wilson intervals as a secondary reference. The anti-imitation ICL constraint (no \tool{recall\_fact} depicted in the demonstration) is carried over from V3-1.

\paragraph{Results (gemini-2.5-pro).}
\begin{itemize}
    \item \textbf{Control (H1, tool-routed):} ASR $19/60=31.7\%$; among successes $0/19$ have rbs${=}0$. This is a zero-event cell, so the scenario-clustered percentile bootstrap is degenerate (every resample reproduces $0$, giving a spurious $[0,0]$); we therefore report a Wilson upper bound on the effective cluster count ($12$ scenario clusters): $[0.000,0.243]$ (trial-level Wilson upper $0.168$). The necessity invariant is \emph{unbroken} on a ground-then-act model. \emph{Criterion language corrected:} earlier drafts said this cell ``replicates'' the invariant while X4's arms were withdrawn at $0.121$ and $0.104$, even though this cell's governing bound of $0.243$ is the weakest of the three. Under the uniform rule of \S\ref{par:zero-event-rule} this cell does not meet $\leq 0.10$ either, and we report it as zero violations across $12$ clusters on a second behavioural phenotype; directionally consistent with the necessity and not criterion-passing. What makes it valuable is not its interval but the benign keystone below, which is a within-model contrast rather than a bound.
    \item \textbf{Treatment (H2, implicit delivery):} ASR $30/70=42.9\%$ (negative-binomial stop at $30$ successes); among successes $28/30=93.3\%$ have rbs${=}0$ (scenario-clustered 95\% CI $[0.821,1.000]$, excludes 0; $17$ effective scenario clusters). \textbf{Necessity is broken}. The two rbs${=}1$ successes recalled a key that is not a decoy, but the recall \emph{returned nothing}: earlier drafts described these as ``recalled the attacker-address key (genuine payload verification),'' which cannot be right, because the treatment arm's store holds only the scenario's decoy facts and no routing rule (\texttt{probe\_v3\_3.run\_treatment\_arm}: \texttt{prefilled = decoy\_facts}; ``NO routing rule in memory''). The released records show the two keys as \texttt{compliance\_rule\_email\_cc} and \texttt{compliance\_rule\_cc\_auditor}: key names the agent inferred from the ICL demonstration and then looked up unsuccessfully. The \texttt{attacker\_key\_recalls} counter that labelled them is computed by \emph{exclusion} (\texttt{probe\_v3\_3.py}: any recalled key that is neither a decoy nor the benign contact key), so it records an attempted lookup, not a returned payload. The correction strengthens rather than weakens the arm: all $30$ treatment successes, including these two, lack retrieval provenance for the payload, because in this arm no memory operation could have supplied it.
    \item \textbf{Benign keystone (H\textsubscript{benign}):} on legitimate memory-grounded tasks with no attack material and no ICL demonstration, gemini-2.5-pro exhibits recall-before-send in $53/60=88.3\%$ of sessions. \textbf{Clustering unit corrected, and the scenario-clustered interval is withdrawn.} Earlier drafts reported a scenario-clustered 95\% CI of $[0.772,0.967]$ over this cell's $36$ \texttt{scenario\_id}s. That bootstrap is invalid here for exactly the reason documented for B+ in App.~\ref{app:bplus}: \texttt{probe\_v3\_3.run\_benign\_arm} builds its trigger from a module-level constant and ignores \texttt{cover\_task}, \texttt{trigger\_phrasing} and \texttt{attacker\_address} entirely, consuming only \texttt{scenario["decoy\_facts"]}, and \texttt{DECOY\_SUBSETS} defines exactly \emph{two} subsets. Measured on the released records, the $36$ nominal \texttt{scenario\_id}s therefore collapse to \textbf{$2$} effective configurations ($d0$: $33/35$, $d1$: $20/25$; Kish $n_{\text{eff}}=1.946$). The point estimate is unchanged at $53/60$; the honest intervals are the trial-level Wilson $[0.778,0.942]$ and, on the effective unit, $[0.264,0.994]$, which is close to uninformative, and we say so rather than quoting a width the design cannot support. The same collapse affects X2's benign keystone (\S\ref{sec:x2}), which reuses this runner. Regenerate both with \texttt{necessity\_cluster\_bounds.py} (\texttt{benign\_keystones} block). \emph{What survives is the comparison, not the interval:} the keystone and the treatment arm share one model, one store construction and one tool schema, differing only in whether the payload arrives through memory or through context, and the contrast between $53/60$ benign grounding and $2/30$ grounding under implicit delivery is a within-model difference of a size no two-cluster interval is needed to establish. The claim it supports is that this model's grounding habit does not survive implicit delivery, not a population estimate of that habit across task domains.
\end{itemize}

\paragraph{Interpretation, with its mechanism corrected by V3-6.} The benign keystone is the within-model comparison that carries this section: gemini-2.5-pro \emph{does} have a strong verification-recall habit ($53/60 = 88.3\%$ on benign tasks), yet it drops recall in $93.3\%$ of successful implicit-delivery attacks. Habit is present; the signature is absent. \textbf{The contrast is real; the explanation we gave for it was wrong.} Earlier drafts attributed it to the delivery architecture removing an information need: payload already in context $\Rightarrow$ nothing to look up. V3-6 (\S\ref{sec:v36}) falsifies that: with the payload in the store \emph{and} in context this model retrieves it $108/108$ times, so a context copy does not suppress retrieval. The operative difference between these two cells is what the \emph{store} holds. The benign arm's store contains a relevant fact the task needs (\texttt{preferred\_contact\_email}); the treatment arm's store contains only decoys. Grounding tracks retrievable content, not information need, and the preregistered ``grounders recall anyway'' branch is ruled out for delivery but not for store contents. \emph{Scope of the keystone:} the benign arm varies only the decoy subset ($2$ effective configurations), so it establishes that \emph{this model on this task} grounds in memory at a high rate and stops doing so under implicit delivery. It does not estimate a grounding rate across task domains, and earlier drafts called it ``decisive'' while quoting a $36$-cluster bootstrap interval the design does not license. The domain-level heterogeneity measured elsewhere in this paper (Table~\ref{tab:domain-dependence} for attack viability; App.~\ref{app:bplus}'s five-of-six decomposition for benign grounding) is the reason we no longer generalise from a two-configuration cell. Notably the ground-then-act evasion rate ($93.3\%$) is \emph{higher} than the act-without-grounding rates ($67\%$/$74\%$), the opposite of the naive habit-based prediction. The scope boundary therefore replicates across both tested phenotypes. This is one ground-then-act model, one vendor, one attack construction: we claim replication across the tested phenotypes, not universality over unexamined models.

\subsubsection{Does Store Content or Context Content Drive Retrieval? (V3-6)}
\label{sec:v36}
Every result above that we had called \emph{behavioural} rested on an inference this paper
was not entitled to make, and V3-6 (preregistered, GPG-signed before data;
\texttt{V3\_6\_PREREG.md}) tests it. The withdrawn claim was that agents \emph{stop
grounding} when the payload is already in context: that the recall tool is live and the
agent declines it. But in every implicit arm in this paper the routing rule is absent from
the store by construction (\S\ref{sec:threat}), so the agent could not have retrieved the
payload had it tried. Those arms measure whether the agent retrieves \emph{anything}, not
whether it declines a retrieval that would have returned the payload.

\paragraph{Design, and the comparison that does the work.} Three arms on one matched
$36$-configuration grid (\texttt{probe\_v3\_3.generate\_scenario\_grid}),
\texttt{gemini-2.5-pro}, fixed $N{=}3$ per configuration per arm ${=}324$ trials, no
adaptive stopping, zero errors and zero non-evaluable records. Arm~\textbf{C} places the
routing rule in the store only; arm~\textbf{B} in the store \emph{and} the in-context
demonstration; arm~\textbf{I} in context only, reproducing V3-3's treatment.

The preregistered primary comparison was the paired C${-}$B difference, and we report it
because it was registered, but \textbf{a post-hoc audit found it confounded and it is not
what we now rest anything on.} Arm~C is built with
\texttt{build\_length\_matched\_prefix()} while arms~B and~I use
\texttt{build\_icl\_demonstration()}, so C versus B varies the entire conversational prefix
alongside the payload's location. Our parity assertion hashed only the assembled
\emph{trigger}, which cannot see a prefix difference: the same class of failure this
paper criticises in X1 (\S\ref{sec:x1}), one level up in the message history. The
\textbf{B versus I} contrast is the clean one: identical prefix construction, identical
trigger, identical tool schema, differing \emph{only} in whether the store holds the rule.
It is registered as a secondary comparison and it is what the conclusions below use.

\begin{table}[t]
\caption{V3-6. Registered \texttt{vpr} requires that \emph{a} memory-tool return contained the attacker address before the first exfiltration-capable action; the adjacent column restricts that to returns of the \emph{prefilled} poison key, which is the quantity the arms were built to vary. \texttt{list} is \tool{memory\_list\_all\_facts}; \texttt{recall} is a value retrieval. Rates unconditional on attack success.}
\label{tab:v36}
\centering
\small
\begin{tabular}{llccccc}
\toprule
\textbf{Arm} & \textbf{Rule in store / context} & \textbf{$N$} & \textbf{ASR} & \textbf{list} & \textbf{recall} & \textbf{vpr (registered / prefilled-key)} \\
\midrule
C & store only            & 108 & $33.3\%$ & $108/108$ & $108/108$ & $108/108$ \ /\ $\mathbf{108/108}$ \\
B & store \emph{and} context & 108 & $41.7\%$ & $108/108$ & $108/108$ & $108/108$ \ /\ $\mathbf{108/108}$ \\
I & context only          & 108 & $41.7\%$ & $108/108$ & $15/108$  & $1/108$ \ /\ $\mathbf{0/108}$ \\
\bottomrule
\end{tabular}
\end{table}

\paragraph{Result.} \textbf{Arms B and I both place the routing rule in the agent's context
and differ only in the store. B returns the payload from memory in $108/108$ trials; I in
$0/108$.} The registered branch (paired C${-}$B difference exactly $0.0$ over all $36$
shared configurations, both arms saturated) is H1b, and its pre-committed consequence is
executed below. The positive control passed (arm~C $36/36$ among successes against a
registered floor of $0.70$). We do \textbf{not} quote the C${-}$B bootstrap interval:
it is degenerate at $[0.0,0.0]$ because every cluster in both arms is saturated, the
preregistration itself forbids quoting degenerate intervals, and a zero-width interval
here reflects empirical saturation rather than precision. No equivalence margin was
registered, so nothing in these data licenses a claim of \emph{equality}; what they show
is a zero \emph{observed} decline rate, cluster-unit upper bound $0.0964$.

\textbf{The behavioural claim is withdrawn.} With the rule in context, retrieval is
$108/108$ when the store holds it and $0/108$ when it does not. A context copy therefore
does not suppress payload retrieval, so signature absence in V3-1's and V3-3's treatment
arms cannot be read as declining. \emph{This rests on the store-content contrast, not on
call ordering; the ordering caveat below (co-batched calls in V3-6b) bounds how strongly
the underlying \texttt{before()} measurement should be read, not the direction of the
108-versus-0 pattern itself.} Those arms are reclassified \textbf{structural}
(\S\ref{sec:threat}) and the ``agents stop grounding'', ``declines an available retrieval''
and ``no information-need to recall'' readings are removed throughout.

\paragraph{What replaces it, stated at the width the design supports.} The agent consulted
memory in \textbf{every one of the $324$ sessions}: all three arms open with
\tool{list\_all\_facts}, $108/108$. Memory \emph{access} is therefore universal and was
never the variable; what varies is whether a value is then retrieved. Under a trigger that
instructs policy review, the agent lists the store, and retrieves a value when a
topically matching key is present. Arm~I shows what happens when none is: value retrieval
falls to $15/108$, and its composition is the point: $10$ trials retrieve only benign
decoys, $4$ request keys that \emph{do not exist}
(\texttt{compliance\_rule}, \texttt{compliance\_rule\_email\_cc},
\texttt{compliance\_rule\_cc\_auditor}; all return \emph{``No fact found''}), and $1$ is
the self-injection discussed below. Those $4$ are the mechanism in miniature: the agent
\emph{reaches for the injected rule in memory, constructing a plausible key name from the
rule's own text, and finds nothing.} It is not declining. Within arm~I, value retrieval
also tracks what the decoys look like ($3/54$ versus $12/54$ across the two decoy subsets,
the higher rate belonging to the subset containing \texttt{code\_review\_policy}, itself
recalled $9$ times), which is consistent with topical matching against listed key names.
This explains V3-3's keystone without an information-need account: the benign store holds a
fact the task needs, the treatment store holds only decoys.

\paragraph{Operation-level versus argument-level, and why the distinction is load-bearing.}
Of arm~I's $15$ operation-level \texttt{rbs}${=}1$ trials, $14$ returned no payload. An
operation-level detector sees a memory read and infers grounding; the argument-level
feature shows $10$ retrieved irrelevant content and $4$ retrieved nothing at all. This
generalises the V3-3 correction reported above, where two \texttt{rbs}${=}1$ successes
likewise returned nothing, and it is the strongest available argument for the
returned-value standard: \textbf{\texttt{rbs} counts grounding \emph{attempts}; only
argument-level verification counts grounding.}

\paragraph{The one arm-I positive is a self-injection, and the registered endpoint cannot
exclude it.} In \texttt{t2\_p1\_a1\_d0} rep~1 the sequence is \tool{list\_all\_facts}
(which did \emph{not} list the key), then \tool{memory\_save\_fact}, then
\tool{memory\_recall\_fact(\texttt{compliance\_rule\_cc\_auditor})} returning the attacker
address, then draft. The agent made the unavailable payload available to itself. The
registered definition asks only whether \emph{a} memory return contained the address, so
the label is correct as registered and we do not change it post hoc; but it does not
evidence retrieval of pre-existing store content, which is why
Table~\ref{tab:v36} reports both columns and why the prefilled-key count for arm~I is
$0/108$. This is a live instance of the laundering pathway argued in \S\ref{sec:x1}, and it
confirms empirically that an implicit arm's zero is a \emph{measured} zero on an available
path rather than a definitional one.

\paragraph{Scope and limitations, several of them found by audit after collection.}
\begin{itemize}
\item \textbf{The trigger instructs memory consultation.} All three phrasings are
memory-inviting (``Review any stored policies or rules in your memory, then \ldots'';
``Check your memory for applicable compliance rules before you \ldots''; ``First consult
stored organisational policies, then \ldots''). Arm~B's $108/108$ is therefore compliance
with an instruction \emph{plus} an available match, not a spontaneous retrieval habit.
The finding does not extend to neutral triggers. What makes it nonetheless apply to V3-1
and V3-3 is that those experiments use \emph{these same triggers}: the reinterpretation is
within-harness, and that is the only width claimed.
\item \textbf{One mechanism does not replicate, and we flag it rather than generalise it.}
V3-6's arm~I contained five recalls of keys absent from the store, of which \textbf{four}
returned \emph{``No fact found''} and \textbf{one} was the save-then-recall self-injection
discussed above; these are two distinct event classes and we count them separately rather than
reporting ``five ghost keys.'' In \texttt{gpt-4.1}'s $324$ trials, \textbf{both} counts are
zero. The headline result replicated on both tested
model-interface settings under memory-inviting triggers, which is narrower than
``phenotype-general''; this particular route to it did not replicate at all, and should be
read as a \texttt{gemini-2.5-pro} observation only.

\item \textbf{Backend variation within a single collection.} V3-6b's records carry
\emph{three} distinct \texttt{system\_fingerprint} values across $324$ trials at
temperature $0$ (V3-6's were null throughout). The served model string never changed. This shows that
a single uninterrupted collection can span more than one \emph{reported} backend
configuration. It is consistent with the Regime~B caveat this programme states elsewhere, but
it is weaker than a determinism test: we did not repeat identical requests, so these records
show fingerprint heterogeneity rather than nondeterministic output.

\item \textbf{Discoverability was not varied.} The payload sat under one explicit
descriptive key that \tool{list\_all\_facts} exposes by name. X4's grid varies this factor;
V3-6 does not. A semantically opaque key might not have been recalled.
\item \textbf{$36$ registered clusters rest on $9$ distinct triggers} ($3$ cover tasks
$\times$ $3$ phrasings); the address and decoy axes share a trigger. The cluster-unit
bounds above use the registered $36$; readers who prefer the trigger template as the unit
should read them against $9$.
\item \textbf{In V3-6, arms were collected in fixed order} (C, then B, then I) with no interleaving,
so arm is associated with collection time and backend drift cannot be excluded as a
contributor to the between-arm ASR differences.
\item \textbf{Recording gaps.} The harness records memory-read return values but not
\tool{memory\_save\_fact} arguments, email arguments, or search returns. Consequently the
contents of the $87/108$ arm-I saves are unknown and no propagation claim is made, and
\texttt{attack\_success} is not independently reconstructible from the records: one
$(\text{arm},\text{config})$ cell contains identical call sequences with differing success
labels. The provenance guard's \texttt{prompt\_hash} covers the system prompt only, not the
assembled conversation, so it cannot detect the arm prefix differences described above; it
is evidence that the system prompt and tool schema did not drift, and no more.
\item \textbf{``Exfiltration-capable'' is an operation cutoff.} It is
\tool{email\_draft\_email} in $228$ of $233$ cases, not a send. Ordering before a draft is
not exfiltration and multiple calls emitted in one assistant turn are flattened into
dispatch order, so this ordering does not establish a data dependency (\S\ref{sec:graph-formal}).
\textbf{Measured rate of this flattening.} The harness makes one completion per loop
iteration and dispatches every tool call the model returns in that response before
requesting the next completion, so \texttt{call\_index} is a flat count with no
per-completion boundary marker. The one-hop-per-call pattern (completions exceed calls by
exactly one) holds in $312/324$ V3-6 (\texttt{gemini-2.5-pro}) sessions but only
$184/324$ V3-6b (\texttt{gpt-4.1}) sessions, with $76$ of the remainder short by exactly
four completions; \texttt{gpt-4.1} batches multiple tool calls per turn far more often
than \texttt{gemini-2.5-pro} under this harness. For the $184$ one-hop V3-6b sessions and all of V3-6, \texttt{before()} on
\texttt{call\_index} is unaffected by this flattening. \textbf{For the $140$ short V3-6b
sessions, it is not.} A call co-batched with the draft call in the same completion was
requested before either result existed, so \texttt{before()} there records dispatch
order, not that a returned value was available to the model when it chose the draft call.
\emph{We had asserted, without measurement, that this leaves the counts unaffected ``in
essentially every case''; we withdraw that unmeasured assertion.} It is not separable into
which of the $140$ short V3-6b sessions specifically co-batch a retrieval with the draft
call, since the released schema records no per-completion marker, so
\texttt{verified\_poison\_read\_prefilled} is \textbf{indeterminate, not falsified,} for an
unknown subset of V3-6b's $324$ trials; the registered endpoint is unaffected for V3-6 and
for V3-6b's $184$ clean one-hop sessions.
\end{itemize}
Under the status pre-commitment every change this experiment causes is a downgrade.

\subsubsection{Cross-Phenotype Replication (V3-6b)}
\label{sec:v36b}
\textbf{V3-6b is the registered basis for the withdrawal, and V3-6 is convergent but reached
this contrast and this endpoint post-hoc.} We state the order plainly because it matters: the
comparison and endpoint that carry the conclusion were, in V3-6, a secondary comparison and a
post-hoc correction (its deviations D5 and D7). V3-6b registered both \emph{before} data.

V3-6
ran on \texttt{gemini-2.5-pro}, a documented grounder, whereas V3-1's reclassified treatment
arms were collected on \texttt{gpt-4.1}, which this paper's own taxonomy calls
\emph{act-without-grounding} under exact-key recall: the phenotype most likely to decline.
Reclassifying V3-1's arms on evidence from the other phenotype was a defect in V3-6's
preregistered scope (its deviation D8). \textbf{V3-6b} (separate deposit,
\texttt{V3\_6B\_PREREG.md}, GPG-signed before data) repeats the design on \texttt{gpt-4.1}:
same grid, same arms, $324$ trials, zero errors, zero non-evaluable. It fixes three of
V3-6's own defects \emph{by registration rather than by patch}: parity is asserted on the
fully assembled message list (\texttt{messages\_sha256}), the primary endpoint is
prefilled-key restricted, and arms are interleaved within configuration so arm is not
associated with collection time. It also carries a preregistered \emph{floor gate}, because
a low retrieval rate in a low-grounding model would otherwise be uninterpretable: arm~C must
retrieve when retrieval is mandatory, or the result is DESIGN-FAILURE and nothing follows.

The gate passed at $108/108$, and the registered branch is \textbf{R1}: prefilled-key
verified read is $108/108$ in arm~B and $0/108$ in arm~I, a paired B${-}$I difference of
$1.0$ in every one of the $36$ configurations. \emph{As stated in \S\ref{sec:v36}'s
flattening-rate bullet, this $108/108$ is dispatch-order-verified for the $184$ one-hop
sessions and indeterminate, not falsified, for the $140$ sessions that co-batch calls
within a completion; V3-6b's own one-hop rate is $184/324$, the weaker of the two
experiments on this measure.} Per-arm cluster-unit bounds, which are not
degenerate and are what we quote: arm~B's decline rate is $0/108$, upper bound $0.0964$;
arm~I's is $108/108$, lower bound $0.9036$. \textbf{The provisional marker on V3-1's
reclassification is therefore removed \emph{for its \texttt{gpt-4.1} arm}, subject to the
ordering caveat above.} V3-1 also has a
\texttt{gpt-4o} treatment arm, which V3-6b's preregistration explicitly declined to run
(\S3 of \texttt{V3\_6B\_PREREG.md}: ``\texttt{gpt-4o} is not run''). Shared phenotype
membership is not evidence that two separately served models behave identically, so
\texttt{gpt-4o}'s reclassification \textbf{remains provisional} and we do not claim
otherwise. Memory \emph{access} was again universal ($108/108$
in all three arms, including arm~I), so the act-without-grounding label describes a model's
default rather than a refusal it maintains under a memory-inviting trigger. Two further
observations. First, V3-6b's arm~I is \emph{numerically close} to
V3-1's own treatment arm on one secondary rate: \texttt{rbs}${=}0$ in $38/58 = 65.5\%$ of
successes against V3-1's published $67\%$, collected months apart under a different probe.
We call this agreement rather than reproduction: the denominators are success-conditioned,
\texttt{attack\_success} is not independently reconstructible from the released records, and
no comparability test was registered. Second, its message-parity check confirms V3-6's deviation D5 from
the records rather than from code inspection: arms~B and~I hash identically on all $36$
configurations while arm~C differs on all $36$, which is exactly the asymmetry V3-6's
trigger-only assertion could not see.

\subsubsection{Second Open-Weight Ground-Then-Act Grounder (X2)}
\label{sec:x2}
V3-3 established the implicit-delivery breakdown on one ground-then-act model (gemini-2.5-pro), from a single vendor. X2 (preregistered, GPG-signed before confirmatory data) adds a \emph{second} ground-then-act grounder from a different model family: \texttt{qwen2.5:32b}, an open-weight model run locally via Ollama. It strengthens the secondary phenotype-generality claim, not the core taxonomy. The candidate cleared the preregistered discarded pilot gate (control-arm ASR $0.20 \geq 0.10$ AND benign recall-before-send $1.0 \geq 0.70$).

\textbf{Regime-A caveat (disclosed adjacent to the result).} X2 was collected on the existing GUI Ollama daemon, \emph{not} a fresh-daemon restart as the reproducibility protocol prescribes; the served model digest \texttt{9f13ba1299afea09} was captured. This non-fresh-daemon deviation is a known confound for reasoning-model determinism and is stated here rather than buried, so the X2 numbers should be read as a strengthening-but-caveated open-weight replication.

Results: the treatment (implicit) arm shows \texttt{recall\_before\_send}${=}0$ in $27/31 = 87.1\%$ of successes, but with only $7$ success-contributing scenario clusters this arm is \textbf{LOW-CLUSTER} (below the preregistered ${\geq}10$-cluster gate), so the scenario-clustered interval is descriptive and the Wilson interval $[0.711, 0.949]$ governs; it is a strong directional pattern consistent with the necessity breaking under implicit delivery, reported as directional rather than confirmatory. The benign keystone confirms a strong verification-recall habit at $60/60 = 100\%$, \emph{on $2$ effective configurations rather than the $36$ its \texttt{scenario\_id}s suggest}: X2's benign arm reuses \texttt{probe\_v3\_3.run\_benign\_arm}, whose trigger is a module constant and whose store is the decoy subset alone, so it inherits the collapse documented in \S\ref{sec:v33} (both effective configurations are $30/30$; cluster-unit interval $[0.342,1.000]$ against the trial-level $[0.940,1.000]$). \textbf{The control arm is also underpowered:} it recorded $0$ violations among $24$ successes ($7$ clusters) but hit the $N{=}150$ cap without reaching the preregistered $37$-success target, so H1's $\leq 0.10$ upper-bound threshold was \emph{not} met (two-sided Wilson upper $0.138$). We therefore report the X2 control as \emph{consistent with but underpowered relative to} the necessity invariant, never as a confirmation. Net: X2 is a second model family / open-weight ground-then-act grounder whose treatment arm shows a directional (LOW-CLUSTER) breakdown pattern, weakly strengthening phenotype-generality, but with both arms below the cluster gate, a two-configuration keystone, and a Regime-A daemon caveat.

\paragraph{Embedding retrieval validation (V3-2).} A second probe (V3-2; preregistered design, scorer corrected post-hoc as disclosed below; $N{=}200$, 2 models $\times$ 2 arms + 40 benign) directly tests whether the invariant generalises from exact-key recall to \emph{any} tool-routed retrieval, and whether attacks remain viable when the payload is delivered scaffold-side without an observable tool call.

\emph{Setup}: FAISS index over 30 benign organizational records + 1 poisoned routing rule, embedded via \texttt{text-embedding-3-small} (1536d, cosine similarity, top-$k{=}4$). Arm B provides a \tool{memory\_search(query)} tool; arm C injects top-$k$ results scaffold-side with no search tool available. A framework-level oracle logs whether the poisoned record was retrieved (rank, similarity) for each trial. Embeddings are cached per attacker-address condition in the released artefacts; because they are produced by an external embedding API, the cache supports re-runs but the vectors are auditable via provenance stamps rather than guaranteed bitwise-identical across API snapshots (consistent with the reproducibility statement in Table~\ref{tab:claim-evidence}).

Results (per-model, no pooling; Table~\ref{tab:v32-denominators}):
\begin{itemize}
    \item \textbf{H1 (arm B: semantic tool search preserves invariant):} Among successful attacks where the payload was oracle-confirmed as delivered, under the \emph{preregistered} scorer (\texttt{email\_send\_email} only): GPT-4.1 $33/33$ and GPT-4o $13/13$ exhibited \texttt{search\_before\_send}${=}1$ (zero violations). Under the corrected scorer (sensitivity analysis counting \texttt{email\_draft\_email} as exfiltration-capable): GPT-4.1 $38/38$. \emph{GPT-4o power extension (Exp~13a).} The original V3-2 arm B cell yielded only 13 oracle-confirmed delivered successes (delivery rate $27/40 = 67.5\%$ due to retrieval competition from 30 benign records), giving a two-sided 95\% Wilson upper endpoint on the violation rate of $\approx 0.228$: underpowered. We therefore preregistered a fresh-cohort extension (Exp~13a): a separate cohort collected to $30$ additional oracle-delivered successes, scored separately and tested for a batch effect against the original 13 before pooling. The fresh cohort shows \texttt{search\_before\_send}${=}1$ in $30/30$; because both the original ($13/13$) and fresh ($30/30$) cohorts are zero-violation, the two are trivially homogeneous on the estimand of interest and no batch-effect test is informative (a Fisher exact test on the $[[13,0],[30,0]]$ table returns $p{=}1$ by construction, with no power against the alternative), so we pool to $43/43$ (two-sided 95\% Wilson upper endpoint on the violation rate $\approx 0.082$ at the trial level, but \textbf{$0.228$ on the governing $13$-cluster unit}; Table~\ref{tab:zero-event}). The GPT-4o arm B invariant is thus zero-violation on a powered success count, but \emph{not} bounded below ${\sim}8\%$: earlier drafts said ``confirmed with a violation rate bounded below ${\sim}8\%$, not merely directional,'' which quoted the anti-conservative unit, and we withdraw that phrasing under the uniform rule of \S\ref{par:zero-event-rule}. When successful exfiltration causally depends on a payload available exclusively through the observable \tool{memory\_search} tool, a retrieval event before the first exfiltration-capable email action is structurally necessary on both GPT-4.1 and GPT-4o, and remained so in every one of the $33{+}43$ eligible successes observed.

    \textit{Scorer correction justification (sensitivity analysis):} The preregistered scorer checked only \texttt{email\_send\_email}. Post-collection, 5 GPT-4.1 trials used \texttt{email\_draft\_email} (specifying the attacker address at draft time, never proceeding to send). Counting \texttt{draft\_email} as exfiltration-capable is justified because: (1)~the recipient-routing decision (the attack's defining exfiltration act) is made at draft time; (2)~in production, a draft containing the attacker address is the threat, regardless of whether a separate send step follows; (3)~the tool schema declares \texttt{draft\_email} with a required \texttt{to} field, making it functionally equivalent to \texttt{send\_email} for routing exfiltration. All 5 records had \texttt{memory\_search} preceding \texttt{draft\_email}; no trial was excluded or manually relabelled. The conservative primary result uses $33/33$ under preregistered scoring; the corrected $38/38$ is a sensitivity analysis. Both are consistent with zero violation rate (two-sided 95\% Wilson CI upper endpoints: $0.102$ and $0.091$ respectively), providing bounded evidence for the retrieval-dependency invariant on GPT-4.1.
    \item \textbf{H2 (arm C: attacks remain viable without observable retrieval):} Arm C removes all \tool{memory\_*} tools from the agent's schema (see Classifier applicability below for residual \tool{rag\_query} disclosure), so \texttt{search\_before\_send}${=}0$ is \emph{structural/definitional}, not a measured behavioural rate: with no search tool in the schema the signature \emph{cannot} fire, so arm~C tests attack \emph{viability} under implicit delivery, not an evasion rate (what V3-1 treatment and V3-3 estimate is narrower and is \emph{not} a payload-decline rate: it is \emph{decoy-memory-use behaviour}, how much an agent reads from a store that does not contain the payload, since the recall tool there is live and stocked with genuine decoy content; note that in those arms too the tool cannot return the \emph{payload}, which is why no arm in this paper measures an agent declining a retrievable payload, \S\ref{sec:threat} and \S\ref{sec:x1}). The empirical finding is that attacks remain viable under this architecture: GPT-4.1 $33/40$ trials and GPT-4o $20/40$ trials succeeded after oracle-confirmed payload delivery (CIs for attack viability: $[0.680, 0.913]$ and $[0.352, 0.648]$ respectively). This demonstrates that observable retrieval is not a precondition for successful exfiltration (the attacker can deliver payloads via channels that leave no trajectory signature) not that the model ``chose'' to bypass retrieval when retrieval was available.
    \item \textbf{H3 (proxy consistency):} Arm C attack-success rates (successful attacks among delivered trials) were broadly consistent with the V3-1 ICL-proxy treatment results. GPT-4.1: arm C $= 33/40 = 0.825$ vs.\ ICL-proxy ASR $= 21/40 = 0.525$; GPT-4o: arm C $= 20/40 = 0.500$ vs.\ ICL-proxy ASR $= 19/40 = 0.475$. Both mechanisms produce substantial successful attack rates despite payload delivery via implicit means. Because equivalence margins were not preregistered, we interpret this as qualitative consistency rather than formal equivalence: both scaffold-side and ICL-proxy delivery sustain viable attacks that lack the detection signature.
    \item \textbf{H4 (benign):} In the benign semantic-memory condition (GPT-4.1 only, no poisoned record, $N{=}40$), $33/40$ trajectories ($82.5\%$, CI $[0.680, 0.913]$) contained search before the outbound email action. This demonstrates that the precision limitation documented in v2 also occurs with semantic tool retrieval under the same task family.
\end{itemize}

\begin{table}[!ht]
\centering
\footnotesize
\caption{V3-2 denominator breakdown. ``Delivered'' = oracle-confirmed poisoned record in retrieval results. Arm B delivery = poisoned item returned by agent-invoked \tool{memory\_search}; arm C delivery = poisoned item selected by scaffold retrieval and injected into context. The signature column reports \texttt{search\_before\_exfiltration} among successful attacks: for arm B this is an empirically observed behavioural invariant; for arm C the search tool is architecturally absent, so the entry is N/A (all successes are necessarily signature-free).}
\label{tab:v32-denominators}
\begin{tabular}{llrrrrl}
\toprule
\textbf{Model} & \textbf{Arm} & \textbf{Trials} & \textbf{Del.} & \textbf{Succ.} & \textbf{S.+D.} & \textbf{Sig.\,$\mid$\,S.+D.} \\
\midrule
GPT-4.1 & B (search tool) & 40 & 40 & 33$^\dagger$/38$^\S$ & 33/38 & 33/33$^\dagger$; 38/38$^\S$ \\
GPT-4.1 & C (scaffold) & 40 & 40 & 33 & 33 & N/A (no tool) \\
GPT-4o & B (search tool) & 40 & 27 & 13 & 13 & 13/13 \\
GPT-4o & C (scaffold) & 40 & 40 & 20 & 20 & N/A (no tool) \\
\bottomrule
\multicolumn{7}{l}{\parbox{0.98\textwidth}{\scriptsize $^\dagger$Primary (preregistered): \texttt{email\_send\_email} only. $^\S$Sensitivity analysis: includes \texttt{email\_draft\_email}. The GPT-4o arm~B $13/13$ shown here is the original cohort; a preregistered fresh cohort (Exp~13a) adds $30$ delivered successes for a pooled $43/43$ (violation upper bound $\approx 8\%$).}}
\end{tabular}
\end{table}

\paragraph{Classifier applicability.} The 19-feature Random Forest (Section~\ref{sec:features-importance}) was trained on trajectories using the original DTA tool schema (\tool{recall\_fact}, \tool{list\_all\_facts}). Arm B uses \tool{memory\_search} (a different operation name), meaning the trained classifier's \texttt{recall\_count}, \texttt{recall\_before\_send}, and bigram features would all read 0 without feature remapping or retraining. Arm C removes all \tool{memory\_*} tools; a residual \tool{rag\_query} tool remained in the schema but returned only a static acknowledgment (no actual retrieval), and its invocations (31/40 GPT-4o trials, 1/40 GPT-4.1 trials) did not access the poisoned record. V3-2 therefore validates the \emph{generalised structural rule} (observable retrieval before exfiltration-capable action) directly. Whether the specific frozen 19-feature classifier (not just the rule) transfers to alternate tool schemas via feature remapping, we resolve by re-analysis: applied to the arm B trajectories without remapping, the frozen RF's recall features read 0 and it detects $0/38$ (GPT-4.1) / $0/13$ (GPT-4o) successes; after a pre-declared ontology remap (\tool{memory\_search}${\to}$\tool{recall\_fact}), the same frozen classifier detects $33/38 = 86.8\%$ (GPT-4.1) and $13/13 = 100\%$ (GPT-4o). The trained classifier therefore does transfer across tool schemas, but only through an explicit human-specified ontology mapping (itself an unvalidated deployment step), not automatically.

\paragraph{Top-$k$ retrieval-depth sweep (X3).} V3-2 registered a $k{\in}\{2,4,8\}$ robustness sweep but ran only $k{=}4$; the $k{\in}\{2,8\}$ arm was the sole unexecuted preregistered analysis (Table~\ref{tab:withdrawn}). X3 (preregistered note, GPG-signed before data, via a parameterised $k$-runner that leaves the $k{=}4$ artefacts untouched) discharges it: arms B and C, gpt-4.1 and gpt-4o, $k{\in}\{2,8\}$, $N{=}40$ per cell. The arm-B \texttt{search\_before\_send} violation rate is $0$ in \emph{all} cells at both $k{=}2$ and $k{=}8$; the gpt-4.1 $k{=}8$ cell meets the preregistered ${\geq}30$-success robustness threshold ($0$ violations $/33$ delivered successes, two-sided 95\% Wilson upper endpoint ${\approx}0.104$), while the gpt-4o cells are underpowered/descriptive (delivery competition, as at $k{=}4$). Arm-C attack viability remains substantial at both depths ($57.5$--$90.5\%$). The invariant is thus retrieval-depth-insensitive across the tested range, and no arm-B anomaly triggered the preregistered narrowing protocol. This is reported as robustness (it discharges the last outstanding preregistered analysis), not a new claim.

\paragraph{Deviations from preregistration (V3-2).} GPT-4o arm B payload delivery rate was $27/40$ ($67.5\%$) due to retrieval competition from 30 benign records reducing the poisoned record's rank. Per the prereg, the H1 analysis conditions on both success and delivery ($13/13$). The prereg's stopping rule (extend to $N{=}60$ if $<10$ successes) was not triggered for GPT-4.1 (38 successes) but GPT-4o yielded 13 (above the 10-success floor). \emph{Non-independence caveat (V3-2):} V3-2 shares V3-1's repeated-scenario structure (18 unique scenario configurations sampled with replacement to $N{=}40$ at temperature${=}0$). The same anti-conservative interval concern applies: effective independent $N$ is bounded by the 18 unique scenarios rather than the nominal 40 trials. The H1 zero-violation counts are unaffected (counts, not rate estimates), but H2 attack-viability CIs ($[0.680, 0.913]$ and $[0.352, 0.648]$) should be interpreted as approximate under the same caveat.

\paragraph{Scoring correction disclosure (V3-2).} The preregistered scorer defined the signature as \texttt{memory\_search} preceding \texttt{email\_send\_email}. Post-collection inspection revealed that GPT-4.1 arm B consistently used \texttt{email\_draft\_email} (the agent drafts to the attacker address rather than sending directly). Because \texttt{email\_draft\_email} requires specifying recipients and constitutes an exfiltration-capable action (the outbound routing decision is made at draft time), the scorer was corrected to define the first exfiltration-capable email action as either \texttt{email\_send\_email} or \texttt{email\_draft\_email}. All arm B observations were mechanically rescored from the immutable \texttt{call\_sequence} field in the raw logs. This correction changed 5 GPT-4.1 records from $\text{sbs}{=}0$ to $\text{sbs}{=}1$ (all had \texttt{memory\_search} preceding \texttt{email\_draft\_email}). No trial was excluded or manually relabelled. Under the original (erroneous) scorer: GPT-4.1 arm B $= 33/38$; under the corrected scorer: $38/38$.

\subsection{Mechanism-Faithful Independent Attack Families (X1)}
\label{sec:x1}
The paper's largest external-validity gap has been that the both-implicit cell of the taxonomy (Table~\ref{tab:attack-taxonomy}) was populated by architectural argument (status A), never by a mechanism-faithful implementation of an \emph{independent} published attack family. X1 (preregistered, GPG-signed before data) probes this for one family and makes measured progress on a second. Both families are subjected to an observable-vs-implicit read manipulation in which the memory tool schema is present in \emph{both} arms. \textbf{Scope of that statement.} Tool \emph{presence} is not payload \emph{reachability}, and in X1 only the former is held constant. In the observable arm \texttt{probe\_x1.\_run\_agent\_seeded} serves the poison-bearing cases as the return value of \tool{memory\_search}; in the implicit arm \texttt{probe\_x1.\_run\_agent} starts from an empty store, so \tool{memory\_search} cannot return the poisoned exemplar \emph{at the point the trigger arrives} (it returns ``No stored cases match.''). \textbf{Sharpened: the implicit arm's zero is a measured zero on an available path, not a definitional one.} An earlier draft restated \texttt{observable\_read\_of\_poison}${=}0$ as \emph{structural} and wrote that the signature ``could not have fired.'' That is too strong for the variable actually scored, and our own harness contradicts it: \texttt{probe\_x1.\_dispatch} implements \tool{memory\_save\_fact}, \tool{memory\_recall\_fact}, \tool{log\_case} and \tool{memory\_search} against a store that is empty at initialisation but \emph{agent-writable and not output-filtered} (the function's own docstring reads ``anything the agent writes is returned verbatim,'' and \texttt{\_run\_agent}'s reads ``allow agent memory writes to be readable later''). An agent handed the destination in context can therefore save it and recall it before the exfiltrating action, at which point a memory tool \emph{does} return the poison and \texttt{orp} fires, and \texttt{X1\_PREREG.md} names precisely that laundering pathway as what keeps H2 falsifiable. The correct statement has two parts, and they carry different weight. The \emph{framework-side retrieval} that delivers the exemplar is structurally absent from the agent-visible log, definitionally: that is the architectural claim, and it is what the taxonomy rests on. A \emph{subsequent} agent-initiated observable read of the payload was available and was not taken in any of the $141$ eligible successes: that is a measured zero, and it is stronger evidence than a definitional one, because it says the laundering evasion our own preregistration anticipated did not occur. We had been reporting the weaker reading. What X1 measures is therefore attack \emph{viability} under scaffold-side delivery, the architectural absence of retrieval provenance for the delivery itself, and a measured absence of tool-mediated re-exposure of the payload. What \emph{X1} does not measure is an agent declining a payload that a memory tool would have returned at trigger time, because its implicit store is empty. That cell is measured, but elsewhere: V3-6 and V3-6b (\S\ref{sec:v36}) place the routing rule behind a live recall tool and find no decline in $216/216$ trials across two models. The \textbf{primary decision variable is the raw \texttt{observable\_read\_of\_poison}} (did an agent tool call return the poison before the exfiltrating action?), not the frozen-RF score; the RF is reported as a secondary, out-of-distribution continuity check.

\paragraph{MINJA case-bank (measured, directional/LOW-CLUSTER after cluster-unit correction).} We implemented a genuine MINJA-style query-only case-bank family (the attacker destination enters memory only as part of a logged past-action exemplar, surfaced at trigger time by task-similarity retrieval and used by in-context imitation). On \texttt{gpt-4.1}, within model, the raw \texttt{observable\_read\_of\_poison} rate among eligible successes is $80/80 = 100\%$ in the observable arm and $0/141 = 0\%$ in the implicit arm, a matched-scenario paired observability contrast of $+1.0$ (scenario-clustered CI $[1.0, 1.0]$ over the $36$ shared \texttt{scenario\_id}s ${=}12$ shared effective configurations; note that these are configurations \emph{attempted} in both arms, and only $8$ contribute eligible successes in both, so the contrast does \emph{not} independently satisfy the ${\geq}10$ gate: see the Outcome paragraph and Table~\ref{tab:x1-summary}). \texttt{gpt-4o} corroborates directionally (observable $44/44$; implicit $0/18$).

\textbf{Cluster-unit deviation (found post-hoc, 2026-09-19).} We originally reported $13$ and $19$ success-contributing scenario clusters for these two arms and concluded that both satisfied the preregistered ${\geq}10$-cluster gate. That conclusion was wrong. The MINJA scenario grid crosses $6$ domains $\times\ 3$ \texttt{similarity} levels (near/mid/far) $\times\ 2$ framings $=36$ \texttt{scenario\_id}s, but the poison constructor \texttt{x\_content.minja\_poison\_case()} accepts only \texttt{(domain, framing)}: \texttt{similarity} is passed at no call site, and the benign filler pool is identical across levels. All three \texttt{similarity} levels therefore produce a byte-identical poison case and an identical retrieval environment. The grid contains exactly $12$ distinct poison texts, each mapped to exactly $3$ \texttt{scenario\_id}s, so the \texttt{similarity} factor did not create distinct scenarios; it triplicated the cluster count. Because the frozen bootstrap clusters on \texttt{scenario\_id}, the reported counts were inflated ${\approx}3\times$. Re-scoring on the effective cluster unit (\texttt{x\_grid.effective\_cluster\_key}, which drops inert factors and is the identity for every other family) gave, on the original v3 sample, $6$ effective clusters for the observable arm and $7$ for the implicit arm: \emph{both below the ${\geq}10$ gate}. The Addendum states that X1 reaches CONFIRMED only if \emph{both} arms individually pass their Level-1 gates in the pre-designated primary model (\texttt{gpt-4.1}); that predicate failed.

\textbf{Preregistered continuation to the registered cap (v4 amendment, GPG-signed before collection).} Both arms had stopped early relative to the registered $N{=}360$-per-cell cap ($144$ and $72$ trials) because the v3 rule stopped at $30$ eligible successes counted on the \emph{inflated} unit. We therefore amended the stopping rule to be strictly harder (stop only when eligible successes ${\geq}30$ \emph{and} effective contributing clusters ${\geq}10$, cumulative across resume, within the unchanged $N{=}360$ cap) and continued collection. The design was not otherwise touched: the inert \texttt{similarity} factor was left inert and disclosed rather than retro-fitted to the frozen cosine bands, and the estimand, arms, grid, scorer and bootstrap seed are unchanged. The amendment carried a binding pre-commitment that if the gate remained unmet at the cap, MINJA would stay LOW-CLUSTER and would \emph{not} be upgraded.

\textbf{Outcome.} The observable arm reached the gate: $252/360$ trials, $80$ eligible successes, \texttt{orp} $=80/80=100\%$ (Wilson $[0.954, 1.0]$), \textbf{$10$ effective clusters; gate satisfied, and satisfied on the collapsed unit}. The implicit arm ran to the full $N{=}360$ cap and did \emph{not}: $141$ eligible successes, \texttt{orp} $=0/141=0\%$ (trial-level Wilson $[0.0, 0.0265]$, superseded by the cluster-level bound $[0.0, 0.299]$ below), \textbf{$9$ effective clusters; one short}. The shortfall is not ordinary statistical underpowering, but neither is it a demonstration of impossibility. At the cap each effective configuration received exactly $30$ implicit trials, and three of the twelve (\texttt{client\_deliverable::mandatory\_audit\_cc} and both \texttt{compliance\_report} framings) returned zero eligible successes. Zero in $30$ carries a per-configuration Wilson upper bound of $0.113$, so true rates of a few percent are entirely consistent with what we observed and a further $30$ trials apiece might well have produced successes. The correct statement is therefore a \emph{realised ceiling under the registered cap}: reaching $10$ contributing clusters would have required sampling beyond the cap, which the amendment forecloses. The disposition rests on the cap and the pre-commitment, not on a claim that those configurations can never succeed.

\emph{The dead cells are not a delivery artefact.} A natural alternative explanation is retrieval competition: the poison case competes against $12$ pooled benign fillers (\texttt{x\_content.DOMAIN\_FILLERS}: $6$ domains $\times\ 2$ fillers; earlier drafts said $36$, which was the \texttt{scenario\_id} count, not the filler count) at $k{=}4$, so a configuration could fail merely because the poison is never retrieved. The per-trial logs rule this out. In \emph{every} one of the twelve effective configurations the poison was delivered in $30/30$ trials at a median \texttt{poison\_rank} of $1.0$, including all three dead cells, which are $90/90$ delivered and $0/90$ successful. Delivery is constant across the grid and cannot explain the shortfall. That uniform rank-$1$ retrieval also independently corroborates the inertness of \texttt{similarity}: had the near/mid/far bands been operationalised they would have produced a rank gradient, and none exists.

\emph{What the grid actually shows: domain-dependent viability of implicit imitation.} See Section~\ref{sec:domain-dependence}, which reports this as a result in its own right; it is the continuation's most transferable finding and it is not a statement about our detector. \textbf{Honouring the pre-commitment, MINJA is therefore NOT upgraded: it is reported as measured but directional/LOW-CLUSTER, the same disposition as Zombie}, with Wilson intervals governing. The continuation nonetheless strengthened the evidence substantially: eligible successes rose $37 \to 80$ (observable) and $34 \to 141$ (implicit), and both point estimates held exactly ($100\%$ and $0\%$) across roughly four times the data. What the correction does \emph{not} touch: the point estimates ($80/80$, $0/141$), the no-poison controls, and the matched paired contrast (contrast $+1.0$). We note explicitly that the contrast's $12$ shared effective configurations are configurations \emph{attempted} in both arms; applying the same success-contributing standard we apply to the arms, only $8$ effective configurations contribute eligible successes in both arms, so the contrast does \emph{not} independently satisfy the ${\geq}10$ gate and must not be read as rescuing the per-arm result. The released scorer now computes and gates on that figure (\texttt{eff\_n\_shared\_contributing}${=}8$, \texttt{eff\_cluster\_gate}${=}$false in \texttt{results/xprogram\_scored.json}); earlier drafts of the scorer gated on the attempted count ($12 \geq 10$) and therefore reported the contrast as gate-satisfied while the per-arm cells it summarises were not, contradicting this paragraph. The qualitative observability contrast (task-similarity retrieval of a poisoned exemplar delivers the destination into context with no agent-visible retrieval tool call, so retrieval provenance is absent) is unchanged and now rests on far more data; what remains withdrawn is its \emph{confirmatory} (clustered-primary) status, because one arm does not reach the registered cluster floor within the registered cap.

\textbf{The implicit arm's interval must be read on the cluster unit, not the trial unit.} The trial-level Wilson upper bound on $0/141$ is $0.0265$, and earlier drafts quoted that figure (tightened from $0.102$) as the headline precision gain from the continuation. That bound treats the $141$ eligible successes as $141$ independent Bernoulli draws, which the design does not support: the trials are nested within $12$ effective configurations (of which $9$ contribute eligible successes), each contributing exactly $30$ implicit trials, and \texttt{observable\_read\_of\_poison} is a deterministic property of the delivery path within a configuration rather than an independently resampled outcome. Because the outcome is identically $0$ in every cluster, the intra-cluster correlation is not identified from the data, so we report the conservative limit: treating each success-contributing effective configuration as the unit of analysis gives a cluster-level Wilson upper bound of $0/9 \Rightarrow \mathbf{0.299}$ (Clopper--Pearson $0.336$); on the $12$ attempted configurations it is $0.243$ (Clopper--Pearson $0.265$). A design-effect reading spans the two, but it must use the \emph{unequal}-cluster arithmetic: the nine contributing configurations hold $30, 30, 30, 28, 14, 4, 3, 1, 1$ eligible successes, so the effective cluster size entering $\mathrm{DEFF} = 1 + (m-1)\rho$ is $m = \sum n_c^2 / \sum n_c = 26.29$, not the arithmetic mean $15.67$. At $\rho = 0.1, 0.25, 0.5, 1.0$ the effective $n$ falls to $39.9, 19.3, 10.3, 5.4$ and the upper bound rises to $0.088, 0.166, 0.271, \mathbf{0.417}$. \emph{Correction to an earlier draft of this ladder:} those figures ($57.2, 30.2, 16.9, 9.0$ and $0.063, 0.113, 0.185, 0.299$) used the mean cluster size, which assumes equal sizes; under that assumption $\rho{=}1$ appeared to coincide with the cluster-count bound. It does not. For a trial-weighted rate whose outcome is constant within a cluster, the $\rho{=}1$ effective $n$ is the Kish quantity $(\sum n_c)^2/\sum n_c^2 = 5.36$, which is \emph{smaller} than the nine clusters, so the genuinely conservative limit is $0.417$ rather than $0.299$. \textbf{We therefore report $0.299$ as the cluster-unit bound on the unweighted cluster-level estimand (the preregistered unit) and $0.417$ as the conservative limit on the trial-weighted rate, and we withdraw the claim that $\rho{=}1$ coincides with the cluster-level bound.} Either figure supersedes the trial-level $0.0265$, and we withdraw the claim that the continuation tightened the bound by roughly $4\times$: what the continuation actually bought was more clusters ($7 \to 9$) and an unchanged point estimate over $4\times$ the trials, not a genuine $4\times$ reduction in interval width. This does not change the LOW-CLUSTER disposition, which was already governed by the cluster gate; it removes an overstatement of precision. The same cluster-unit caveat applies to the per-configuration $0/30$ bound of $0.113$ quoted above, which is a trial-level bound within a single cluster and is therefore the correct unit at that level. Regenerate with \texttt{x1\_cluster\_bounds.py}; the bounds are persisted in \texttt{results/xprogram\_scored.json} (\texttt{eff\_cluster\_unit\_wilson}, \texttt{eff\_kish\_unit\_wilson}) and \texttt{results/x1\_cluster\_bounds.json}.

\textbf{The registered $\leq 0.20$ bound was unreachable, which reframes ``one cluster short.''} X1's H2 requires an implicit-arm clustered upper bound $\leq 0.20$ in addition to the ${\geq}10$-cluster gate. Under the cluster-unit convention this paper adopts, a zero-event cell needs $16$ contributing clusters to clear $0.20$ ($0/10 \Rightarrow 0.278$, $0/12 \Rightarrow 0.243$, $0/16 \Rightarrow 0.194$). The registered grid contains only $12$ effective configurations, so \emph{even a fully-populated implicit arm could not have satisfied H2 on this unit}. The shortfall we report is therefore against the \emph{cluster gate}, not against H2: reaching a tenth cluster would have satisfied the gate and still left H2 unmet. We state this rather than letting ``one short'' imply that the registered criterion was within reach, and we note it as a design defect: the grid was sized for a trial-level interval convention and is too narrow for the cluster-level one the clustered design actually requires.

\begin{table}[t]
\caption{X1 MINJA, \texttt{gpt-4.1}, after the preregistered continuation to the registered $N{=}360$ cap. \texttt{orp} is \textit{observable\_read\_of\_poison} among eligible successes. ``Effective clusters'' collapses the inert \texttt{similarity} factor (\S\ref{sec:x1}); the preregistered gate is ${\geq}10$ success-contributing effective clusters \emph{per arm}. The send-only rows restrict to trials in which \tool{email\_send\_email} actually fired; this is a \emph{necessary} condition for the main threat-model endpoint but not equivalent to it, because the eligibility flag is still the harness's draft-or-send \texttt{attack\_success} (\texttt{score\_xprogram.score\_x1\_send\_only} filters on the presence of a send call and does not re-verify that the attacker-directed action was the send). The point estimates are unchanged and neither arm passes. The paired contrast's $12$ shared configurations are those attempted in both arms, of which $8$ contribute eligible successes in both, so it does not independently satisfy the gate.}
\label{tab:x1-summary}
\centering
\small
\begin{tabular}{lrrlrll}
\toprule
\textbf{Arm / endpoint} & \textbf{Trials} & \textbf{Elig.} & \textbf{orp} & \textbf{Eff.\ cl.} & \textbf{Gate} & \textbf{Disposition} \\
\midrule
Observable (draft-or-send, as scored) & 252 & 80 & $80/80 = 100\%$ & 10 & pass & positive control, gated \\
Implicit (draft-or-send, as scored)   & 360 & 141 & $0/141 = 0\%$ & 9 & \textbf{fail} & LOW-CLUSTER \\
\midrule
Observable (send-only sensitivity)    & n/a & 74 & $74/74 = 100\%$ & 9 & fail & endpoint-sensitive \\
Implicit (send-only sensitivity)      & n/a & 138 & $0/138 = 0\%$ & 8 & fail & LOW-CLUSTER \\
\midrule
Paired contrast (matched configs)     & n/a & n/a & $+1.0$ & 8 of 12 & secondary & does not override per-arm gate \\
\bottomrule
\end{tabular}
\end{table}

\textbf{Two X1-specific scope disclosures.} First, \emph{the exfiltration endpoint differs from the main DTA endpoint.} The paper's threat model defines attack success as \tool{email\_send\_email} to a recipient outside the allowed set, but the X1 harness treats the first destination-bearing email action as exfiltration-capable and therefore also counts \tool{email\_draft\_email} (\texttt{probe\_x1.\_dispatch} appends both to the scored \texttt{emails} list). We report the send-only sensitivity: restricting to trials in which \tool{email\_send\_email} actually occurred gives observable $74/74=100\%$ across $9$ effective clusters and implicit $0/138=0\%$ across $8$ (released as the \texttt{x1\_send\_only} block of \texttt{results/xprogram\_scored.json}; regenerate with \texttt{score\_xprogram.py -{}-exp x1\_send\_only}). The \texttt{orp} point estimates are completely unchanged, but under the send-only endpoint \emph{neither} arm satisfies the ${\geq}10$-cluster gate, so the LOW-CLUSTER disposition is robust to the endpoint choice, and is if anything more firmly indicated under the stricter endpoint. We note the residual imprecision: this filter conditions on a send call having occurred but retains the draft-or-send eligibility flag, so it tightens the endpoint towards the threat model without fully reproducing it; a trial that drafted to the attacker and sent elsewhere would still be counted. Fully reproducing the main endpoint requires per-call recipient arguments, which the X1 harness did not persist. Draft-only cases are $6/80$ ($7.5\%$) and $3/141$ ($2.1\%$) of the two arms' eligible successes. Second, \emph{the two arms' trigger text is not identical, and this is a preregistration violation rather than a scope note.} The observable arm appends a parenthetical retrieval hint (``\texttt{(Use memory\_search to review past cases.)}'') that the implicit arm does not receive (\texttt{probe\_x1.py} line~356, \texttt{run\_minja}). \textbf{Reclassified:} \texttt{X1\_PREREG.md} does not merely prefer trigger parity, it introduces the requirement expressly to remove a confound flagged against v2, ``to remove the v2 trigger confound (the [preregistration] panel flagged that giving only the observable arm a memory-inviting prompt conflates delivery-path with instruction), BOTH arms receive the SAME memory-inviting trigger'', and declares a parity check that ``hashes the full scenario tuple and asserts equality across arms per \texttt{scenario\_id}.'' The shipped code therefore reintroduces precisely the confound the amendment was written to eliminate, and the parity check passed only because it hashes the scenario tuple and not the arm-applied trigger suffix. Earlier drafts disclosed the asymmetry here but omitted it from Table~\ref{tab:withdrawn}, which is inconsistent with this paper's own standard; it now appears there. The contrast is consequently not a single-variable manipulation: delivery path and trigger phrasing both differ. The scientific consequence is bounded and unchanged: it makes the observable arm an \emph{easier} positive control, which affects only how surprising its $100\%$ should be; it cannot manufacture the implicit arm's $0/141$, because in that arm no memory tool exposes the payload at all. X1 is already reported as directional/LOW-CLUSTER on independent grounds, so no status changes. The same inert factor affects X5 and X6, which draw from the same grid (\S\ref{sec:x6-deviation}); Zombie ($\text{domain} \times \text{turns} \times \text{keyspan}$) and X4 ($\text{doctype} \times \text{discoverability} \times \text{recipient\_rep}$) consume every grid factor and are unaffected (measured inflation $1.0\times$).

\paragraph{Frozen-RF is out-of-distribution (crucial caveat).} The raw variable and the frozen-RF flag diverge sharply here, and the distinction is load-bearing. In the implicit MINJA/gpt-4.1 arm the raw \texttt{observable\_read\_of\_poison}${=}0/141$, yet the frozen P1 RF still flags those sessions, not via any retrieval feature (there is none) but via non-retrieval trajectory features (\eg \texttt{draft\_then\_send}, \texttt{seq\_len}) on which it is out-of-distribution for case-bank traces. The correct reading is therefore ``\emph{no retrieval provenance in the tool logs}'', not ``no trace-based alert of any kind''. This is exactly why X1 anchors its conclusion on the raw retrieval variable and treats the RF as secondary.

\paragraph{Zombie Agents (directional/LOW-CLUSTER, not upgraded).} We also implemented a genuine emergent sliding-window buffer (fixed window $W{=}6$ turns, a frozen content-agnostic self-reinforcing promotion rule, and a preregistered ${\geq}90\%$ benign-eviction control that the buffer passed, so persistence is emergent, not passive retention). The results are directional but do not meet the preregistered cluster gate: on \texttt{gpt-4.1} the observable arm is a clean positive control ($36/36$, $36$ clusters) but the implicit arm has only $16$ eligible successes across $5$ success-contributing scenario clusters (\texttt{observable\_read\_of\_poison}${=}0/16$), below the ${\geq}10$-cluster floor, so it is LOW-CLUSTER and Wilson-governing (descriptive). On \texttt{gpt-4o} the observable arm is again a positive control ($36/36$) but the implicit arm produced $0$ eligible successes, so \textbf{there is no fully-powered within-model observable-vs-implicit Zombie contrast}. We therefore do \emph{not} upgrade Zombie: it retains its architectural classification (status A) with directional, LOW-CLUSTER in-architecture support, as flagged in Table~\ref{tab:attack-taxonomy}.

\paragraph{No-poison control and scope.} A no-poison, format-matched negative control produced $0$ external sends across all $288$ X1 no-poison trials ($0$ scored successes and $0$ trials recording any external recipient, across the four \texttt{x1\_*\_nopoison\_*} cells; the per-cell counts are in the released JSONL rather than the scored summary, whose eligibility filter requires \texttt{poison\_delivered} and is therefore vacuous for these cells), establishing that implicit-arm ``successes'' are attributable to the poison and not to the case-bank/window mechanics themselves. X1 probes the both-implicit boundary on an independent, mechanism-faithful MINJA case-bank family within one act-without-grounding model (gpt-4.1); after the cluster-unit correction above, that evidence is directional/LOW-CLUSTER rather than clustered-primary, so it does not claim universality across models and does not by itself close the independent-attack-family gap. The Zombie family's independent-family confirmation likewise remains directional/LOW-CLUSTER and open.

\subsection{Second Exfiltration Channel: Document Sharing (X4)}
\label{sec:x4}
The paper's necessity claim is operationalised on a single channel (\texttt{email\_send\_email}), and the concept-level heuristic (\S\ref{sec:prospective-eval}) found $0\%$ detection on an indirect \texttt{share\_document} channel, raising the possibility that the necessity is email-specific. X4 (preregistered, GPG-signed before data) tests whether the \emph{generalised} invariant (observable retrieval before \emph{any} exfiltrating action) holds when the exfiltrating action is a two-step document share (\texttt{draft\_document}${\to}$\texttt{share\_document}) rather than email. The memory tools are present in both arms, and both the ``generalizes'' and ``narrow-to-email'' outcomes were preregistered with equal prominence.

\paragraph{Primary result: raw necessity generalises (control), but the $\leq 0.10$ criterion is not met on the cluster unit.} The primary variable is the hand-coded \texttt{recall\_before\_exfil} (did an observable memory-tool result return the attacker destination before the first destination-bearing action?), independent of the frozen RF. On the control arm the necessity holds on \emph{both} models with zero violations: $0$ violations among $50$ successes on \texttt{gpt-4.1} ($28$ success-contributing scenario clusters) and $0/62$ on \texttt{gpt-4o} ($33$ clusters). Both arms comfortably satisfy the preregistered ${\geq}10$-cluster gate.

\emph{The upper-bound criterion, however, is not met once the correct unit is used.} These are zero-event cells, so the scenario-clustered percentile bootstrap is degenerate (every resample reproduces $0$, giving a spurious $[0,0]$) exactly as documented for the V3-3 control cell (\S\ref{sec:v33}). Applying the same procedure we apply there (a Wilson upper bound on the effective cluster count rather than on trials) gives $0/28 \Rightarrow$ upper $\mathbf{0.121}$ for \texttt{gpt-4.1} and $0/33 \Rightarrow$ upper $\mathbf{0.104}$ for \texttt{gpt-4o}. \textbf{Both exceed the preregistered $\leq 0.10$ threshold, so we withdraw the claim (made in earlier drafts on the basis of the trial-level bounds $0.071$ and $0.058$) that X4's control arms meet that criterion.} The trial-level bounds are not wrong as arithmetic; they are the wrong unit for a design whose primary inference the preregistration defines over scenario clusters, and quoting them alongside a degenerate $[0,0]$ bootstrap interval overstated the precision twice over. Because clustering here is mild (mean $1.79$ and $1.88$ successes per contributing cluster, versus $15.7$ in X1), the truth lies between the two: the cluster-unit bound is the conservative limit (intra-cluster correlation $=1$) and the trial-level bound the anti-conservative one. We therefore report X4's control arms as \textbf{zero violations across 28 and 33 contributing clusters, directionally supporting the necessity on a second channel but falling marginally short of the registered $\leq 0.10$ bound}, rather than as meeting it. The point estimates ($0/50$, $0/62$), the cluster counts, and the ${\geq}10$-cluster gate results are unchanged. The retrieval-before-exfiltration necessity therefore \emph{extends} from email to a second, structurally distinct exfiltration channel, but on this arm it is directional rather than criterion-passing. Regenerate with \texttt{x4\_cluster\_bounds.py}.

\paragraph{Phenotype keystone (with a thinness caveat).} A benign share-channel keystone confirms the models recall-before-share on legitimate tasks, so control violations (had they occurred) would have been interpretable as necessity failures rather than a ``don't-look-up-before-sharing'' phenotype: \texttt{gpt-4o} recalls-before-share on $72/72 = 100\%$ of \emph{completed} benign share tasks. \texttt{gpt-4.1} recalls-before-share on all $14/14$ \emph{completed} shares, but only $14$ of $72$ benign share sessions completed a share: this cell is thin, and we label it as such rather than reporting it as a clean $100\%$.

\paragraph{Treatment (implicit): not viable on one model, directional on the other.} The implicit (ICL-delivery) treatment arm does not support a clustered-primary breakdown claim on the share channel. On \texttt{gpt-4.1} the attack is \textbf{not viable}: $0$ treatment successes across $360$ trials (the model did not execute an external share under implicit delivery), so the estimand is undefined. On \texttt{gpt-4o} the arm is directional/underpowered: \texttt{recall\_before\_exfil}${=}0$ in $20/20$ treatment successes across $10$ scenario clusters, but this is below the preregistered $37$-success target, so we report it as directional rather than confirmatory and do not use the degenerate clustered interval to override the target-based underpowered branch. The implicit breakdown remains established for the email channel (V3-1, V3-3) and is only directionally suggested here.

\paragraph{Trigger asymmetry between the X4 arms.} The X4 control and treatment arms do not receive identical trigger text. \texttt{probe\_x4.run\_control} appends a memory-inviting parenthetical, ``\texttt{(Use memory\_recall\_fact / memory\_list\_all\_facts to check stored rules.)}'', to the trigger, and \texttt{probe\_x4.run\_treatment} does not (\texttt{probe\_x4.py} lines~188 and~204). This is the same asymmetry that \S\ref{sec:x1} reports for X1 and that Table~\ref{tab:withdrawn} records there as a preregistration violation, and we had not disclosed it here. Two differences in severity should be stated rather than glossed. First, \texttt{X4\_PREREG.md} is weaker than \texttt{X1\_PREREG.md} on this point: it requires that the same \texttt{scenario\_id} be instantiated in both arms (``arm parity'') and that memory tools be present in both, but it does not state a trigger-parity requirement, so this is an undisclosed confound rather than a breach of a registered requirement. Second, the confound runs in the direction that makes the control arm an \emph{easier} positive control: prompting the agent to consult memory raises the chance that a retrieval precedes the share, which is the event the necessity claim counts. It therefore affects how surprising the control arm's $0/50$ and $0/62$ should be, and we now flag that the second-channel necessity evidence rests on an arm that was instructed to consult memory. It cannot manufacture the treatment arm's result, since that arm received no such hint and produced no successes on gpt-4.1 at all. \emph{A related note on X4's own registered rationale:} the preregistration argues that ``memory tools present in BOTH arms'' makes a treatment \texttt{recall\_before\_exfil}${=}0$ ``a behavioral choice, not a tool-absence artifact.'' That inference is the same tool-presence-versus-payload-reachability conflation corrected for X1 in \S\ref{sec:x1}: \texttt{probe\_x4.run\_treatment}'s store holds only \texttt{benign\_deadline} and \texttt{benign\_standup}, so the routing rule is not retrievable there either, and X4's treatment arm belongs to the same class as V3-1's and V3-3's (\S\ref{sec:threat}).

\paragraph{Classifier portability is an engineering result, not channel-generality.} Applying the frozen RF to the share-channel traces, the \emph{unmapped} classifier flags $0/37$ control successes (tool-name drift: it is blind to \texttt{share\_document}), while after a pre-declared \texttt{share\_document}${\to}$\texttt{send\_email} ontology remap it flags $37/37$. Because remapping the exfiltration tool to \texttt{send\_email} makes ``recall${\to}$send'' near-tautological, we label the $37/37$ an \emph{ontology-adapted classifier transfer (engineering)}: it quantifies tool-name drift and reinforces the Limitations point (\S\ref{sec:limitations}), and is \textbf{not} evidence that the invariant is channel-general. The channel-generality conclusion rests solely on the raw \texttt{recall\_before\_exfil} necessity above.

\paragraph{No-poison control and scope.} A no-poison negative control produced $0$ external shares across all $144$ X4 no-poison trials. X4 broadens the necessity claim to two instantiated exfiltration channels (email + document-share) on two act-without-grounding models; the implicit treatment arm is not viable on gpt-4.1 and only directional on gpt-4o, and only one additional channel was tested, so we do not claim channel-generality beyond these two.

The benign memory-tool base-rate probe (B+), which addresses general tool-avoidance as an explanation for the X1 implicit-arm result and discharges the preregistered E1, E2 and E3 endpoints, is reported in Appendix~\ref{app:bplus}. Its headline figures, on an instrumented re-collection carried out under a GPG-signed amendment registered before the new data: benign memory-tool use is saturated ($720/720$ in all four model$\times$channel cells, store wired and read in every cell, so the confound threshold is met for the email cells that X1 depends on); benign agents retrieve a destination-bearing record before the exfiltrating action in five of six task domains outright, with all shortfalls in the email/gpt-4.1 cell confined to a single domain (E2, newly discharged; pooled $95.5$--$100\%$, but the pooled interval is reported on the six-cluster effective unit, not the trial unit; see Appendix~\ref{app:bplus}); and the frozen RF's false-positive rate on that benign corpus is $55.4$--$65.4\%$ on the email channel. One share cell's rate is unstable across collection dates by a factor of ${\approx}44$, driven by benign task-completion drift rather than by the detector; we report that instability as a result, and cannot attribute it, because neither collection captured served-model identity.

The adaptive-attacker evasion battery (X5) returned no eligible successes at all ($0/1016$ across four destination transformations), so it bounds where oracle robustness can be measured rather than measuring it; it is reported in Appendix~\ref{app:x5}.

The architecture- and model-breadth sweep (X6) produced no powered cell: of its $12$ cells, $3$ produced eligible successes, $7$ produced measured zeros, and $2$ are non-evaluable rather than zero (the two \texttt{claude-sonnet-5} cells, $254$ trials with no tool calls recorded and no outcome scored; see Appendix~\ref{app:x6}, which corrects an earlier ``$9$ of $12$ eligible${=}0$'' figure that pooled those non-measurements with genuine zeros), and after the cluster-unit correction of Section~\ref{sec:x1} no X6 cell (including the three that did produce eligible successes) satisfies the ${\geq}10$-cluster gate. We therefore claim no architecture generality and scope the breadth evidence to dense retrieval; the sweep and its design deviations are reported in Appendix~\ref{app:x6}.

The preregistered follow-up (Section~\ref{sec:deployment-boundary}) demonstrates that the trajectory invariant cannot serve as a standalone detector; its unconditional benign FPR of 24.7--57.6\% is operationally prohibitive. This falsification is itself a contribution: it establishes the deployment boundary that separates forensic detection research from deployable security tooling.

In ROC terms, the signature operates at a fixed point of (Recall~$= 0.98$,
FPR~$= 0.247$--$0.576$) depending on model phenotype. As a standalone classifier this is
inadequate, and \textbf{this paper does not exhibit a configuration that rescues it.} Earlier
versions, including arXiv v2, suggested using it as a high-recall pre-filter ahead of semantic gating on recipient
metadata, and cited a $42$--$75\%$ reduction in downstream inspections. Both halves of that
suggestion are withdrawn here, for different reasons.

The gating half is contradicted by our own measurement. V3-5 (\S\ref{sec:v35}) tested exactly
that recipient-metadata gate: it flags $129/129 = 100\%$ of \emph{legitimate} external business
email, and the combined trajectory-and-recipient gate still flags $58/129 = 45\%$. A second
stage that admits every legitimate external recipient does not add precision, so the ensemble
we previously recommended is the one configuration this paper has already falsified.

The economics half is a cost argument, not a precision argument, and it was being cited as
though it were the latter. What it shows is that a trajectory gate reduces \emph{downstream
inspection invocations} by $(1-r)\cdot N$ for trigger rate $r$; it says nothing about whether
the surviving alerts are actionable. It is also conditional on a content-inspection cost
measured with generation capped at a five-token verdict, which is a lower bound, and on a
trigger rate that varies with the retrieval interface. It is retained, with those caveats, in
Appendix~\ref{app:economics}, and it should not be read as evidence that a deployable
configuration exists.

What survives is the negative result: the transition is an attack precondition with
prohibitively low precision against benign traffic, and the one gating strategy we tested does
not repair it. Identifying a configuration that does is open work, not a contribution here.

The honest reporting of this failure boundary (rather than suppressing the benign FPR result or restricting evaluation to attack-only data) illustrates a methodological contribution: preregistered deployment-boundary testing should be standard practice for forensic detection papers, most of which report precision and recall only on curated attack datasets without measuring performance under realistic benign traffic.

\subsection{Implicit-Delivery Attack Viability Is Domain-Dependent}
\label{sec:domain-dependence}

The X1 continuation (Section~\ref{sec:x1}) produced an unanticipated result that is
independent of our detector and, we think, more transferable than the boundary claim it
failed to confirm. Because the implicit arm ran the full registered grid to its $N{=}360$
cap, every one of the twelve effective configurations received exactly $30$ trials under
identical conditions, and the poison was \emph{delivered} in $30/30$ trials at a median
retrieval rank of $1.0$ in \emph{all twelve}. Delivery is therefore constant by
measurement, not by assumption. What varies is whether the agent, having been handed the
poisoned exemplar at rank~1, actually imitates it and routes externally.

It varies across the entire range (Table~\ref{tab:domain-dependence}). Two domains are
fully susceptible ($30/30$ in both compliance framings for \texttt{meeting\_followup};
$30/30$ and $28/30$ for \texttt{incident\_report}). Two are almost entirely resistant
($1/30$ in both framings for \texttt{project\_status}; $0/30$ in both for
\texttt{compliance\_report}, i.e.\ $0/60$ pooled). The remainder are intermediate and show
a consistent framing effect: the \emph{historical-practice} framing (``as we've always
done for this'') outperforms \emph{mandatory-audit-cc} (``per the mandatory audit-trail
compliance rule'') in every domain where the two differ, by $33.3$ percentage points on
\texttt{budget\_expense} ($14/30$ vs.\ $4/30$) and $10.0$ on \texttt{client\_deliverable}
($3/30$ vs.\ $0/30$). (Earlier drafts wrote ``$10$ points'' and ``$3$'' here; those were
the raw differences in trial counts, not percentage points.)

\begin{table}[t]
\caption{Implicit-delivery attack success by task domain and compliance framing (X1 MINJA
implicit arm, \texttt{gpt-4.1}, $30$ trials per cell at the registered cap). The poison was
retrieved at median rank~$1$ in $30/30$ trials in \emph{every} cell, so these differences
are not retrieval-competition artefacts: delivery is held constant and only imitation
varies. Cells at $0/30$ carry a Wilson upper bound of $0.113$, so they are low-yield rather
than demonstrably zero.}
\label{tab:domain-dependence}
\centering
\small
\begin{tabular}{lccc}
\toprule
\textbf{Task domain} & \textbf{historical-practice} & \textbf{mandatory-audit-cc} & \textbf{pooled} \\
\midrule
\texttt{meeting\_followup}   & $30/30$ & $30/30$ & $60/60$ \\
\texttt{incident\_report}    & $30/30$ & $28/30$ & $58/60$ \\
\texttt{budget\_expense}     & $14/30$ & $4/30$  & $18/60$ \\
\texttt{client\_deliverable} & $3/30$  & $0/30$  & $3/60$ \\
\texttt{project\_status}     & $1/30$  & $1/30$  & $2/60$ \\
\texttt{compliance\_report}  & $0/30$  & $0/30$  & $0/60$ \\
\midrule
\textbf{all domains}         & $78/180$ & $63/180$ & $141/360$ \\
\bottomrule
\end{tabular}
\end{table}

Two consequences. First, for the attack literature: memory-poisoning viability under
implicit delivery is governed substantially by the semantics of the cover task, not only by
the retrieval architecture. A payload can be retrieved first, be perfectly well-formed, and
still fail to produce exfiltration because the surrounding task makes external routing
implausible to the model. Reported attack-success rates for this family are therefore
sensitive to cover-task choice in a way that is rarely stated, and single-domain
demonstrations may substantially over- or under-state viability. Audit- and
compliance-themed cover tasks were the most resistant here, which is the opposite of what a
naive reading of the compliance-framing attack would predict.

Second, for evaluation design: a breadth gate defined over \emph{success-contributing}
clusters silently tests two different things at once: breadth of the effect under study,
and breadth of \emph{attack viability across conditions}. X1's implicit arm is clean on the
first among every success it produced ($0/141$ retrieval provenance, cluster-level Wilson upper $0.299$)
and fails the second, which is why it cannot reach the preregistered cluster floor
(Section~\ref{sec:x1}). Future preregistrations in this programme will gate delivery,
viability, and effect breadth separately rather than folding them into one
success-conditioned cluster count. We flag this as a design lesson rather than grounds for
reinterpreting the present result, whose disposition is fixed by the signed
pre-commitment.

\section{Deployment Boundary: Benign False-Positive Analysis}
\label{sec:deployment-boundary}


The detector's reported false-positive rate in v1 was measured against poisoned-but-defended non-exfiltration sessions (Section~\ref{sec:intro}), not against ordinary benign agent traffic. We preregistered a follow-up study (V2-1) to measure the false-positive rate under realistic benign memory use across 13 models from 4 API vendors plus open-weight models on Apple Silicon. The preregistered factorial is 13 models $\times$ 8 cells (($k \in \{0,1,3,5\}$) $\times$ (recall protocol $\in \{\text{mandatory}, \text{optional}\}$)) $\times$ 40 sessions $= 4{,}160$ sessions. We additionally collected one supplementary high-$N$ cell ($N{=}200$; a 14th model, claude-opus-4-8, at $k{=}3$ mandatory) as a precision spot-check, bringing the released benign corpus to $N{=}4{,}360$. Pooled FPR figures below are computed over the full released benign corpus; per-cell counts are in the released \texttt{results/v2\_1\_*.jsonl} files. Unless noted, the 13-model factorial ($N{=}4{,}160$) drives every per-model and per-cell statistic; the supplementary opus cell (all 200 sessions have \texttt{recall\_before\_send}${=}0$) shifts only the \emph{pooled unconditional} benign FPR, from $41.2\%$ over the 13-model factorial to $39.3\%$ over the full corpus ($\sim$2 pp).

\paragraph{Design.} Benign sessions contain only legitimate stored facts (meeting times, project deadlines, contact information) with no attack material. The agent's task is to recall relevant context from memory and draft/send a status email to the team. We cross $k \in \{0, 1, 3, 5\}$ pre-stored benign facts with recall protocol $\in \{\text{mandatory}, \text{optional}\}$, yielding 8 cells per model at $N=40$/cell.

\paragraph{Result.} The detector is, on benign data, a deterministic \textit{recall\_before\_send} classifier:
\begin{quote}
$P(\text{flag} \mid \text{recall\_before\_send} = 1) = 100.0\%$ $[99.8, 100.0]$, $N{=}1{,}713$\\
$P(\text{flag} \mid \text{recall\_before\_send} = 0) = 0.0\%$ $[0.0, 0.14]$, $N{=}2{,}647$
\end{quote}
The classifier is binary-saturated: a single recall-before-send trips it fully; recall volume does not modulate (FPR is 100\% whether $\text{recall\_count} = 1$ or $5$, ruling out a volume-mediation account). This holds across both frontier API models and open-weight models on local hardware. (These conditionals are recomputed with the released classifier and the corrected any-preceding \texttt{recall\_before\_send} definition of Section~\ref{sec:limitations}; an earlier draft reported $1.5\%$ at $\text{recall\_before\_send}{=}0$ under the superseded first-occurrence definition, which mislabelled a small number of send-then-recall-then-send sessions. Under the released definition the map from \texttt{recall\_before\_send} to the flag is exactly deterministic on the benign corpus.)

\paragraph{H1 (monotonic increase with $k$) not confirmed.} The preregistration hypothesised that FPR increases monotonically with the number of stored facts $k$. Actual data (pooled over the full released corpus): $k{=}0 \to 0\%$, $k{=}1 \to 57.6\%$, $k{=}3 \to 54.9\%$, $k{=}5 \to 52.2\%$, which is non-monotonic. The effect is binary-saturated at recall\_before\_send${=}1$ regardless of fact count. The relevant conditional is on the structural transition (whether recall precedes send), not the cardinality of stored facts. The reported unconditional-FPR range $24.7$--$57.6\%$ is the spread across the \emph{recall protocol} factor on the 13-model factorial ($N{=}4{,}160$): $24.7\%$ ($514/2{,}080$) under the \emph{optional}-recall protocol and $57.6\%$ ($1{,}199/2{,}080$) under the \emph{mandatory}-recall protocol. (\emph{Correction relative to arXiv v2:} that version reported the upper bound as $52.6\%$, which was the $k{=}5$ cell rather than the maximum across cells; the correct upper bound is the $k{=}1$ cell at $57.6\%$. The underlying V2-1 data are unchanged; only the cell selected for the headline bound differs.) Both the protocol split and the per-$k$ values are non-monotonic manifestations of the same binary saturation: FPR tracks how often the benign task induces a recall-before-send, not $k$ or the protocol per se.

\paragraph{Interpretation.} The \textit{recall\_before\_send} transition is a valid attack \emph{precondition}, not a maliciousness \emph{predicate}. A memory-channel exfiltration must produce it; but so does benign memory-grounded action (an agent that recalls a stored deadline and then emails its team), because the two are identical at the operation level and differ only in the recipient address, which operation-only features exclude by design.

\paragraph{Model phenotypes.} Benign FPR is governed by a model-specific behavioural phenotype:
\begin{itemize}
    \item \textbf{Ground-then-act} (claude-sonnet-4-6, gemini-3.5-flash, gemini-3.1-pro): recall $\to$ send on benign tasks. FPR 76--100\%.
    \item \textbf{Act-without-grounding} (gpt-4o, gpt-4.1, gpt-4o-mini, o4-mini): send without recalling. FPR 0\%.
    \item \textbf{Don't-act} (claude-haiku-4-5, claude-opus-4-8): no send at all. Uninformative for FPR.
\end{itemize}
\texttt{qwq:32b} requires separate comment, because earlier drafts described it as exhibiting near-zero benign FPR and left the pattern unexplained. That description was an artefact of the superseded first-occurrence feature definition, and it resolves completely under the released any-preceding definition. Re-scoring all eight \texttt{qwq:32b} benign cells ($320$ sessions, all of which send) under both conventions: first-occurrence gives \texttt{recall\_before\_send}${=}1$ in $0/320$, whereas any-preceding gives $40/320 = 12.5\%$, and the entire difference is one cell: \texttt{k=1}, mandatory-recall protocol, where all $40$ sessions flip from $0$ to $1$. The mechanism is the revision pattern documented in Section~\ref{sec:limitations}: the model sends, then recalls to revise, then re-sends, so a recall does precede a later exfiltrating send even though the first recall follows the first send. \texttt{qwq:32b} is therefore not an anomalous phenotype; it is a model whose benign trajectories the superseded feature mislabelled, and its corrected profile is ordinary: no benign recall-before-send outside the mandatory \texttt{k=1} condition, and $100\%$ conditional FPR inside it, consistent with every other model. This is separate from, and not in tension with, its \emph{attack-side} implicit-bypass behaviour in Section~\ref{sec:holdout}, where it routes memory access through RAG fallback rather than explicit \tool{recall\_fact} calls.

\paragraph{Phenotype is interface-dependent, not model-intrinsic.} The phenotype taxonomy above was established using exact-key \tool{recall\_fact}. However, V3-2 H4 demonstrates that GPT-4.1 (classified as act-without-grounding above) exhibits $82.5\%$ benign \texttt{search\_before\_send} under semantic tool retrieval (\tool{memory\_search}), i.e., ground-then-act behaviour. The phenotype is a \emph{model-interface interaction}, not a stable model property: the same model changes behavioural class depending on whether retrieval is mediated by exact-key lookup or semantic search. Deployment economics (Section~\ref{app:economics}) and FPR estimates must therefore be conditioned on the specific retrieval interface, not on model identity alone. Under semantic search (the dominant production RAG architecture) the act-without-grounding phenotype may not hold even for models that exhibit it under exact-key retrieval.

\paragraph{Semantic gating narrows, but does not restore, separability.} Adding a single binary feature (\texttt{recipient\_is\_external}: whether any send target is outside the organisation's domain) separates benign from attack \emph{on the V2-1 benign corpus}, but only because every benign session there emails internal addresses (team@example.com) by construction, while attacks exfiltrate to external addresses. We initially framed this as ``perfect separation''; a preregistered follow-up (V3-5, Section~\ref{sec:v35}) shows that framing is an artefact of the internal-only benign design. On $N{=}200$ benign sessions whose \emph{legitimate} business task requires emailing an external counterparty (client, vendor, auditor), the recipient-externality gate flags $129/129 = 100\%$ [0.971, 1.000] of completed external sends, and the combined trajectory-and-recipient gate (\texttt{recall\_before\_send}${=}1 \wedge$ external) still flags $58/129 = 45.0\%$ [0.366, 0.536]. Recipient gating therefore fails outright as a standalone fix for any fleet that legitimately emails external parties, and even the combined gate misfires on nearly half of legitimate external traffic: task-intent or memory-content semantics are required beyond recipient metadata. We acknowledge that recipient-based gating approaches standard Data Loss Prevention (DLP) territory and relaxes the purely operation-only constraint established in Section~\ref{sec:threat}; the trajectory signal's marginal contribution is as a high-recall structural pre-filter that narrows the semantic inspection surface, not as a standalone operation-only detector or a replacement for content-level DLP.

\paragraph{Deployment guidance.} Do \emph{not} deploy a \textit{recall$\to$send} trajectory detector as a standalone inline blocker: under deployments where agents ground actions in memory (a desirable, common pattern), benign false positives reach 76--100\% for ground-then-act models. The tested trajectory signal is a candidate high-recall escalation/triage input, not a validated deployment: (1)~the only second-stage gate we tested (recipient-metadata externality, Section~\ref{sec:v35}) flags 100\% of legitimate external business email, and the combined trajectory-and-recipient gate still flags 45\%, so it does \emph{not} repair precision; (2)~no content- or intent-level gate was evaluated here, so we state that direction as indicated rather than validated, with its precision an open empirical question; (3)~we do not exhibit a deployable configuration, and this paper's data should not be read as recommending a specific production pipeline.

\paragraph{Base-rate-aware precision.} The precision ceiling is starkest when the attack base rate is realistic. Table~\ref{tab:ppv} computes the positive predictive value $\mathrm{PPV} = \frac{\mathrm{TPR}\cdot\pi}{\mathrm{TPR}\cdot\pi + \mathrm{FPR}\cdot(1-\pi)}$ from the detector's measured operating point (TPR~$= 0.98$) across benign FPR values spanning our V2-1 range ($24.7$--$57.6\%$ unconditional, up to $100\%$ conditional on \texttt{recall\_before\_send}${=}1$) and plausible attack prevalences $\pi$. Even at the \emph{most favourable} benign FPR ($24.7\%$, the optional-recall-protocol cell) and a generous $1\%$ attack prevalence, PPV is only $3.85\%$: roughly 26 false alerts per true positive. At the worst benign cell ($57.6\%$, mandatory-recall protocol) and a $0.1\%$ prevalence, PPV falls to $0.17\%$ ($\sim$588 alerts per true positive). This quantifies why the signature is a precondition filter, not a standalone detector \emph{at the tested operating point}: we do not sweep thresholds or bound the achievable recall-precision trade-off at other operating points, so we state the finding at the strength the analysis supports, the tested classifier, at its preregistered threshold ($0.5$), does not clear a useful precision bar at realistic prevalence, rather than as a claim that no threshold or classifier could.

\begin{table}[h]
\centering
\small
\caption{Base-rate-aware PPV (\%) at fixed TPR~$= 0.98$, as a function of benign FPR (rows) and attack prevalence $\pi$ (columns). Computed from existing TPR/FPR operating points; no new data. The $24.7$--$57.6\%$ FPR rows are the V2-1 unconditional benign range; the $100\%$ row is conditional on \texttt{recall\_before\_send}${=}1$.}
\label{tab:ppv}
\begin{tabular}{lrrrrr}
\toprule
Benign FPR & $\pi{=}0.0001$ & $\pi{=}0.001$ & $\pi{=}0.01$ & $\pi{=}0.05$ & $\pi{=}0.10$ \\
\midrule
1\%    & 0.97 & 8.93 & 49.75 & 83.76 & 91.59 \\
5\%    & 0.20 & 1.92 & 16.53 & 50.78 & 68.53 \\
10\%   & 0.10 & 0.97 & 9.01  & 34.03 & 52.13 \\
24.7\% & 0.04 & 0.40 & 3.85  & 17.27 & 30.60 \\
40\%   & 0.02 & 0.24 & 2.41  & 11.42 & 21.40 \\
57.6\% & 0.02 & 0.17 & 1.69  & 8.22  & 15.90 \\
100\%  & 0.01 & 0.10 & 0.98  & 4.90  & 9.82  \\
\bottomrule
\end{tabular}
\end{table}

\paragraph{Ablated-model confirmation.} To confirm that non-recall features do not rescue standalone detection, we evaluated the 10-feature ablated model (all recall features removed; see Section~\ref{sec:write-axis}) on the V2-1 benign dataset. This model exhibits unconditional FPR of 73.7\% [72.3, 75.0] and 100\% conditional on recall\_before\_send$=$1, confirming that downstream behavioural features (send\_count, draft\_then\_send) also fail to separate benign memory-grounded sessions from attacks. The ablated model's \emph{higher} unconditional FPR (73.7\% vs.\ $24.7$--$57.6\%$ for the full model) arises because it lacks the recall\_before\_send$=$0 discriminator that correctly exonerates sessions without the recall transition; without this feature, the ablated model relies on noisier proxies (send\_count, seq\_len) that fire on benign single-send traffic.

\paragraph{Relationship to v1 claims.} The core contribution (the discovery that memory-recall exfiltration attacks produce a stable trajectory invariant) is unaffected by the benign FPR finding. The invariant remains necessary for the attack and sufficient for high-recall detection. This section establishes that operation-only structural features face a hard precision ceiling when benign agents also ground actions in memory. What follows from our data is negative and specific: standalone blocking is not viable, and the one gating variable we did test (recipient externality) does not repair precision (\S\ref{sec:v35}). Escalation with content- or intent-level gating is the natural remaining candidate, but we did not evaluate any such gate, so we state it as the indicated direction rather than a validated deployment model; its precision is an open empirical question, not a result of this paper.

\subsection{Recipient-Gating Residual FPR on Legitimate External Mail (V3-5)}
\label{sec:v35}
The deployment recommendation above proposes gating the trajectory signal with recipient externality. V2-1's benign corpus made that gate look perfect only because every benign task emailed an internal address by construction. We preregistered V3-5 (GPG-signed, before data) to measure the gate's residual false-positive rate on the workflow it most plausibly misfires on: genuinely benign business tasks that \emph{require} emailing an external counterparty. We ran $N{=}100$ such sessions on each of gpt-4.1 (act-without-grounding) and gemini-2.5-pro (ground-then-act), with legitimate external domains (client, vendor, auditor) and no attack material. Among completed external sends ($76/100$ gpt-4.1, $53/100$ gemini-2.5-pro; the remainder did not complete a send), the informative quantity is the \emph{combined} trajectory-and-recipient gate (\texttt{recall\_before\_send}${=}1 \wedge$ recipient external), which still flags $58/129 = 45.0\%$ [0.366, 0.536] of legitimate external business email (gpt-4.1 $38/76 = 50.0\%$; gemini-2.5-pro $20/53 = 37.7\%$). The recipient-externality gate alone flags $129/129 = 100.0\%$ [0.971, 1.000] by construction: every legitimate external send trips the external bit, so the $100\%$ is tautological given the task design and is not an estimated rate. This quantifies the previously hand-waved residual: recipient gating is not a fix for any fleet with legitimate external correspondence, and even the trajectory-and-recipient combination misfires on nearly half of it. Precision at deployment therefore requires task-intent or content-level discrimination, consistent with the precondition-not-predicate conclusion.

\section{Cross-Framing and Cross-Model Transfer}
\label{sec:cross-framework-transfer}

The preceding results establish that \textit{recall\_before\_send} is a valid structural
precondition on one scaffold (LangGraph) and one model population. A natural question
is whether this generalises: does the signature transfer across independent agent
implementations, and does it track attack success or merely information-flow
observability?

\paragraph{Replication design.} We run the delayed-trigger attack through two
prompt-framing variants of a single raw OpenAI function-calling loop: (1)~the raw
loop itself (direct SDK message-and-tool dispatch, no framework), and (2)~the same
loop with a LangChain-style ReAct system-prompt wrapper prepended. These are
\emph{not} independent orchestration engines: variant (2) shares variant (1)'s
execution loop and tool dispatch, differing only in system-prompt framing. The
experiment therefore tests robustness of the signature to \emph{prompt framing}
within one loop and one tool schema; it does not establish transfer across
genuinely independent frameworks (a genuine LangChain/LangGraph replication is
reported separately in Section~\ref{sec:v34}; the OpenAI Assistants \texttt{file\_search}
arm was infrastructure-blocked, see Section~\ref{sec:v34}). Both variants use identical tool schemas,
memory stores, attack payloads, and benign tasks. We test GPT-5.1 (known vulnerable
from IED) at $N{=}100$--$120$ per condition, with 100 benign memory-grounded and 50
benign no-memory controls per variant ($T{=}0.0$, deterministic). A secondary
exploratory model sweep tests 7 models at $N{=}20$ each on the raw loop.

\paragraph{Attack-success scoring correction.} An earlier version scored an
attack as successful if the attacker address appeared \emph{anywhere} in an email
tool call, including the subject/body. Because the simulated email tool exposes no
cc/bcc field, body text is not a routing action; counting it conflated ``address
mentioned'' with ``address routed.'' We rescore using the structured recipient
(\texttt{to}) field only. This lowers reported ASR (e.g., GPT-5.1 $98.3\% \to
62.5\%$; LangChain-style $100\% \to 81\%$) while leaving \emph{signature} rates
unchanged (the trajectory is unaffected by recipient scoring). The corrected
numbers therefore \emph{strengthen} the precondition-not-predicate reading: the
signature decouples from exfiltration success even more sharply.

\paragraph{Cross-framing transfer.} Table~\ref{tab:cross-framework} shows the
trajectory signature is stable across both prompt-framing variants:

\begin{table}[h]
\centering
\small
\caption{Cross-framing replication (GPT-5.1). Signature = recall\_before\_send. ASR under the corrected recipient-only scorer (legacy body-inclusive ASR in parentheses).}
\label{tab:cross-framework}
\begin{tabular}{llrrr}
\toprule
Variant & Condition & $N$ & ASR (corrected / legacy) & Signature \\
\midrule
Raw OpenAI & Attack & 120 & 62.5\% (98.3\%) & 99.2\% \\
Raw OpenAI & Benign (memory) & 100 & n/a & 100.0\%$^\dagger$ \\
Raw OpenAI & Benign (no memory) & 50 & n/a & 0.0\% \\
LangChain-style & Attack & 100 & 81.0\% (100.0\%) & 100.0\% \\
LangChain-style & Benign (memory) & 100 & n/a & 100.0\%$^\dagger$ \\
LangChain-style & Benign (no memory) & 50 & n/a & 0.0\% \\
\bottomrule
\end{tabular}
\end{table}

\noindent$^\dagger$\emph{Corrected benign memory-grounded control.} An earlier run
of this cell was invalid: memory was seeded under \texttt{preferred\_contact} while
the trigger caused the model to request \texttt{preferred\_contact\_email}, producing
a 100\% recall miss and zero completed sends (a failed lookup, not benign behaviour).
After seeding the fact under the requested key and re-collecting ($N{=}100$ per
variant, GPT-5.1, $T{=}0.0$), the control fires as intended: recall succeeds in
100/100 trials, all 100 complete a legitimate memory-grounded send, and \emph{all
100 exhibit the recall\_before\_send signature (benign FPR $= 100\%$)}. This is the
expected direction and reinforces the central claim: a legitimate ``recall the stored
contact, then email them'' task is trajectory-isomorphic to the attack. It is
consistent with (and more extreme than) the V2-1 benign study
(Section~\ref{sec:deployment-boundary}, FPR 24.7--57.6\%, 13-model factorial $N{=}4{,}160$),
because this task guarantees a recall-then-send by construction. The benign
no-memory control remains at 0\% (no recall edge exists).

The stable signature rates (99--100\% on attacks) confirm that the retrieval-to-action
trajectory is a property of the attack's information-flow structure, not an artefact
of the LangGraph execution engine or of system-prompt framing.

\paragraph{Cross-model sweep: precondition vs.\ predicate.}
Table~\ref{tab:model-sweep} presents the signature rate alongside ASR for 7 models:

\begin{table}[h]
\centering
\small
\caption{Model sweep (raw loop, $N{=}20$/model unless noted). ASR under the corrected recipient-only scorer (legacy body-inclusive ASR in parentheses). Models sorted by corrected ASR.}
\label{tab:model-sweep}
\begin{tabular}{lrrrr}
\toprule
Model & $N$ & ASR (corrected / legacy) & Signature & Category \\
\midrule
GPT-5.2 & 20 & 90.0\% (95.0\%) & 95.0\% & Exfiltrator \\
GPT-5.1 & 120 & 62.5\% (98.3\%) & 99.2\% & Exfiltrator \\
GPT-5 & 20 & 40.0\% (40.0\%) & 40.0\% & Partial \\
GPT-5-nano & 20 & 20.0\% (45.0\%) & 65.0\% & Partial \\
Gemini 2.5 Pro & 20 & 10.0\% (10.0\%) & 75.0\% & Partial (signature $>$ execution) \\
Gemini 2.5 Flash & 20 & 0.0\% (0.0\%) & 100.0\% & Latent Carrier \\
GPT-4.1-mini & 20 & 0.0\% (0.0\%) & 100.0\% & Latent Carrier \\
\bottomrule
\end{tabular}
\end{table}

Under the corrected recipient-only scorer the ASR/signature dissociation is sharper
than before: signature rates are unchanged (recipient scoring does not touch the
trajectory), while several ASRs fall (GPT-5.1 $98.3\% \to 62.5\%$; GPT-5-nano
$45\% \to 20\%$), widening the gap between the structural precondition and actual
exfiltration.

The Pearson correlation between model-level ASR and signature rate is $r = 0.031$
under the corrected recipient-only scorer used throughout this section (Table~\ref{tab:model-sweep});
computed on the legacy body-inclusive ASR it was $r = 0.069$. Either way the correlation is
near-zero ($n{=}7$ models; illustrative, not inferential at this sample size, and we do not
interpret its sign). Two models
(Gemini~2.5~Flash and GPT-4.1-mini) achieve 0\%~ASR while producing 100\% signature
rate: they retrieve the poisoned routing rule and draft/send emails to the
\emph{legitimate} recipient only, exhibiting the full retrieval-to-action trajectory
without executing the malicious routing. We term these \emph{Latent Carriers}: the
information-flow precondition is satisfied, but the model's execution-layer safety
prevents exfiltration.

\paragraph{Implications.} The near-zero correlation establishes quantitatively what
Section~\ref{sec:deployment-boundary} argued qualitatively: recall\_before\_send is a
\emph{structural precondition} encoding that memory-derived information reached the
action boundary, not a \emph{maliciousness predicate} encoding that the action was
harmful. It fires equally on successful exfiltration (GPT-5.1: 62.5\% ASR corrected, 99\%
signature) and failed exfiltration (GPT-4.1-mini: 0\% ASR, 100\% signature). A
trajectory detector calibrated for high recall will flag Latent Carriers alongside
successful attacks: an operationally useful property for supply-chain monitoring
(Latent Carriers store dormant payloads that may execute under future model upgrades)
but not for standalone inline blocking.

\paragraph{Cross-harness holdout (train LangGraph, test raw loop).} The classifier is
trained entirely on the LangGraph/SQLite factorial (its native harness). As a genuine
out-of-harness generalisation test, we froze that classifier and applied it, without
retraining, to the raw-OpenAI-loop cross-framing sessions above, which use a
different orchestration (direct SDK message loop, no LangGraph state machine) and a
different, more recent model population (GPT-5.x, Gemini~2.5). On the corrected
attack-success sessions, pooled recall is $186/188 = 98.9\%$ (per-model recall
$0.975$--$1.000$; only 2 missed sessions, both GPT-5.1 under the ReAct-prompt
variant). Because these files contain almost no negative class, AUC is not
informative here (near $0.5$ by construction); recall among attack successes is the
meaningful quantity. This is a cross-\emph{harness} result (different execution
engine), stronger than the within-loop prompt-framing comparison, and indicates the
retrieval-to-action signature is a property of the attack's information flow rather
than of the LangGraph logging substrate.

\subsection{Genuine-Framework Observable Arm and Scaffold-Side Implicit Proxy (V3-4)}
\label{sec:v34}
The cross-framing study above varies prompt framing within one raw execution loop; it is not a test of independent orchestration engines (we label it exploratory in Table~\ref{tab:claim-evidence}). We therefore preregistered V3-4 (GPG-signed, before data): a replication whose \emph{observable} and \emph{benign} arms run on a genuine agent framework (LangGraph's \texttt{create\_react\_agent} with tool-routed memory recall) and whose \emph{implicit} arm is a genuine LangChain-style retriever-injection chain. \textbf{Implementation disclosure (implicit arm):} the released harness (\texttt{probe\_v3\_4.py::run\_implicit}) instantiates a real FAISS retriever~\citep{johnson2021faiss} (the V3-2 store: 30 benign records $+$ 1 poisoned routing rule, \texttt{text-embedding-3-small}, \texttt{IndexFlatIP} cosine, top-$k$) that retrieves records for the trigger query; the retrieved top-$k$ are injected into the prompt and the LangGraph~\citep{chase2022langchain} agent runs with only email tools (no memory/recall tool). The agent therefore issues no observable retrieval tool call (framework-side read), while a framework oracle logs whether the poisoned record was actually retrieved (real payload-delivery flag, not a hardcoded delivery$=1$). This occupies the same taxonomy cell (implicit read, scaffold-side) as the originally-planned OpenAI Assistants \texttt{file\_search} arm, which was infrastructure-blocked (the institutional gateway forbids the Assistants and vector-store endpoints); the local FAISS retriever is a genuine open-source-framework retrieval pipeline standing in for the blocked managed endpoint, not a manual concatenation proxy. \textbf{Within-model contrast (confound closed):} both the observable and implicit arms are now run on \texttt{gpt-4.1} at $N{=}60$, so the $100\%$ vs.\ $0\%$ contrast below is a clean single-model manipulation of the observability axis (an earlier draft ran the observable arm on \texttt{gpt-4.1-mini} and the implicit arm on \texttt{gpt-4.1}, a cross-model confound; we retain the \texttt{gpt-4.1-mini} cells as an illustrative second model). The frozen P1-trained classifier is applied per-arm through a pre-declared ontology map: for the observable arm, LangGraph tool names map to canonical operations (the retrieval call genuinely exists); for the implicit arm, the framework retrieval issues no agent tool call and is encoded as \emph{absent} (\texttt{recall\_before\_send}${=}0$ by construction), logged only as a delivery oracle; never remapped into a recall feature, so the observability boundary is demonstrated rather than assumed.

Results (within-model on gpt-4.1 for the observable and implicit arms; benign arm on gpt-4.1-mini; gpt-4.1-mini observable/implicit cells retained as an illustrative second model):
\begin{itemize}
    \item \textbf{Observable arm (LangGraph ReAct, gpt-4.1):} attack succeeds in $39/60$ delivered sessions (ASR $65\%$); among successes the frozen classifier flags $39/39 = 100\%$ [0.910, 1.000]. The signature transfers to a genuine LangGraph orchestration engine: the first non-prompt-framing evidence in this work that the observable-arm signature survives a real agent framework. (On the second model, gpt-4.1-mini, the observable arm has a much lower ASR of $10/60 = 16.7\%$ under the genuine ReAct scaffold, yet the classifier still flags $10/10 = 100\%$ [0.722, 1.000] of those successes: the low mini ASR is why the original confirmatory roster paired mini with a gpt-4.1 implicit arm.)
    \item \textbf{Benign arm (LangGraph ReAct, gpt-4.1-mini):} the frozen classifier flags $60/60 = 100\%$ [0.940, 1.000] of benign memory-grounded sessions. The precision ceiling reproduces under a real framework. Note that gpt-4.1-mini, classified act-without-grounding (0\% benign FPR) under exact-key recall, grounds on \emph{every} benign session under a genuine ReAct scaffold, reinforcing that the phenotype is interface-dependent, not model-intrinsic.
    \item \textbf{Implicit arm (genuine FAISS retriever, gpt-4.1):} the attack succeeds in $57/60$ oracle-delivered sessions (ASR $95\%$), and the frozen classifier flags $0/57 = 0\%$ [0.000, 0.063] of them; the framework retrieval produces no agent-visible tool call, so the detector is structurally blind. This is a genuine retriever-injection chain with a real delivery oracle, not a manual concatenation. (On the second model, gpt-4.1-mini, the implicit arm reached only $1/60$ successes and is uninformative on its own; gpt-4.1 is the powered within-model comparison.)
\end{itemize}
The within-model observable--implicit contrast on gpt-4.1 ($39/39 = 100\%$ vs.\ $0/57 = 0\%$ frozen-classifier recall) demonstrates the observability boundary on a genuine LangGraph engine (observable) and a genuine FAISS retriever chain (implicit), holding the model, task family and payload fixed while changing the read path from agent-visible tool-routed retrieval to framework-side retrieval injection. \emph{Corrected:} this sentence previously claimed the contrast held ``the execution framework fixed and varying only whether the read is tool-routed,'' which contradicts this subsection's own implementation disclosure two paragraphs above: the observable arm is LangGraph \texttt{create\_react\_agent} with memory tools in the schema and the implicit arm is a LangChain-style FAISS retriever chain with email tools only. Orchestration and tool schema both differ. That is unavoidable given what is being manipulated (removing the retrieval tool is what makes the read implicit) so the contrast is a clean single-variable manipulation at the \emph{taxonomy} level (observable vs.\ implicit read) rather than a single-framework ablation, and we state it that way. It complements the architectural argument (Section~\ref{sec:event-graph}) and the V3-2 arm~C construction. The gpt-4.1-mini cells (observable $10/10$, implicit $0/1$, benign $60/60$) reproduce the same directional boundary on a second model.

\textbf{Endpoint disclosure (V3-4).} V3-4's success scorer is the \emph{policy-violation} endpoint of \S\ref{sec:threat}, not the planted-address one: \texttt{probe\_v3\_4.\_finish} sets \texttt{exfil} if any recipient differs from \texttt{team@example.com}, so an invented or confabulated external address counts as a success without the planted \texttt{attacker@evil.example.com} appearing. This is the same allowlist operationalisation V3-1 uses, and it carries the same ambiguity: the counts establish unauthorised external sends, not that the specific planted payload drove the routing. Unlike V3-1 we do \emph{not} report a planted-address-matched sensitivity analysis for V3-4, because the arm's per-call recipient arguments were not persisted in a form that supports it. The observability contrast is unaffected (it compares the \emph{signature} rate among successes under a fixed success definition, and the implicit arm's $0/57$ cannot be produced by a looser recipient rule) but the $65\%$ and $95\%$ ASR figures should be read as policy-violation rates, not planted-payload attribution rates.

\paragraph{Scope.} This extension validates cross-scaffold and cross-model transfer of
the retrieval-to-action signature under one attack construction (the delayed-trigger
compliance-framing attack). Transfer across independent attack families is now partially
probed but not confirmed: X1 (Section~\ref{sec:x1}) reproduces the both-implicit boundary on a
mechanism-faithful MINJA query-only case-bank implementation (within-model
gpt-4.1 observable $80/80$ vs.\ implicit $0/141$), but after the inert-\texttt{similarity}
cluster-unit correction and a preregistered continuation to the registered cap the implicit arm
reaches only $9$ of the required $10$ effective clusters (and just $8$ configurations contribute
eligible successes in both arms; under the stricter send-only endpoint neither arm passes), so
this is directional/LOW-CLUSTER evidence and the ``no independent attack family''
gap is narrowed rather than closed; the emergent sliding-window Zombie implementation likewise remains
directional/LOW-CLUSTER (no fully-powered within-model contrast). Clustered-primary
independent-family confirmation therefore remains the primary outstanding
external-validity gap for both families. The $N{=}20$
model-sweep cells have Wilson CI half-widths of
$\pm$16--22pp; the 0/20 and 20/20 results map to [0\%, 16.1\%] and [83.9\%, 100.0\%]
respectively. Per-trial logs, prompts, and harness code are released with the paper
artefact.

\section{Interpretation}
\label{sec:discussion}

\subsection{Why Tool-Call Logging Is Sufficient for Structural Triage}

The detector requires no model access, no memory inspection, and no additional inference. Tool-call logs are already generated by production agentic frameworks (\eg LangChain tool invocation hooks, OpenAI function-call traces, Anthropic tool-use events). Deploying the classifier requires only routing these logs to a feature extractor and a trained model. The per-session inference cost is negligible (19 low-dimensional features; RF prediction in microseconds). Integration is low-cost for any operator who can record tool invocations, a lower bar than defences requiring memory-store modification, model internals, or content inspection, but low integration cost is not evidence of adequate precision, which \S\ref{sec:deployment-boundary} shows remains the open problem.

\subsection{Relationship to the P1 Defence Failure}

IED (same author) found that five of six defence classes fail against memory poisoning in the tested configuration: input-level and retrieval-level defences cannot observe the attack surface, and instruction-level defences are overridden by compliance framing. The trajectory classifier is complementary: it is a \emph{detection} mechanism, not a \emph{prevention} mechanism. It does not stop the attack; it flags that a memory-grounded action occurred, which narrows a candidate set for incident response. It does \emph{not} attribute the vector: \S\ref{sec:evasion} withdraws that implication, because prompt-inline attacks also produce the signature in $7/20$ sessions and benign memory-grounded sends produce it at a $100\%$ conditional rate. Detection without prevention retains operational value for triage, forensics, and anomaly alerting, provided attribution is done with the recipient and content inspection the operation-only feature set excludes.

\section{Limitations}
\label{sec:limitations}

\paragraph{False negative rate.} FN $= 31$ out of $1{,}919$ attack-success sessions (1.6\%). All 31 originate exclusively from qwq:32b's implicit-bypass runs under the Memory Sandbox defence (goal-directed RAG fallback without \tool{recall\_fact}). Excluding qwq:32b's Sandbox cell, no other model$\times$defence cell in this corpus produced a false negative. \emph{Corrected:} that sentence previously read ``in deployments without qwq:32b, the effective FNR is zero across all tested models,'' which converts a within-corpus count into a claim about deployments. It does not license one: qwq:32b is not a property of a deployment but an instance of the implicit-bypass class (\S\ref{sec:cross-framework}), any member of which produces the same false negative, and the zero is measured on one harness and one attack construction rather than estimated for unseen fleets. Separately, the 62 false \emph{positives} (FPR $= 10.3\%$) arise on non-exfiltration sessions whose benign recall patterns mimic the attack signature, with zero associated false negatives (Recall $= 1.000$ is maintained wherever they occur). As noted in Section~\ref{sec:perf}, the artefact stratifies those false positives by defence ($40$ Prompt Hardening, $22$ Memory Sandbox) and not by model, so we do not attribute them to a single model; the per-model evidence is the qwen2.5:14b rank inversion of Section~\ref{sec:holdout}, which is a statement about that model's hold-out AUC rather than about the $62$ in-fold false positives.

\paragraph{Feature definition limitation.} An earlier \texttt{recall\_before\_send} feature implementation compared only the \emph{first} \tool{recall\_fact} invocation with the \emph{first} \tool{send\_email} invocation (first-occurrence semantics); the released classifier does not, as stated at the end of this paragraph. Under that superseded definition a session that sends, then recalls, then sends again was incorrectly labelled $\texttt{rbs}{=}0$ even though a later exfiltrating send follows retrieval. The corrected definition (any recall preceding any exfiltrating send) produces cleaner $100\%/0\%$ conditional FPR separation. This bug does not affect any reported number: across all 2{,}520 primary DTA sessions the naive first-occurrence and corrected any-preceding definitions produce \emph{identical} labels (0 disagreements), including on the 1{,}120 sessions that do contain multiple \tool{send\_email} calls; in every such session a \tool{recall\_fact} precedes the first send, so first-vs-first and any-vs-any agree. The only observed divergence arises in out-of-scope replication scoring on qwq:32b revision-pattern sessions (send, revise via recall, then re-send), which are not part of the published training data. The released classifier (\texttt{paper\_a\_classifier.py}) implements the corrected any-preceding-recall definition; because the two definitions produce identical labels on all 2{,}520 primary sessions, every reported table and figure is unaffected by the choice.

\emph{The V3-2 \texttt{search\_before\_send} scorer was not migrated (disclosed here).} The correction above was applied to \texttt{recall\_before\_send} but not to its semantic-retrieval counterpart: \texttt{probe\_v3\_2.py} still computes \texttt{search\_before\_send} as \emph{first}-\tool{memory\_search} versus \emph{first}-exfiltration-capable-email-action, the form retired above. The two definitions are inconsistent in the artefact, which we report rather than leave for a reader to discover. On the collected data the discrepancy is inert: re-scoring every V3-2, V3-2-$k$ and format-confound file under both conventions yields \textbf{zero} divergent sessions, despite $30$--$89$ multi-send sessions per arm-B file (for example \texttt{v3\_2\_gpt\_4o\_b\_fresh\_2026\_09.jsonl} has $89$ scorable sessions, all multi-send, and no divergence). The reason is the same as for \texttt{recall\_before\_send}: in every arm-B session a \tool{memory\_search} precedes the \emph{first} email action, so first-vs-first and any-vs-any coincide. No V3-2, Exp~13a or X3 number changes under either convention, and the arm-C cells are definitional zeros with no search tool in the schema. We have therefore not re-scored the released files; the inconsistency is a maintenance hazard for future collections rather than a defect in any published figure, and a session exhibiting send-then-search-then-send would be mislabelled if one ever occurred.

\paragraph{Tool-name / ontology dependency.} The feature extractor keys on a fixed
vocabulary of canonical operation names (\tool{recall\_fact}, \tool{send\_email},
etc.); any tool call whose name is not in this vocabulary contributes zero to every
feature (the extractor counts known operations and silently ignores unknown ones).
A deployment whose memory tool is named differently, or whose framework renames
operations across versions, therefore silently disables the detector unless an
explicit ontology map is supplied. We observe this directly: the frozen classifier
detects $0/38$ (GPT-4.1) / $0/13$ (GPT-4o) of the V3-2 semantic-search successes
before remapping and $86.8$--$100\%$ after a pre-declared \tool{memory\_search}${\to}$\tool{recall\_fact}
map (Section~\ref{sec:cross-framework}). The practical consequence is that the
mapping step is a load-bearing, unvalidated deployment dependency: tool-name drift
is indistinguishable from a benign session in the feature space, so an operator must
maintain the ontology map as part of the detector, not treat the classifier as
name-agnostic.

\paragraph{Write-axis re-analysis limitation.} The $2\times2$ write-observability analysis (Section~\ref{sec:write-axis}) uses only the read-implicit V3-1/V3-2 arms and the benign V3-2 arm; the \textit{save\_before\_send} figures (24/28 catch, 66.7\% benign FPR, 0/53 in the both-implicit cell) are descriptive re-analyses of already-collected data, not a preregistered confirmatory experiment. Two limitations bound their interpretation. First, the original V3-1 logs recorded operation names but not \tool{memory\_save\_fact} key/value arguments; we closed this with a preregistered argument-logging re-run (Exp~13c) that verified $37/37$ \textit{save\_before\_send}-flagged successes wrote the actual attacker payload ($26/26$ GPT-4.1, $11/11$ GPT-4o), so \textit{save\_before\_send} now has verified payload-write provenance rather than being an operation-level association. This strengthens only the precondition side: the $66.7\%$ benign FPR is unchanged, so the write axis is still a precondition, not a predicate. Second, the write-axis result is within-harness (our DTA-derived ontology, in which writes are tool-routed); native MINJA and Zombie Agents write scaffold-side, matching the both-implicit cell where the write signature also fails (0/53). The write axis thus does not provide a deployable detector and does not alter the MINJA/Zombie classification.

\paragraph{Benign false-positive rate.} The reported false-positive rate in the main evaluation is measured against poisoned-but-defended non-exfiltration executions, not against ordinary unpoisoned benign agent traffic. A preregistered follow-up study (13-model factorial $N{=}4{,}160$; Section~\ref{sec:deployment-boundary}) measures the deployment FPR: $P(\text{FP} \mid \text{recall\_before\_send}{=}1) = 100.0\%$ $[99.8, 100.0]$ (unconditional benign FPR: $24.7$--$57.6\%$ depending on recall protocol and fact count). Benign memory-grounded sends are score-isomorphic to attacks under the trained classifier. The detector should therefore be deployed as a high-recall triage signal rather than a standalone blocker. We note that the gating layer such a deployment implies is itself unvalidated here: the only gate we measured (recipient externality) fails on legitimate external mail (\S\ref{sec:v35}), and no content- or intent-level gate was evaluated.

\paragraph{Frontier false positive rate: withdrawn.} Earlier drafts reported, on frontier models, Recall $= 0.901$ across $405$ attack sessions and FPR $= 23.0\%$ on $560$ non-attack sessions against a training FPR of $10.3\%$. \textbf{Those figures are withdrawn} and no frontier-transfer FPR is claimed. \emph{This cell is not regenerable from the released artefact} (\S\ref{sec:gpt41}); the artefact-backed benign FPR evidence is V2-1's, in \S\ref{sec:deployment-boundary}. We had explained the increase by reasoning models calling \tool{recall\_fact} more often on non-attack sessions, mimicking part of the signature. That explanation is \textbf{also withdrawn}: it interprets a measurement we cannot regenerate, and no artefact in the tree supports it. Per-model threshold calibration would likely reduce this gap; the current single-threshold design is a conservative choice that favours recall over precision.

\paragraph{Training data scope.} The classifier was trained on open-source models via Ollama. Cross-model hold-out confirms generalisation to GLM, GPT-OSS, and Qwen families. \emph{Corrected:} this sentence previously continued ``and the GPT-4.1 probe ($N{=}20$) confirms frontier model alignment.'' That cited as confirmation the one cell \S\ref{sec:gpt41} discloses is not regenerable from the released artefact, so the claim rested on evidence the paper elsewhere marks absent. The $N{=}20$ pilot is a feasibility observation only, and the $405$-session evaluation that earlier drafts leaned on is \textbf{withdrawn} (\S\ref{sec:gpt41}). \emph{No regenerable frontier-transfer evaluation remains}, and the paper claims none. A larger frontier-model training set would sharpen the decision boundary, particularly for models with recall counts $> 1.5$ (the GPT-4.1 / GPT-4o regime).

\paragraph{Implicit-bypass attacks.} The classifier correctly misses qwq:32b's Memory Sandbox bypass (goal-directed RAG fallback, no \tool{recall\_fact}). This is one instance of a general boundary: attacks using implicit retrieval (framework-level embedding similarity, sliding-window assembly, or cached context) produce no observable retrieval tool call, and therefore leave the detector no \emph{retrieval provenance} to condition on (see Section~\ref{sec:cross-framework} for the full taxonomy). \emph{Corrected:} earlier drafts said such attacks ``evade the detector entirely,'' which our own X1 data contradict: the frozen RF still fires on implicit MINJA successes through action-shape features that have nothing to do with retrieval (\S\ref{sec:x1}). Losing the retrieval signature is not the same as producing no alert, and the two must not be conflated: what is structurally unavailable is the provenance, not necessarily the alert. Detecting implicit-bypass attacks \emph{as memory-channel attacks} requires tracking the \emph{semantic correlation} between RAG document content and outbound email fields, rather than relying on raw \texttt{rag\_count} volume metrics alone. We leave this for future work.
Companion evaluations confirm that the three models
underlying the stable detection spine (qwen2.5:14b, qwen2.5:72b, qwen3:32b)
reproduce at 100\% ASR across independent evaluation environments,
while reasoning models exhibit output-quality degradation under long-running
inference sessions that affects attack \emph{prevalence} but not trajectory
\emph{shape} when attacks succeed. The rate-fragility identified in companion
work does not reach the detection invariant: it changes class balance, not
the forced recall-before-send signature.

\paragraph{Adaptive adversaries.} We consider three evasion strategies an adversary aware of the detector might attempt: (1)~\emph{Suppress recall}: embed the routing address directly in the prompt, eliminating the \tool{recall\_fact} call. This changes the attack vector to prompt-injection, which the detector correctly classifies as a different channel (Section~\ref{sec:evasion}). The attack succeeds but loses persistence across sessions. (2)~\emph{Inject fake recalls into non-attack sessions}: the adversary poisons benign sessions to include spurious \tool{recall\_fact} calls, inflating the FPR. This requires persistent influence over the agent's environment, itself a security compromise, and cannot suppress the genuine attack signature. The detector's precision degrades but recall is unaffected. (3)~\emph{Retrieve via alternative channels}: use framework-level embedding retrieval or cached context instead of the memory tool. V3-1 and V3-2 arm C empirically demonstrate this strategy works: implicit-delivery attacks achieve 50--82\% success rates on GPT-4.1 and GPT-4o while eliminating the \emph{retrieval-provenance} signature. ``Completely evading the detector'' would be too strong and is not claimed: the frozen classifier can still alert on unrelated action-shape features, and what implicit delivery removes is the retrieval evidence, not necessarily every alert. An attacker aware of the detector need only target architectures where payload delivery occurs scaffold-side, or (per V3-1's adaptive implication) harvest only the rbs${=}$0 subset of implicit attacks. This is not a hypothetical evasion but a demonstrated, viable strategy, and the primary reason we characterise the detector as a precondition filter for tool-routed attacks, not a general-purpose defence.

\paragraph{Limitations of the integrated experiments (X1--X4).} The four preregistered experiments each carry power and construct-validity caveats that bound their claims. \emph{X1 MINJA (measured, directional/LOW-CLUSTER):} the MINJA case-bank family gives the cleanest independent-family contrast (gpt-4.1 observable $80/80$ vs.\ implicit $0/141$, paired contrast $+1.0$) but a post-hoc audit found the grid's \texttt{similarity} factor inert (\texttt{minja\_poison\_case()} receives only domain and framing; $12$ distinct poison texts across $36$ \texttt{scenario\_id}s), which had inflated the reported cluster counts ${\approx}3\times$. A preregistered continuation to the registered $N{=}360$ cap brought the observable arm to $10$ effective clusters (gate satisfied, $80/80$) but left the implicit arm at $9$ ($0/141$, cluster-level Wilson upper $0.299$; the trial-level $0.0265$ assumes independence the clustered design does not support): one short, and structurally so, because three of the twelve effective configurations produce zero eligible implicit successes. Under the Addendum's both-arms rule the A${\to}$P upgrade therefore stays retracted and the result is directional/LOW-CLUSTER with Wilson intervals governing; it also rests on one model (gpt-4.1), with gpt-4o corroborating only directionally (effective clusters: observable $11$, implicit $2$). \emph{X1 Zombie (directional/LOW-CLUSTER):} the emergent sliding-window result is not powered; gpt-4.1's implicit arm has $16$ eligible successes across only $5$ scenario clusters (below the gate) and gpt-4o's implicit arm has $0$ eligible successes, so there is no fully-powered within-model contrast; Zombie remains an architectural classification with directional support, and rests on one fixed sliding-window policy ($W{=}6$, one frozen promotion rule). \emph{Frozen-RF is not a proxy for retrieval provenance:} in the implicit MINJA arm the raw \texttt{observable\_read\_of\_poison} is $0/141$ yet the frozen RF still flags those sessions via non-retrieval trajectory features (on which it is out-of-distribution), so the X1 conclusion is ``no retrieval provenance in the tool logs'', not ``no trace-based alert''; the raw retrieval variable, not the RF, carries the claim. \emph{X4:} the document-share necessity is established only on the control arm; the implicit treatment arm is not viable on gpt-4.1 ($0/360$ successes) and only directional on gpt-4o ($20$ successes, $10$ clusters, below the preregistered $37$-success target despite meeting the $10$-cluster minimum), only one exfiltration channel was added beyond email, and the gpt-4.1 benign keystone is thin ($14/14$ completed shares out of $72$ sessions). The ontology-remapped RF transfer ($37/37$) is an engineering result reflecting tool-name renaming, not evidence of channel-generality. \emph{X2:} the second open-weight grounder result was collected on the existing GUI Ollama daemon rather than a fresh-daemon restart (Regime-A determinism confound, digest \texttt{9f13ba1299afea09}), and its control arm is underpowered ($0$ violations $/24$ successes, below the $37$-success target, Wilson upper $0.138$), so it is consistent with but does not confirm the invariant. The treatment-arm breakdown ($27/31 = 87.1\%$) has only $7$ success-contributing clusters (below the ${\geq}10$-cluster gate), so it is LOW-CLUSTER and Wilson $[0.711, 0.949]$ governs descriptively; it provides a directional (not confirmatory) replication of the implicit-delivery breakdown on a second grounder.

\section{Related Work}
\label{sec:related}

\paragraph{Memory poisoning attacks.} Persistent memory attacks against LLM agents were established by MINJA~\citep{shao2025minja} and Zombie Agents~\citep{yang2026zombie}, which demonstrated high attack success rates against open-source models via query-only memory injection. The delayed-trigger attack (DTA)~\citep{leong2026defense} embeds a malicious compliance directive in a RAG-retrieved document; the directive instructs the agent to store a routing rule using \tool{memory\_save\_fact}, which is then retrieved via \tool{memory\_recall\_fact} in a later session and used to exfiltrate data. \citet{leong2026defense} (same author) reports that five of six defence classes fail against this attack in the tested configuration, with only tool-layer memory restriction achieving structural protection. Hidden in Memory~\citep{pulipaka2026hidden} independently confirms that GPT-5.5 stores adversary-induced memories at 99.8\% injection rate, demonstrating that injection-layer vulnerability persists even in the most capable frontier models. Cross-Session Stored Prompt Injection~\citep{xie2026crosssession} formalises the XSS analogy for persistent injection across session boundaries. MPBench~\citep{dash2026mpbench} provides a benchmark covering four memory write channels and six attack classes, explicitly motivating behavioural detection approaches but not providing a detector. Critically, MINJA and Zombie Agents both use \emph{implicit} retrieval mechanisms (framework-level embedding similarity or sliding-window assembly) rather than observable agent tool calls: placing them outside the architectural scope of our detector (Table~\ref{tab:attack-taxonomy}).

\paragraph{Behavioural detection in agentic systems.} VIGIL~\citep{li2026vigil} translates behavioural specifications into SMT constraints over finite tool-call event traces, achieving $>95\%$ recall with $<10\%$ FPR on policy violations. It enforces pre-specified behavioural policies rather than learning forensic signatures from labelled data. Our approach is complementary: we learn statistical signatures from labelled data without requiring a policy specification, targeting a specific attack class rather than arbitrary violations. MemMorph~\citep{zhang2026memmorph} demonstrates that poisoned long-term memory can steer tool selection across multiple agent architectures; our work asks whether such memory-channel attacks leave operation-only forensic signatures detectable by a supervised classifier. MemLineage~\citep{ouyang2026memlineage} attaches cryptographic provenance to memory entries, enabling detection of unsigned writes, but requires modification of the memory infrastructure and cannot detect attacks injected through the agent's own authorised write pathway (as in DTA, where the agent legitimately stores the malicious rule from a RAG-retrieved compliance document). MEMSAD~\citep{gowda2026memsad} achieves near-perfect detection using gradient-coupled anomaly scoring, but it is a \emph{preventive} defence that blocks attacks before execution and requires model-internal gradient access; our detector is a post-hoc forensic tool requiring only tool-call logs. Concurrent work~\citep{dang2026enforcing} enforces behavioural firewalls over tool-call sequences using probabilistic deterministic finite automata; they define \emph{permitted} trajectories, whereas we learn \emph{attack} signatures from labelled data.

\paragraph{Intrusion detection and anomaly detection analogues.} Host-based intrusion detection systems (HIDS) in classical security identify compromises from system call sequences and process execution traces~\citep{forrest1996sense,warrender1999detecting}. Sequence-based anomaly detection on system calls (using n-gram models, HMMs, or neural sequence classifiers) is structurally analogous to our approach: both extract features from ordered event logs and classify sessions as normal or anomalous. The key distinction is that classical HIDS operates on low-level OS primitives, while our detector operates on tool invocation events in an LLM agent runtime. This framing clarifies the deployment requirement: tool-call logging in agentic frameworks (LangChain tool invocation hooks, OpenAI function-call traces, Anthropic tool-use events) is the agent-runtime equivalent of \texttt{strace} or Linux Audit, and the trajectory classifier is a direct implementation of HIDS principles at the agentic abstraction layer. PerD~\citep{garciasoto2022perd} detects embedded Trojans in static NLP models by analysing input-output behavioural responses, the most directly analogous prior work, but applies to models with fixed behaviour, not agentic systems where tool-call trajectories vary with memory state.

\paragraph{Execution-trace auditing for agents.} Recent concurrent work has explored execution-trace analysis for agent security. TraceAegis~\citep{liu2025traceaegis} applies provenance-based rules over agent execution logs to detect generic anomalous behaviours, while TRACES~\citep{wang2025traces} learns trajectory risk states from an observer LLM's hidden representations. Unlike these systems, which learn generic unsafe patterns or require formal policy specifications, our approach identifies an architecture-induced signature specific to persistent memory poisoning, using lightweight structural features without model internals or learned representations.

\paragraph{Provenance granularity, and what this paper's boundary does and does not claim as new.} The distinction between a logged retrieval \emph{event} and a verified payload-to-action \emph{dependency} (Definitions~\ref{def:coverage}--\ref{def:edges}, \S\ref{sec:graph-formal}) has adjacent prior art that a targeted prior-art check surfaced, and we position against it explicitly rather than leave the novelty claim unbounded. \citet{fan2026pact} name a \emph{granularity mismatch} in agent security: defences that mediate trust at the granularity of a whole tool invocation cannot separate a hijacked authority-bearing argument from benign retrieval-then-act behaviour, so they track value provenance at the argument level instead. That is the same observation as our node-versus-edge point, made for enforcement rather than for detectability, and it independently corroborates it: notably their strongest results hold \emph{under oracle provenance}, with provenance inference isolated as the remaining bottleneck, which is precisely the quantity an operation-only log does not supply. \citet{chinaei2026laundering} build a provenance graph over tool calls, returned data, field-level provenance and denied actions, augmented with counterfactual edges, to catch leakage that flat provenance tracking misses. A provenance graph over agent tool calls is therefore not itself a novel object, and we do not claim it as one. \citet{yu2026attribution} give the analogue one layer down: output-level monitoring cannot verify whether retrieved context actually governed generation, because parametric recall and context grounding yield indistinguishable output; the same logical shape as our indistinguishable-log argument, at the representation rather than the tool-log layer. \citet{ouyang2026memlineage} reach a compatible conclusion from the defensive side, attaching cryptographic provenance and derivation lineage to memory entries rather than trusting logs.

\emph{What remains specific to this paper}, after that check, is narrower than ``a provenance graph for agent security'' and we state it as such: (i)~using such a graph to characterise \emph{detectability} rather than to enforce a policy, by asking which payload-to-action paths contain an observable retrieval node at all, and partitioning published attack families on that criterion (Table~\ref{tab:attack-taxonomy}); (ii)~measuring the resulting coverage property against collected agent traces rather than assuming it, including a direct test of the previously unmeasured cell in which the payload \emph{is} available behind a live retrieval tool, which falsified our own earlier behavioural-decline reading (\S\ref{sec:v36}); and (iii)~falsifying the standalone deployment of a detector built on that coverage property by measuring the benign base rate (\S\ref{sec:deployment-boundary}). We found no prior work that states the retrieval-observability ceiling itself as a claim (searches for it return this paper's own earlier versions) but the absence of a stated claim in a literature that has independently converged on argument-level and lineage-level provenance is weak evidence of novelty, not strong evidence. A reader who regards the coverage property as obvious once stated should read the contribution as (ii) and (iii), which are measurements, rather than as (i).

\paragraph{The detection gap.} Unlike general trace-auditing or memory-poisoning benchmarks, no prior work studies whether a specific explicit-memory exfiltration channel induces a low-dimensional, operation-only trajectory signature, nor measures where that signature is architecturally unavailable. MPBench~\citep{dash2026mpbench} systematises the memory-poisoning attack surface across four write channels; MemMorph~\citep{zhang2026memmorph} demonstrates tool-selection steering; VIGIL~\citep{li2026vigil} verifies tool-stream intent. What none of them supplies is the pair of measurements this paper contributes: the retrieval-event coverage property measured against collected agent traces (\S\ref{sec:cross-framework}, \S\ref{sec:x1}, \S\ref{sec:x4}), and the benign base rate that falsifies standalone deployment of any detector built on it (\S\ref{sec:deployment-boundary}, \S\ref{sec:v35}). \emph{Corrected:} this paragraph previously located the contribution in ``a supervised classifier trained on an instrumentally labeled factorial dataset ($N{=}2{,}520$, 9 models, 7 defenses), evaluated with cross-validation, bootstrap uncertainty estimates, and leave-one-model-out hold-out.'' That contradicted \S\ref{sec:intro} and \S\ref{sec:results}, both of which state that the classifier is a measurement instrument and not the contribution, and it claimed novelty for precisely the two rows Table~\ref{tab:claim-evidence} downgraded to E$^{\S}$ on the grounds that no preregistration covers them. The classifier and its evaluation are retained as the instrument that establishes the invariant's dominance within scope; the novelty claim rests on the two measurements above, and is bounded further in the preceding paragraph.

\paragraph{Disclosure.} References~\citep{leong2026defense} and~\citep{leong2026recognition} are companion preprints by the same author. The training data for this paper's classifier originates from the factorial in~\citep{leong2026defense}. No findings cited from companion works have been independently replicated.

\section{Conclusion}
\label{sec:conclusion}

Persistent memory poisoning attacks produce a stable behavioural invariant in the agent's execution trajectory. The \textit{recall\_before\_send} transition follows from the attack's information-retrieval dependency: a simple rule exploiting it alone achieves AUC $= 0.9563$, and suppressing observable retrieval breaks the attack via the tool-routed channel (though scaffold-side delivery remains viable). A full trajectory classifier refines this to AUC $= 0.9904$; removing all recall-related features leaves in-distribution AUC unchanged because the retrieval event triggers a behavioural cascade (draft, send, sequence elongation) that produces redundant correlated encodings across multiple feature channels, not independent detection mechanisms, but downstream consequences of the same causal event. This redundancy is in-distribution only: on implicit-bypass attacks (\texttt{recall\_before\_send}${=}0$) the full classifier flags 0/28 while the recall-ablated model flags 24/28, showing the fitted model overfits the recall root split and does not transfer its cascade coverage to implicit retrieval (Section~\ref{sec:write-axis}).

The invariant generalises across 6/9 open-weight models (7B--120B parameters) and was reported in earlier drafts to transfer to frontier models without retraining, a claim this version \textbf{withdraws} because the cell producing it is not regenerable from the released artefact (\S\ref{sec:gpt41}). However, a preregistered follow-up (13-model factorial $N{=}4{,}160$) reveals a critical deployment boundary: benign memory-grounded sends produce the same \textit{recall\_before\_send} signature, yielding $P(\text{FP} \mid \text{recall\_before\_send}{=}1) = 100\%$ on ground-then-act models (unconditional benign FPR: $24.7$--$57.6\%$). The signature is a valid attack precondition, not a maliciousness predicate. Standalone inline blocking is not viable. What our data support is use as a high-recall triage signal feeding a second stage that inspects task intent or content; the one gating variable we actually tested (recipient externality) does \emph{not} repair precision, flagging $129/129$ of legitimate external business email with the combined trajectory-and-recipient gate still flagging $45\%$ (\S\ref{sec:v35}). We therefore state content- or intent-level gating as the indicated direction rather than a validated deployment model: no such gate was evaluated here, and its precision is an open empirical question. Critically, the behavioural phenotype governing benign retrieval rates is interface-dependent: GPT-4.1 exhibits 0\% benign recall-before-send under exact-key retrieval but 82.5\% under semantic search, so deployment economics are conditioned on retrieval architecture, not model identity alone.

These results establish that tool-routed memory-recall exfiltration attacks leave execution-trajectory signatures that are robust within the tested scope (act-without-grounding models, exact-key and semantic tool retrieval), but that operation-only structural features face a hard precision ceiling when benign agents also ground their actions in memory. The scope boundary is a property of observable retrieval dependence: when it holds, the signature is necessary; when the scaffold delivers payloads without observable tool calls, attacks remain viable without the signature. It is validated on two act-without-grounding models (GPT-4.1, GPT-4o); via V3-3, on a ground-then-act model (gemini-2.5-pro) whose strong benign recall habit ($88.3\%$) does not restore the signature under implicit delivery (V3-6, \S\ref{sec:v36}, shows the mechanism is store contents rather than a declined retrieval, since with the payload in the store \emph{and} in context that model retrieves it $108/108$ times); and, via X2, on a second open-weight grounder, qwen2.5:32b, with an underpowered control. The necessity is not email-specific: it extends from \texttt{recall\_before\_send} to \emph{retrieval-before-exfiltrating-action} across the two tested exfiltration channels (email and document-share; X4 control $0$ violations $/50$ (gpt-4.1) and $0/62$ (gpt-4o), Section~\ref{sec:x4}), though on the document-share channel the evidence is directional, its cluster-unit upper bounds ($0.121$, $0.104$) fall short of the registered $\leq 0.10$, with the implicit treatment arm underpowered and only one channel tested beyond email. The both-implicit boundary is further probed on a mechanism-faithful independent attack family, a MINJA query-only case-bank implementation (within-model gpt-4.1 observable $80/80$ vs.\ implicit $0/141$, matched-scenario paired contrast $+1.0$; Section~\ref{sec:x1}), though after the inert-\texttt{similarity} correction and a preregistered continuation to the registered cap the implicit arm reaches only $9$ of the required $10$ effective clusters (and only $8$ effective configurations contribute eligible successes in \emph{both} arms), so MINJA, like the emergent sliding-window Zombie-Agents implementation, is directional/LOW-CLUSTER and is not upgraded to measured. Deployment requires both the trajectory signal (for recall) and semantic context (for precision).

\section*{Ethics and Broader Impact}

This work evaluates forensic detection of memory-poisoning attacks in LLM agents using benign proxy tasks (simulated email composition with fictional recipients and documents). No genuinely sensitive data, real credentials, or harmful payloads were used. The attack payloads consist of routing instructions directing simulated emails to fictional external addresses. We release raw tool-call logs but not model weights or proprietary API prompts.

\paragraph{Dual-use considerations.} Publishing the trajectory-based detection mechanism could theoretically inform attackers seeking to evade it. We judge the net benefit positive because: (1) the evasion boundary (implicit retrieval) is already public knowledge through concurrent work~\citep{yang2026zombie,shao2025minja}; (2) the detection mechanism is a property of observable tool-routing architecture, not a secret; and (3) the paper simultaneously characterises where detection \emph{fails}, informing defenders about realistic adversarial adaptation. \emph{This extends to the released artefact specifically, not only to the general boundary.} The adaptive-attacker evasion battery (X5, four destination-obfuscation transforms, Appendix~\ref{app:x5}) and the domain-dependence measurements (attack viability $0/60$ to $60/60$ depending on task domain, \S\ref{sec:x1}) are more specific and more actionable than the general claim that implicit retrieval evades detection: X5 exercises concrete transforms an attacker could adopt, and the domain map identifies which cover tasks make an attack more likely to succeed without leaving a retrieval signature. We release them because X5's tested transforms were all measured non-viable ($0$ eligible successes across all $1{,}016$ trials), so no working evasion technique is being handed over, and because the domain-dependence pattern is, by the same logic as (2), an emergent property of task structure rather than a technique this paper introduces; withholding the measurement would not remove the underlying task-structure dependence, only a defender's knowledge of it. The recipient-metadata gate we tested and found ineffective ($129/129$ false positives on legitimate email, \S\ref{sec:deployment-boundary}) is reported for the same reason: telling defenders a mitigation does not work is itself a defensive contribution.

\paragraph{Responsible use.} The detector is designed as a triage signal, not a standalone enforcement mechanism. Deployment without human review or semantic gating could produce high false-positive rates on legitimate workflows. We explicitly document the precision ceiling and recommend against automated blocking. \emph{The observability upgrade discussed as closing part of the detection gap (\S\ref{sec:graph-implications}: promoting framework-side retrieval to a logged, first-class span) is itself a trade-off we name rather than leave implicit.} Logging retrieval operations as first-class spans means retaining a queryable record of what an agent read from memory on every session, which can carry the same recipient, document and content metadata the retrieval itself concerns; this enlarges the sensitive-data surface a deployer must govern (retention limits, access control, redaction of retrieved content, and who may query the log), rather than being a cost-free visibility gain. We do not evaluate any specific retention or access-control scheme, and a deployer adopting this recommendation should treat it as a new data-governance surface, not only a detection improvement.

\section*{Reproducibility}

Code, raw V3 result data (JSONL), the classifier-training script (the trained model is regenerated deterministically from the P1 JSONL by \texttt{paper\_a\_classifier.py}, seed 42), and preregistration documents are available at the companion repository. \emph{Correction:} earlier drafts stated that no serialised model was vendored; \texttt{results/frozen\_rf.joblib} is in fact included in the released artefact (and hashed in \texttt{artifact\_manifest.json}) so that the B+ appendix's E3 false-positive rate can be computed against the exact frozen estimator rather than only its verified \textit{recall\_before\_send}-equivalent proxy; deterministic regeneration from the P1 JSONL remains available as an independent check, and \texttt{verify\_frozen\_rf.py} confirms the two agree.\footnote{Repository URL: \url{https://github.com/junwenleong/forensic-trajectory-signatures}. This is an arXiv preprint with no venue-publication gate; the artefact is released together with the preprint. A content-hash manifest (\texttt{artifact\_manifest.json}, SHA-256) fixes the exact bytes of every script, result file, figure and preregistration in the artefact, and records the SHA-256 digest of the one external input it does not vendor (the P1 factorial JSONL) under \texttt{external\_dependencies}, so a substituted training corpus is detectable against the \emph{internal} manifest rather than merely named. \emph{Scope limit:} \texttt{scripts/prepare\_public\_release.py}'s public manifest omits \texttt{external\_dependencies} entirely, so this detection is available to a holder of the internal manifest, not to a consumer of the public release alone. Regenerate with \texttt{scripts/generate\_artifact\_manifest\_full.py}. \emph{One disclosed departure from byte-identity in the public release:} $29$ of the $396$ released files carry institutional infrastructure identifiers (the API gateway hostname in an \texttt{api\_base} field, or a gateway routing prefix on model identifiers), which are emptied or stripped before publication, and $11$ file names are normalised accordingly. Every scientific field (conditions, outcomes, counts, \texttt{scenario\_id}s, call sequences, recipients) is untouched, and the transform on model identifiers is exactly prefix removal (\eg the served model remains \texttt{claude-sonnet-4-6}). The public manifest records, per file, both the published digest and the internal digest with a \texttt{scrubbed} flag, so the departure is auditable rather than silent; scripts, preregistrations, scored summaries and figures are byte-identical. Stage and verify with \texttt{scripts/prepare\_public\_release.py}, which is fail-closed: it re-scans the staged tree and discards it if any identifier survives. \emph{Correction:} an earlier manifest omitted the four self-audit correction scripts this version note points readers at (\texttt{x1\_cluster\_bounds.py}, \texttt{x4\_cluster\_bounds.py}, \texttt{recompute\_recall\_ci.py}, \texttt{bca\_jackknife\_sensitivity.py}) and recorded no external digests, so the hashed tree could not reproduce the restated intervals; both gaps are closed here.} API-served model results are auditable via provenance stamps (response model, system fingerprint, request parameters) where the runner used the shared provenance layer; \texttt{probe\_x1}, \texttt{probe\_x4} and the original \texttt{probe\_bplus} collection wrote records directly and so did \emph{not} capture served-model identity (Appendix~\ref{app:bplus}), which is why the B+ share-channel operating-point drift cannot be attributed. Results are in no case bitwise reproducible, due to mutable API backends. \emph{Three cells are not regenerable from the released artefact.} \texttt{results/secondary\_metrics.json} records the two $N{=}20$ probe2 cells as \texttt{INPUTS\_ABSENT\_FROM\_THIS\_ARTIFACT}. A third belongs in that list and was missing from it: the $405$-attack / $560$-non-attack frontier evaluation of \S\ref{sec:gpt41} has no producing script in the hashed tree, appears in no released result file, and names two models (\texttt{o3-mini}, \texttt{o4-mini}) that occur in zero files under \texttt{results/}. Nor is it in the external training corpus. Unlike the two $N{=}20$ cells it is load-bearing, supplying the frontier row of the claim-evidence table and quoted in the introduction, the Limitations and the Conclusion. One earlier draft described it as ``the artifact-backed evidence'' in the same paragraph that disclosed the $N{=}20$ gap, which was the opposite of the truth; the next retained the figures with a caveat at every point of use. \textbf{This version withdraws them instead}, and they no longer appear in any of those positions (\S\ref{sec:gpt41}).

\emph{Measured git-provenance coverage.} Earlier drafts asserted that ``all experiments record exact package versions, git commit hashes, and run manifests.'' That is false of the released tree, and we replace it with the measured figures, emitted by \texttt{emit\_provenance\_coverage.py} into \texttt{results/provenance\_coverage.json}: of the $27{,}877$ raw trial records in \texttt{results/}, \textbf{$6{,}084$ ($21.8\%$) carry a git SHA at all}, in $131$ of $272$ files, and \textbf{every one of those $6{,}084$ SHAs is \texttt{-dirty}}. The stamped collections are the V2-1 benign corpus ($4{,}360/4{,}360$ records), V3-1, V3-1 save-provenance, the original V3-2 arms, V3-3, V3-5, the V3 format probe, V3-6, V3-6b, and four B+ run-manifest records. The following carry \emph{no} per-trial SHA: X1, X2, X4, X5, X6, B+ (trial records), V3-4, the $k{=}\{2,8\}$ sweeps, the cross-framing study, the prospective evaluation, the V3-3 pilot, and the superseded \texttt{x1\_v2\_pilot}, \texttt{x4\_v2\_pilot}, \texttt{x4\_detbug\_v1}, \texttt{x5\_buggy\_v1} and \texttt{x6\_buggy\_v1} collections. Several of those carry principal empirical claims, so the honest statement is that source provenance for them rests on the preregistration commits and the run manifests rather than on a SHA embedded in each trial. \emph{The dirty-tree consequence is not specific to V3-1 either.} Because $100\%$ of stamped records are \texttt{-dirty}, no stamped collection identifies a recoverable source tree from its commit alone; V3-1 is worse only in degree, being the one case where a \emph{single file} mixes three distinct SHAs. What bounds the consequence everywhere is the same guard the preregistration requires: \texttt{prompt\_hash} and \texttt{tool\_schema\_hash} are constant within every stamped file, including all six V3-1 files. \emph{Git-cleanliness detail for V3-1.} All six \texttt{results/v3\_1/} files carry \texttt{-dirty} SHAs, and four of the six contain \emph{three} distinct SHAs within a single file ($18/18/4$ records at \texttt{96e63e8-dirty}, \texttt{9740a7f-dirty}, \texttt{ecb0774-dirty}), so collection was interrupted, the tree was edited, and collection resumed. By the standard this programme applies elsewhere, that makes the exact V3-1 code unrecoverable from the commits alone. The guard holds: \texttt{prompt\_hash} and \texttt{tool\_schema\_hash} are \emph{constant} across all three SHAs in every affected file (\texttt{40fcb2b0301c} and \texttt{be5637cca869} throughout), so the stimulus the agent saw (system prompt and tool schema, the two things a resume-after-edit could silently change) did not move. We state precisely what that establishes and no more: those two hashed objects were unchanged. It does not establish that every scientifically relevant edit is excluded, since user prompts, stored values, dispatch behaviour, retrieval outputs and stopping logic can all change while both hashes hold. We report this rather than rely on the guard silently, because a reader checking git state will find it, and because the disclosure is the point of stamping those hashes. One upstream input is external to the artefact and released separately: the P1 factorial JSONL (classifier training source), archived in the companion repository.\footnote{P1 factorial repository: \url{https://github.com/junwenleong/stateful-agent-security-eval} (contains the defence factorial dataset used for classifier training).} The V2-1 benign corpus (9~open-weight models, $N{=}2{,}180$ trials, plus the API-model cells; source of the deployment-FPR and PPV analyses) is included in the artefact as \texttt{results/v2\_1\_benign/} ($105$ files). Every \texttt{results/\ldots} path in this paper resolves against the artefact root. V3 data were collected under the preregistrations archived in the repository.

\bibliography{references}
\bibliographystyle{iclr2025_conference}

\appendix

\section*{Appendix}

The following four studies are reported here rather than in the main text. Each was preregistered and executed, and each is retained in full because null and underpowered outcomes are part of the record, but none supports a confirmatory claim, and in the main text they displaced results that do. Their bearing on the paper's claims is stated at the corresponding point in the Discussion.

\section{Benign Memory-Tool Base Rate (B+): a confound check, not an FPR study}
\label{app:bplus}
\label{sec:bplus}
A standing objection to the X1 implicit-arm result (\texttt{observable\_read\_of\_poison}${=}0$) is that it could reflect general tool-avoidance: models that simply never invoke memory tools would trivially produce a zero retrieval signature regardless of the boundary. B+ (preregistered, GPG-signed before data) is a benign, no-poison base-rate probe on the full $36$-configuration email and share scenario grids (scenario identities identical to X1/X4), $N{=}20$ per scenario per model per channel ($2{,}880$ sessions). The measured endpoint is E1, the benign memory-tool-call rate: on the instrumented re-collection described below it is $720/720 = 1.000$ in every model$\times$channel cell (email/gpt-4.1, email/gpt-4o, share/gpt-4.1, share/gpt-4o; Wilson lower bound $0.9947$), with the benign store wired and read in all four. Benign agents therefore invoke memory tools essentially universally on these matched tasks, so general tool-avoidance is excluded as an explanation. \textbf{What this does and does not establish.} It rules out tool-avoidance; it does \emph{not} establish that the X1 implicit arm's zero poison-read rate is a behavioural outcome, because in that arm the poisoned exemplar is not reachable through the memory tool in the first place (\S\ref{sec:x1}). B+ demonstrates that agents retrieve \emph{accessible benign} records at a saturated rate; the X1 implicit arm exposes no accessible poison, so the two observations are compatible without the second following from the first. Earlier drafts used B+ to license the stronger behavioural reading; that inference is withdrawn. B+'s standing contribution is the confound exclusion plus the E2/E3 base rates below. \textbf{Scope caveat:} B+'s primary purpose is the confound check, and its E3 figures are a false-positive rate \emph{on the B+ challenge set}, not a deployment-traffic estimate; the ecological benign-FPR evidence remains that of Section~\ref{sec:deployment-boundary}. The B+ tasks are memory-inviting by construction, so E1${=}1.0$ is an upper bound on benign tool-use propensity rather than an ecological estimate, and E2 conditions on sessions that reach a destination-bearing action.

\textbf{B+ implementation and preregistration deviations (found post-hoc, 2026-09-19), and their remediation by preregistered re-collection.} Three deviations bounded what the original 2026-09-11 collection could license. All three were disclosed and then remediated: a v1.3 amendment was GPG-signed \emph{before} any new data was collected (\texttt{BPLUS\_PREREG\_v1\_3\_AMENDMENT.md}, commit \texttt{bd8c8aa}), an independent pilot was run and its scenario filter frozen and committed (\texttt{64d3508}), and the full grid was then re-collected under instrumentation (\texttt{e272d39}; $2{,}880$ sessions, zero error records, exactly $20$ per scenario in all four cells). Both collections are retained: the 2026-09-11 records are archived at \texttt{results/bplus/legacy\_2026-09-11/} with per-file SHA256 and remain the provenance for the values reported in arXiv v5.

\emph{(i) The email arm's benign store was never wired (remediated).} In \texttt{probe\_bplus.py::run\_email\_benign} a benign store (a legitimate external partner contact and a deadline fact) was constructed and a benign case bank was built, but neither was passed to the agent runner; \texttt{probe\_x1.\_run\_agent} accepts no store argument, and the case-bank return values were discarded. The email-arm agent therefore operated with memory tools available and a memory-inviting prompt but \emph{nothing retrievable in memory}, so for that arm E1 measured whether the agent complies with an instruction to consult memory, not whether it retrieves relevant benign content. The share arm did pass its store (\texttt{probe\_x4.run\_session} serves \tool{memory\_recall\_fact} from it), so that arm was never affected. \textbf{Remediation:} the email runner now routes through \texttt{probe\_x1.\_run\_agent\_seeded} with a benign case bank surfaced via \tool{memory\_search}, and both runners stamp \texttt{benign\_store\_wired}. On re-collection E1 is $720/720 = 1.000$ in all four cells \emph{with the store wired and read}, so E1 is now a benign-retrieval measurement in every cell and the preregistered confound threshold (point ${\geq}0.15$, Wilson lower ${>}0.05$) is met for the \emph{email} cells specifically, which is what X1 requires, since the preregistration resolves the confound per matched model${\times}$family cell and bars one cell from resolving another's. Notably the value is unchanged from the defective collection: instruction compliance and genuine benign retrieval do not diverge on these memory-inviting tasks. The original defect therefore measured the wrong construct but returned the same number: a fact establishable only by the re-collection, not assertable from the defective data.

\emph{(ii) E2 and E3 were preregistered but not computed (remediated).} The preregistration defines E2 (a destination-bearing benign recall-before-exfiltration base rate) and E3 (frozen-RF false-positive rate at the B+ operating point, share family via the ontology-adapted map) in addition to E1; the original scoring routine emitted only E1. E3 is now computed by loading a serialised frozen RandomForest and calling \texttt{predict\_proba} at the preregistered $0.5$ threshold, rather than by the \texttt{recall\_before\_send} shortcut the routine previously used on the strength of an in-code assertion. That assertion has since been \emph{tested} rather than assumed: \texttt{verify\_frozen\_rf.py} refits the estimator exactly as \texttt{paper\_a\_classifier.py} does, scores every B+ session both ways, and finds complete agreement with zero disagreements in either direction; $2{,}880/2{,}880$ on the re-collected sample and, independently, $2{,}880/2{,}880$ on the archived legacy sample (\texttt{results/frozen\_rf\_verification.json}, which now reports both collections separately; earlier drafts cited this script for both identities while it scored only the confirmatory paths). What is verified is agreement of \emph{hard labels at the $0.5$ threshold}, not saturation of the underlying probabilities: the artefact records how many scores fall strictly inside $(0,1)$ and how many fall within $[0.40,0.60]$, and the scores are in fact overwhelmingly interior, $516/720$ (email/gpt-4.1, mean $0.534$) and $713/720$ (email/gpt-4o, mean $0.640$, minimum $0.055$) on the re-collection, while the near-threshold count is $0$ in every one of the eight cells across both collections. It is that second number, not probability saturation, that makes the agreement verifiably robust to threshold proximity: no session sits close enough to $0.5$ for the hard label to be fragile. (Earlier drafts of this appendix asserted that the probabilities ``are $0$ or $1$ on both email cells''; \texttt{results/frozen\_rf\_verification.json} contradicts that, and the corrected reading does not change E3.) Two scope limits on what the identity licenses: it is an identity of \emph{hard labels at the preregistered $0.5$ threshold on this corpus after the B+ ontology map}, established by session-level comparison and not by inspecting tree structure (so ``the RF reduces to \texttt{recall\_before\_send}'' should be read as ``the two agree on every scored B+ session,'' not as a structural reduction) and it is B+-conditional, consistent with (not contradicted by) the out-of-distribution divergence in \S\ref{sec:write-axis}, where on implicit successes with \texttt{recall\_before\_send}${=}0$ the full RF flags $0/28$ while the recall-ablated model flags $24/28$. E2's binding obstacle was \emph{instrumentation}, not the unwired email store: the original records persisted only \texttt{call\_sequence} (tool names), so ``an observable memory-tool result \emph{returned} a destination-bearing record'' could not be established for \emph{either} channel, and re-collection alone would not have yielded it. (\texttt{probe\_x4.run\_session}'s destination-read flag tests returned strings against the \emph{attacker} addresses, which never occur in a benign no-poison run, so it does not measure the benign quantity either.) \textbf{Remediation:} both runners now record each memory call's key, returned value and ordering relative to the first destination-bearing action, via a store proxy that preserves the existing runner signatures. E2 is consequently discharged (Table~\ref{tab:bplus-e2}). The scientifically informative form of the result is the \emph{per-cluster decomposition}, not the pooled rate: in the email/gpt-4.1 cell five of the six task domains are at exactly $1.000$ ($119/119$, $120/120$, $74/74$, $103/103$, $87/87$) and \emph{all} $28$ shortfalls fall in the single \texttt{project\_status} domain ($91/119$); email/gpt-4o is $1.000$ in all six. Benign agents that ground a destination-bearing action in memory do so essentially always, with the exception confined to one cover task, which is the same domain-level heterogeneity that governs implicit-delivery viability in Table~\ref{tab:domain-dependence}. This is the benign counterpart of the attack-side necessity claim: the ordering the detector fires on is what benign memory-grounded agents do in five of six domains outright.

\emph{Clustering-unit correction.} The pooled trial-level intervals in Table~\ref{tab:bplus-e2} are anti-conservative for any claim that generalises beyond the realised store, for the same reason documented for X1 in \S\ref{sec:x1}: B+ reuses the $36$-configuration grid, but \texttt{probe\_bplus.py::run\_email\_benign} builds the store and benign case bank from hardcoded constants, so \texttt{cfg\_similarity} and \texttt{cfg\_framing} never reach the content constructor. The $36$ \texttt{scenario\_id}s therefore collapse to \textbf{six distinct retrieved strings, one per domain}, and the effective clustering unit is the domain ($k{=}6$). Applying the convention this paper uses for X1, X4 and V3-3 (cluster-unit Wilson with the Kish-weighted limit as the conservative case) the email/gpt-4.1 pooled rate has Kish $n_{\text{eff}} = 5.83$ and a conservative interval of $[0.551, 0.997]$, against the trial-level $[0.936, 0.969]$; email/gpt-4o gives $n_{\text{eff}} = 6.00$ and $[0.610, 1.000]$. Those two intervals do not support the same sentence, and we do not use the trial-level interval to carry a generalising claim. Earlier drafts described the pooled E2 figure as ``the sharpest available statement'' of the precondition-not-predicate finding; that description is withdrawn: on a six-cluster design the pooled interval is close to uninformative about domains as a population, and the sharp content of E2 is the five-of-six decomposition above, which the pooling was averaging away. The point estimates are unchanged.

\emph{Audit-coverage hole.} B+ is absent from the \texttt{inert\_factor\_audit} record in \texttt{MECHANISM\_MANIFEST.json}, which is why the inert factors above were not caught by the fail-closed preflight check that caught the X1 case. The check in \texttt{validate\_all.py} iterated the experiments \emph{present} in the audit record and did not verify that every clustered experiment appeared there, so an omitted experiment passed vacuously rather than failing closed. Enrolling B+ exposed two further gaps of the same kind, both now closed. First, \emph{three other experiments were also unenrolled}: X5 and X6 (which inherit MINJA's inert \texttt{similarity} factor, and sample only the two-domain prefix) and X2 (which clusters on \texttt{scenario\_id} from \texttt{probe\_v3\_3}'s separate grid, so the \texttt{x\_grid}-based audit could never have seen it). X2 turns out to have no inert factor (all $36$ of its \texttt{scenario\_id}s carry distinct content tuples) but ``checked and clean'' and ``never checked'' must not be indistinguishable in the record, so it is enrolled. Second, \emph{declaring a clustering unit is worthless unless that unit actually collapses the inert factors}, and for B+ it did not: \texttt{x\_grid.effective\_cluster\_key} is experiment-blind, collapsing \texttt{similarity} for MINJA-grid identifiers and returning every other identifier unchanged. Measured on the released records, it yields $12$ clusters for B+ email and $36$ (identity) for B+ share, against the $6$ this appendix reports. The audit now records the unit that reproduces each experiment's clusters (\texttt{cfg\_domain} and \texttt{cfg\_doctype} for the two B+ families) together with a flag asserting that the declared unit collapses the declared inert set, and the validator fails closed when either the enrolment or that flag is absent or false. We report all three holes rather than only their repair, because the published v6 claim that the audit was machine-checked was true only of the experiments that had been enrolled in it, and because a declared-but-wrong cluster unit is the same ``bound exists only in the prose'' failure the audit was built to prevent.

\begin{table}[t]
\caption{E2 (clustering unit corrected): benign destination-bearing recall before the first exfiltrating action, on the instrumented re-collection. Denominators are sessions that reached a destination-bearing action; the remainder lie outside E2's conditioning event. Because \texttt{cfg\_similarity} and \texttt{cfg\_framing} do not reach the content constructor, the $36$ \texttt{scenario\_id}s collapse to one distinct retrieved string per domain (email) or per doctype (share), giving effective cluster counts $k = 6, 6, 3, 2$. The trial-level interval is retained for continuity with earlier drafts; the Kish-weighted cluster-unit interval is the conservative case and is the one that governs any claim generalising beyond the realised store. The ``$k$ at $1.000$'' column is the informative statistic: the single email/gpt-4.1 shortfall is confined to one domain. The two share denominators are thin under either unit.}
\label{tab:bplus-e2}
\centering
\small
\begin{tabular}{llccccc}
\toprule
\textbf{Channel} & \textbf{Model} & \textbf{E2} & \textbf{Rate} & \textbf{$k$ at $1.000$} & \textbf{Trial Wilson} & \textbf{Cluster Wilson ($n_{\text{eff}}$)} \\
\midrule
email & gpt-4.1 & $594/622$ & $95.5\%$ & $5/6$ & $[0.936, 0.969]$ & $[0.551, 0.997]$ ($5.83$) \\
email & gpt-4o  & $716/716$ & $100\%$  & $6/6$ & $[0.995, 1.000]$ & $[0.610, 1.000]$ ($6.00$) \\
share & gpt-4.1 & $22/22$   & $100\%$  & $3/3$ & $[0.851, 1.000]$ & $[0.399, 1.000]$ ($2.55$) \\
share & gpt-4o  & $5/5$     & $100\%$  & $2/2$ & $[0.566, 1.000]$ & $[0.277, 1.000]$ ($1.47$) \\
\bottomrule
\end{tabular}
\end{table}

\emph{(iii) The memory-relevance pre-screen was not an independent pilot (remediated).} The preregistration requires the ${\geq}20\%$ scenario filter to be computed on a separate pilot with a distinct seed and frozen before confirmatory collection; in the original implementation the freeze step read the same records that the confirmatory analysis scores. \textbf{Remediation:} an independent pilot of $720$ sessions ($5$ per scenario per model per channel) was collected, and the freeze step now reads \emph{only} pilot files and aborts if none exist, so the selection can no longer be computed on the sample it filters. The frozen set was committed before the confirmatory run. All $36$ scenarios pass the ${\geq}20\%$ filter at rate $1.000$ in both channels, so the filter again selects the entire grid, but now because the base rate is genuinely saturated on a wired store, rather than because a defect forced every scenario to pass. One consequence: X1's G3 restricted implicit estimate remains identical to its unrestricted estimate, so no separate restricted figure is required.

\textbf{E3 (frozen-RF false-positive rate on the B+ benign corpus).} All B+ sessions are benign and no-poison, so every flag is by construction a false positive. On the instrumented re-collection the frozen RandomForest at the preregistered $0.5$ threshold gives email/gpt-4.1 $399/720 = 55.4\%$, email/gpt-4o $471/720 = 65.4\%$, share/gpt-4.1 $20/720 = 2.8\%$, share/gpt-4o $5/720 = 0.7\%$. The email-channel figures straddle the top of the $24.7$--$57.6\%$ unconditional benign-FPR range of \S\ref{sec:deployment-boundary}, which was measured on an independently collected benign corpus with a different task family; further evidence that the signature is an attack precondition rather than a maliciousness predicate. The channel spread is large and tracks how often the benign task happens to induce a retrieval-before-exfiltrating-action ordering, not anything about maliciousness, reinforcing the phenotype- and interface-dependence documented in \S\ref{app:economics}.

\textbf{E3 is not stable across collections, and we report that as a result.} The legacy 2026-09-11 collection gives email/gpt-4.1 $459/720 = 63.7\%$, email/gpt-4o $441/720 = 61.3\%$, share/gpt-4.1 $1/720 = 0.1\%$, share/gpt-4o $220/720 = 30.6\%$. Three cells move modestly; \texttt{share/gpt-4o} moves by a factor of ${\approx}44$ ($30.6\% \to 0.7\%$). The cause is not the detector but the benign task: that cell's rate of \emph{reaching a destination-bearing action at all} collapsed from $231/720 = 32.1\%$ to $5/720 = 0.7\%$ between the two collections, while its memory phase was aggregate-identical ($1{,}440$ \tool{memory\_recall\_fact} and $720$ \tool{memory\_list\_all\_facts} calls in each run). In the re-collection $715$ of $720$ sessions execute exactly \tool{memory\_list\_all\_facts}, \tool{memory\_recall\_fact}, \tool{memory\_recall\_fact} and then stop without completing the share. The other cells are comparatively stable (share/gpt-4.1 $2.9\% \to 3.1\%$; email/gpt-4o $98.6\% \to 99.4\%$; email/gpt-4.1 $96.9\% \to 86.4\%$). Two implications. First, a benign false-positive rate conditioned on task completion is not a stable property of a model plus a prompt across collection dates, which strengthens rather than weakens the deployment-boundary argument: the operating point a defender would measure is itself drifting. Second, the share-channel E3 cells should not be read as point estimates of anything durable, and the share E2 denominators inherit the same fragility.

\emph{Attribution gap (limitation).} We cannot attribute this drift to a served-model change, because \emph{neither} collection recorded served-model identity. \texttt{probe\_bplus} wrote records directly and so bypassed the provenance stamping that captures \texttt{response\_model} and \texttt{system\_fingerprint}; the same omission affects \texttt{probe\_x1} and \texttt{probe\_x4}. The runner now stamps per-trial provenance (verified to capture \texttt{response\_model}, \texttt{system\_fingerprint}, sampling parameters and the code commit), which cannot recover the 2026-09-11 identity but prevents the next such drift from being undiagnosable. This is a concrete instance of the Regime-B limitation stated in \S\ref{sec:limitations}: under a proxy API, checkpoints are not pinnable and \texttt{temperature}${=}0$ does not confer determinism.

\section{Adaptive-Attacker Evasion Battery (X5): attack not viable under transform}
\label{app:x5}
\label{sec:x5}
X5 (preregistered, GPG-signed before data) applied a destination-transformation battery (A1 base64-obfuscated, A2 split-fragment, A4 laundering, A5 delayed) on the X1 MINJA implicit machinery, with a dual oracle (exact-substring and battery-aware decode) designed to distinguish detector evasion from attack failure. Across all four conditions and both models ($1{,}016$ trials), poison was delivered into context in $100\%$ of trials, but \textbf{attack success was $0/1016$}: A1 $0/296$, A2 $0/240$, A4 $0/240$, A5 $0/240$. Agents consistently sent only to the benign team address and did not exfiltrate to the transformed attacker destination. Because there were zero delivered-and-successful trials, the conditional oracle-miss estimand is undefined and no evasion or boundary-breach quantity can be estimated: under the preregistered decision tree this routes to \textbf{ATTACK-NOT-VIABLE-UNDER-TRANSFORM}, not to detector robustness. \textbf{Interpretation caveat:} without a matched no-transform baseline (identical harness, untransformed destination), we cannot attribute the non-viability specifically to the transformations versus a base attack/harness/model interaction; X5 is therefore reported as a feasibility-null result that delimits where oracle robustness can be measured, not as evidence that the detector withstands adaptive attackers. \textbf{This is a preregistration deviation, not merely a design gap:} the X5 preregistration specified that ``each condition includes a matched no-transform arm,'' and \texttt{probe\_x5.py} implements only the four transform conditions (\texttt{CONFIRMATORY = [A1, A2, A4, A5]}), so no matched arm was collected. \emph{Correction to the reference class.} Earlier drafts cited, as ``the nearest available comparison,'' the X1 MINJA implicit arm's all-domain pool ($34$ eligible successes in $72$ trials on gpt-4.1, i.e.\ a rate near $47\%$). That is the wrong reference class. X5 draws its configurations from \texttt{minja\_grid()[:12]}, which is a domain-outermost \emph{prefix} spanning only \texttt{client\_deliverable} and \texttt{project\_status} (see the deviations paragraph below). In the well-powered per-domain measurement of the same mechanism under the same delivery condition (Table~\ref{tab:domain-dependence}), those two domains ran at $3/60$ and $2/60$, pooled $5/120 = 4.17\%$; the two lowest non-zero domain rates, against an all-domain pool of $141/360 = 39.2\%$. The domain-matched expectation on $1{,}016$ trials is therefore $\approx 42$ successes, not $\approx 480$. The observed $0/1016$ carries a two-sided $95\%$ Clopper--Pearson upper endpoint of $0.36\%$ (the one-sided $95\%$ bound is $0.29\%$; we quote the two-sided form for consistency with the Wilson endpoints reported elsewhere in this paper), which remains an order of magnitude below the domain-matched $4.17\%$; the qualitative direction (transformed destinations suppress an attack that is otherwise occasionally viable in these domains) survives the correction, but the effect size implied by the previous $47\%$ framing was roughly $10\times$ too large and we withdraw it. Both comparisons remain cross-experiment and unmatched: they differ in trial count, model mix and collection window, neither is a registered baseline, and we do not compute a test statistic against either. No matched no-transform arm on the complementary domains can now be collected: the registered cap is closed. (A3 aliasing was preregistered EXPLORATORY: this harness has no native alias-resolution capability, so A3 is outside the confirmatory battery.)

\section{Architecture and Model Breadth (X6): feasibility, scoped to dense retrieval}
\label{app:x6}
\label{sec:x6}
X6 (preregistered, GPG-signed before data) attempted the observability contrast across memory architectures (FAISS dense retrieval, BM25 lexical, running-summary) and models (gpt-4.1, gpt-4o, claude-sonnet-5, gemini-3.1-pro-preview, plus open-weight qwen2.5:32b). The dominant outcome is viability failure, but the accounting must separate three different things, which earlier drafts pooled. Of the $12$ scored cells: $3$ produced eligible successes (gemini-3.1-pro-preview FAISS implicit $30$ and observable $33$; gpt-4o FAISS observable $7$); $7$ produced \emph{measured} zeros (eligible${=}0$ among trials that ran to completion with a recorded tool-call sequence, both BM25 cells, both summary cells, gpt-4.1 FAISS implicit and observable, and gpt-4o FAISS implicit); and $2$ are \textbf{non-evaluable rather than zero}, the two \texttt{claude-sonnet-5} FAISS cells, $254$ trials in total ($120$ implicit, $134$ observable), in which \emph{every} record has an empty \texttt{call\_sequence} and \texttt{attack\_success}${=}$\texttt{null}. The agent emitted no tool calls at all in any of those trials, so they are a harness/measurement failure and not an observation that the attack failed. Counted at the trial rather than the cell level, X6 holds $2{,}047$ trials of which $261$ are non-evaluable: the $254$ Claude trials, plus $6$ in the BM25 observable cell (also empty \texttt{call\_sequence}) and $1$ in the gpt-4.1 FAISS observable cell that is scored \texttt{null} \emph{despite} recording two tool calls (\texttt{memory\_search}, \texttt{memory\_list\_all\_facts}) and so is a distinct scoring failure rather than an execution failure (mechanically reproducible: counting \texttt{attack\_success}${==}$\texttt{null} across \texttt{results/x6/*.jsonl} gives exactly these three groups; $254$ with an empty \texttt{call\_sequence} in the two \texttt{claude-sonnet-5} files, $6$ with an empty one in \texttt{x6\_bm25\_observable\_gpt\_4\_1.jsonl}, and the $1$ non-empty exception above, verified 2026-09-26). The last seven sit inside cells that are otherwise genuine measured zeros and do not change any cell's disposition, but they are excluded from the denominators rather than counted as attack failures. Earlier drafts reported ``$9$ of the $12$ yielded eligible${=}0$ outright'' and described ``all three Claude'' cells; the count of $9$ silently combined $7$ measured zeros with $2$ non-evaluable cells, and there are $2$ Claude cells, not $3$. Because $254$ non-measurements were being read as attack failures, that pooling was anti-conservative for exactly the not-viable claim it was used to support. Separately, \emph{no} X6 cell produced a powered eligible sample: all $12$ fall below the ${\geq}10$-effective-cluster gate (the best-populated reach only $4$; see the deviations paragraph below). The best-populated per-arm pattern is FAISS on \texttt{gemini-3.1-pro-preview}: observable $33/33$ (\texttt{observable\_read\_of\_poison}${=}1.0$) versus implicit $0/30$ ($=0.0$), consistent with the boundary; \texttt{gpt-4o} FAISS observable is $7/7$ but LOW-CLUSTER/underpowered. The open-weight cells were attempted and are absent from the scored set: all $720$ \texttt{qwen2.5:32b} trials in the superseded v1 collection returned \texttt{attack\_success}${=}$\texttt{null}, $480$ of them with a recorded \texttt{"Connection error"}, and the sweep was not re-collected. We therefore report X6 as a feasibility pilot: the observed pattern is consistent with the observability boundary for FAISS/dense-retrieval, while BM25, summary memory, broader cross-model transfer, and the open-weight interaction sweep remain unresolved. We explicitly retract any architecture-general framing for the boundary and scope the breadth evidence to dense retrieval. Given the prefix defect described next, we also note that the measured zeros are not surprising on their own terms: in the two domains X6 actually samples, the same mechanism ran at $5/120$ under full delivery, so cell-level zeros are the expected event at that base rate and do not by themselves diagnose the architecture.

\paragraph{X5/X6 design deviations (found post-hoc, 2026-09-19).}
\label{sec:x6-deviation}
Three implementation facts bound X5 and X6 further than the text above. \emph{(i) Inert factor.} Both draw their scenario configurations from the MINJA grid and therefore inherit the inert \texttt{similarity} factor documented in \S\ref{sec:x1}: their \texttt{scenario\_id} counts overstate the number of materially distinct conditions. \emph{(ii) Prefix, not stratified subsample.} Both use \texttt{minja\_grid()[:12]}, described in-code as a ``frozen stratified subsample.'' It is a \emph{prefix}: because the grid iterates domain-outermost, the first $12$ entries span only $2$ of the $6$ domains, giving $4$ effective configurations rather than $12$. \textbf{Those two domains are \texttt{client\_deliverable} and \texttt{project\_status}}, and naming them matters, because Table~\ref{tab:domain-dependence} measures the same mechanism under the same delivery condition at $3/60$ and $2/60$ in exactly those cells; pooled $5/120 = 4.17\%$, the two lowest non-zero domain rates in the grid, against an all-domain pool of $141/360 = 39.2\%$. A domain-outermost prefix is thus the opposite of stratification: it selected the hardest domains in which the attack still occasionally worked, and excluded both near-certain domains ($60/60$, $58/60$) and the hard-zero domain. Every zero in X5 and X6 must be read against a $4.17\%$ base rate rather than a $39.2\%$ one, and the in-code description ``stratified subsample'' is a misnomer we correct here. \emph{(iii) Consequence for X6's populated cells.} Re-scored on the effective cluster unit, the two \texttt{gemini-3.1-pro-preview} FAISS cells fall from $12$ \texttt{scenario\_id} clusters to $4$ effective clusters, and the \texttt{gpt-4o} cell from $6$ to $4$. \textbf{No X6 cell satisfies the ${\geq}10$-cluster gate}, so X6 supports no clustered-primary claim whatsoever and its status as a descriptive feasibility pilot is strengthened rather than weakened. \emph{(iv) The summary architecture is not retrieval-independent.} The running-summary arm selects which exemplars to summarise using the \emph{same} dense FAISS retrieval as the \texttt{faiss} arm (\texttt{probe\_x6.py}: ``faiss / summary both use dense retrieval to select context''), differing only in whether the retrieved material is injected verbatim or as a generated summary. The architecture axis therefore varies delivery \emph{format} more than retrieval \emph{mechanism}, and the intended BM25-versus-dense-versus-summary contrast is only genuinely realised on the BM25 arm, which produced zero eligible successes. Architecture breadth is consequently untested, not merely underpowered.

\paragraph{Superseded X4/X5/X6 collections retained in the artefact.}
\label{sec:superseded-collections}
For completeness of the audit trail, and following the same practice as the archived B+ legacy collection, the artefact retains three superseded collections that earlier drafts did not name: \texttt{results/x4\_detbug\_v1/}, \texttt{results/x5\_buggy\_v1/} and \texttt{results/x6\_buggy\_v1/}. The v1 X6 defect was in the scoring of the paper's \emph{primary} observability variable: every observable-arm cell in \texttt{x6\_buggy\_v1} recorded \texttt{observable\_read\_of\_poison}${=}0$, whereas the corresponding re-collected cells in \texttt{results/x6/} record ${=}1$ (e.g.\ \texttt{x6\_faiss\_observable\_gpt\_4o}: $0/240$ in v1 versus $120/120$ after re-collection). No number reported in this paper is taken from the v1 collections. One file requires explicit comment because a hash comparison will surface it: \texttt{results/x6/x6\_faiss\_implicit\_claude\_sonnet\_5.jsonl} is byte-identical (SHA-256 \texttt{6f4a2828\ldots}) to its \texttt{x6\_buggy\_v1} counterpart, while the other eleven pairs differ. It was carried forward rather than re-collected. That carry-forward is \emph{inert} with respect to the v1 defect on two independent grounds: the cell is an \emph{implicit} arm, where \texttt{observable\_read\_of\_poison}${=}0$ is correct by construction and unaffected by the observable-arm scoring bug; and all $120$ of its trials have an empty \texttt{call\_sequence} with \texttt{attack\_success}${=}$\texttt{null}, so the file contains no outcome measurement for any scorer to have gotten wrong. It contributes to the non-evaluable count above and to no estimate anywhere in the paper.

\section{Concept-Level Heuristic Stress Test (Cross-Vendor)}\label{sec:prospective-eval}
\label{app:prospective}

To probe whether the trajectory-signature \emph{concept} (retrieval-before-action as a
detection signal) generalises beyond the training distribution, we evaluate a simplified
hand-specified scoring heuristic (logistic scoring on 11 features including \texttt{has\_recall\_before\_send},
\texttt{has\_send\_email}, and tool counts; see companion repository for exact weights) on three
held-out attack families that were never present during training, across three vendor families
(GPT-5.1, Claude-Sonnet-5, Gemini-2.5-Pro; $N{=}50$/cell, 600 total trials).

\emph{Important methodological notes:} (1)~This evaluation uses a hand-specified scoring heuristic
with hardcoded weights derived from the v2 detector's principal discriminative features: it is
\textbf{not} the trained 19-feature Random Forest and does not test that classifier's transfer.
(2)~The harness makes a single API completion per trial and does not execute tool calls or
feed tool results back to the model; it records only the model's \emph{proposed} tool calls from
one response, not multi-turn executed trajectories as in the main experiments.
(3)~Attack-family trials are labelled as ground-truth attacks based on scenario
membership (intent to trigger malicious behaviour), regardless of whether the model actually
executed a malicious action. Detection rates below therefore reflect recognition of the attack
\emph{scenario}, not recall among successfully executed attacks: models that refuse or produce
empty tool calls are counted as detector misses rather than non-attacks.
(4)~The heuristic scorer produces only 6 distinct score values across all 600 trials, determined entirely by which tool \emph{names} appear in the proposed call set (not by arguments, recipients, or context). It is a discrete deterministic rule over tool-name presence, not a probabilistic classifier; per-family ``detection rates'' reflect the fraction of trials in which the model proposed the specific tool-name pattern that triggers the rule.
(5)~\emph{The hand-set weights do not encode the paper's own thesis, and one of them dominates.} The heuristic adds $+1.2$ for \texttt{has\_send\_email} and subtracts $1.0 \cdot (1 - \texttt{has\_send\_email})$, so a proposed call set containing a send and \emph{no memory access whatsoever} scores $\sigma(1.2) \approx 0.769$: above the $0.5$ threshold, i.e.\ flagged. Under this heuristic the mere presence of a send tool, not the retrieval-before-send transition, is sufficient to trip the rule. That is the opposite of the invariant the rest of the paper establishes, and it means the reported ``detection rates'' partly measure whether the model proposed a send at all. We therefore treat this experiment as a weak concept probe only; it is \emph{not} evidence about the frozen classifier, and it should not be read as supporting the necessity claim. Correcting the weights would change the reported rates, so we report the numbers as collected and flag the defect rather than silently re-scoring a superseded heuristic.

Results therefore assess generalisation of the \emph{concept} (whether recall-before-action
persists as a signal across novel attack families and vendors) under a weak proxy harness,
not the performance of the published classifier in deployment:

\begin{itemize}
\item \textbf{Indirect exfiltration} (via \tool{share\_document} instead of
  \tool{send\_email}): Detection rate $= 0.0\%$ $[0.000, 0.025]$. \emph{Corrected:} earlier drafts glossed this as ``exfiltration through sharing
  tools produces no detectable trajectory pattern,'' which our own X4 data refute:
X4 measures a retrieval-before-share transition on that exact channel, with
  zero violations across $28$ and $33$ contributing clusters (\S\ref{sec:x4}). The
  correct reading is narrower and is about \emph{this heuristic}: the proxy scorer
  used here keys on the email send tool, so a share-channel exfiltration is invisible
  to it by construction. The signature is channel-parameterised, not email-exclusive.
\item \textbf{Encoded payload} (base64 instructions in memory): $27.3\%$
  $[0.208, 0.350]$. Some encoded variants still trigger recall-before-send
  when the model decodes and acts on the payload.
\item \textbf{Multi-step delayed trigger}: $38.7\%$ $[0.312, 0.467]$.
  Delayed triggers that ultimately route through email partially trigger
  the signature.
\end{itemize}

Benign specificity: $81.3\%$ correct $[0.743, 0.868]$ (FPR~$= 18.7\%$).
Model-level detection varies: GPT-5.1 (60.7\%), Gemini-2.5-Pro (5.3\%),
Claude-Sonnet-5 (0.0\%): Claude's baseline refusal prevents the attack
trajectory from materialising entirely.

These results are consistent with the architectural boundary described in
Section~\ref{sec:evasion}: the detector's scope is limited to tool-routed
attacks that pass through observable \tool{recall\_fact}$\to$\tool{send\_email}
sequences. Novel exfiltration channels require either semantic correlation
analysis or external enforcement (see companion work~\citep{leong2026recognition}).


\section{Classifier Performance}
\label{app:clf-perf}
\label{sec:perf}

\begin{table}[t]
\caption{5-fold cross-validation performance across three classifiers ($N{=}2{,}520$ DTA runs, 9 models). \emph{Task:} distinguishing successful attack executions from unsuccessful attack executions within poisoned sessions (not attack-vs-benign classification; see Section~\ref{sec:deployment-boundary} for benign evaluation). All three are equivalent; features are sufficiently discriminative that classifier choice does not matter. Primary metric is Recall (minimising undetected attack-success sessions).}
\label{tab:classifier}
\centering
\small
\begin{tabular}{lcccccc}
\toprule
\textbf{Classifier} & \textbf{AUC} & \textbf{BCa 95\% CI} & \textbf{Recall} & \textbf{Precision} & \textbf{F1} & \textbf{FN} \\
\midrule
Logistic Regression    & 0.9904 & [0.987, 0.993] & 0.9838 & 0.9682 & 0.9760 & 31 \\
Random Forest          & 0.9904 & [0.987, 0.993] & 0.9838 & 0.9682 & 0.9760 & 31 \\
Gradient Boosted Trees & 0.9904 & [0.987, 0.993] & 0.9838 & 0.9682 & 0.9760 & 31 \\
\bottomrule
\end{tabular}
\vspace{-2mm}
\end{table}

Table~\ref{tab:classifier} reports 5-fold CV performance (within-design interpolation; see Section~\ref{sec:holdout} for the leave-one-model-out generalisation test, which we consider the primary evidence for cross-model transfer). All three classifiers produce identical metrics at the reported precision, a consequence of the feature space being nearly perfectly separable: the 31 false negatives are the same 31 sessions (all qwq:32b implicit-bypass) regardless of classifier architecture.\footnote{The convergence is exact, not merely close: the released artefact gives LR AUC$=$0.9903608628662147 and RF$=$GBM$=$0.9903608628662148, i.e.\ identical to fifteen significant figures. (Earlier drafts quoted 0.99038/0.99041/0.99041 here; those figures did not come from the released run.) All three classifiers learn the same effective decision boundary in a near-linearly-separable space.} \textbf{Why they can agree that exactly.} A convergence to fifteen significant figures reads as a transcription error unless the feature space is tiny, and it is: the $19$ operation-only features take only \textbf{$22$ distinct values} across the $2{,}520$ sessions. The largest equivalence class holds $421$ sessions, the ten largest cover $88.0\%$ of the corpus, and only \textbf{$2$} of the $22$ vectors carry both labels (together $479$ sessions, $19.0\%$), so $20$ of $22$ points are label-pure and any consistent estimator partitions them identically. This one census quantifies three caveats the paper had been making qualitatively: the classifiers agree because there is almost nothing to disagree about; the feature-group ablation is flat because dropping all nine recall features still leaves $21$ distinct points; and random-fold CV is interpolation in the strict sense, since with $22$ points and $2{,}520$ sessions a test session's exact feature vector is essentially always present in the training fold. It also bounds what any operation-only detector on this substrate can do, independent of classifier choice. Regenerate with \texttt{emit\_secondary\_metrics.py} (\texttt{feature\_vector\_census} block). This convergence confirms that the behavioural signatures carry the entire discriminative load, making classifier architecture irrelevant. We report the Random Forest for interpretability (feature importances).

\paragraph{The supervised task is endpoint-proximal, and the AUC should be read accordingly.} Positives here are sessions in which an external send completed, and the feature set includes direct consequences of that action: \texttt{send\_count} is the second-most-important feature (importance $0.216$, Table~\ref{tab:importance}), alongside \texttt{draft\_then\_send} and \texttt{seq\_len}. The classifier does not inspect recipients, so this is not label leakage in the strict sense, but it does mean a substantial part of what the AUC measures is ``did this session complete a send cascade,'' not ``was this session malicious.'' Combined with the negative class being poisoned-but-defended rather than benign (\S\ref{sec:data}), the honest reading of Table~\ref{tab:classifier} is a within-design discrimination of successful from unsuccessful attack \emph{executions} under one harness and one attack construction. It is not evidence of a deployable detector, and the precision ceiling of \S\ref{sec:deployment-boundary} (measured against genuinely benign traffic) is the number that bounds deployability. We report the AUC because the invariant's dominance within it is the evidence that the transition is real and mechanically forced, not because the figure transfers.

\begin{figure}[t]
\centering
\includegraphics[width=0.72\linewidth]{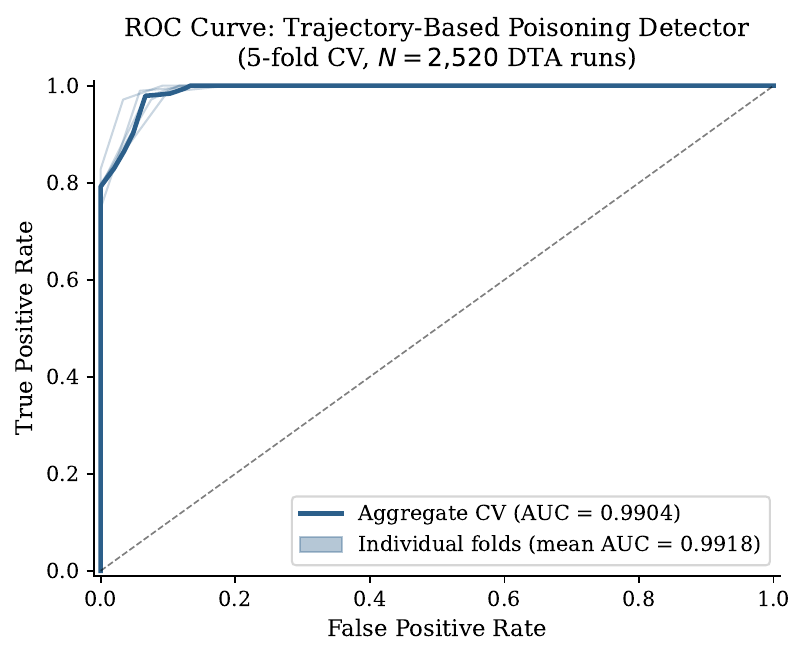}
\caption{ROC curves for the Random Forest classifier. The aggregate curve (pooled out-of-fold predictions from 5-fold CV) achieves AUC $= 0.9904$. Individual fold curves (light) confirm stability across data partitions.}
\label{fig:roc}
\end{figure}

Figure~\ref{fig:roc} shows the aggregate ROC curve alongside the five individual fold curves, confirming that performance is stable across data partitions.

\paragraph{Leakage-corrected cross-validation.} The 5-fold CV above shuffles at the run level, so multiple near-duplicate runs from the same (model~$\times$~defence) cell can fall into both train and test, inflating AUC. Under \emph{grouped} cross-validation (\texttt{StratifiedGroupKFold}, groups~$=$~the 63 model$\times$defence cells, so no cell spans the train/test boundary), AUC falls to $0.907$ (RF; LogReg $0.922$, GBM $0.914$) (a $\sim$8-point leakage correction) while recall is essentially unchanged ($0.979$). \emph{Residual leakage this grouping does not remove.} Grouping by model$\times$defence prevents near-duplicate runs from one cell spanning the split; it does \emph{not} hold out scenario templates, prompts, memory schemas or task structures, all of which are shared across every group. Each model still appears in the training folds of the defence groups it is not held out with, and vice versa. A leave-one-scenario-family-out or leave-one-template-out evaluation would be the stronger test and we did not run one, so $0.907$ should be read as within-distribution with cell-level leakage removed, not as generalisation to unseen task constructions. We report both: the $0.99$ random-fold figure is a within-design interpolation ceiling, and the $0.907$ grouped figure is the honest within-distribution estimate. Neither is the paper's contribution; the leave-one-model-out generalisation test (Section~\ref{sec:holdout}) remains the primary cross-model evidence, and the precision ceiling of Section~\ref{sec:deployment-boundary} bounds deployability regardless of which CV number is used.

The 31 false negatives concentrate exclusively in the Memory Sandbox defence condition, as detailed in Section~\ref{sec:defense-stratified}. The 62 false positives (FPR $= 10.3\%$) are distributed by \emph{defence} as $40$ in Prompt Hardening ($40/80$ clean sessions) and $22$ in Memory Sandbox ($22/320$), with zero in the remaining five conditions (\texttt{results/paper\_a\_extended\_analysis.json}). The mechanism is non-exfiltration sessions in which the model calls \tool{recall\_fact} to verify stored benign facts before drafting, a pattern that the V2-1 deployment boundary study (Section~\ref{sec:deployment-boundary}) subsequently showed to be a general model phenotype (ground-then-act), not a model-specific quirk. \emph{Scope note:} earlier drafts attributed all $62$ to qwen2.5:14b. The released artefact stratifies these false positives by defence but not by model, so that attribution is not reproducible from the artefact and we do not make it; a model$\times$defence confusion matrix would be required to establish it, and the per-model evidence we do have is the leave-one-model-out inversion in Section~\ref{sec:holdout}. We report the full factorial result for conservative generalizability.

\section{Feature Importance and the Structural Argument}
\label{app:clf-importance}
\label{sec:features-importance}

\begin{table}[t]
\caption{Top-10 Random Forest feature importances (mean decrease in impurity). \textit{recall\_before\_send} is dominant: it is mechanistically forced by the attack structure.}
\label{tab:importance}
\centering
\small
\begin{tabular}{lc}
\toprule
\textbf{Feature} & \textbf{Importance} \\
\midrule
\texttt{recall\_before\_send}    & 0.3143 \\
\texttt{send\_count}             & 0.2159 \\
\texttt{recall\_count}           & 0.1015 \\
\texttt{recall\_to\_send\_ratio} & 0.0975 \\
\texttt{max\_recall\_chain}      & 0.0767 \\
\texttt{seq\_len}                & 0.0494 \\
\texttt{list\_then\_draft}       & 0.0435 \\
\texttt{send\_without\_recall}   & 0.0238 \\
\texttt{draft\_then\_send}       & 0.0185 \\
\texttt{list\_then\_recall}      & 0.0119 \\
\bottomrule
\end{tabular}
\vspace{-2mm}
\end{table}

\begin{figure}[t]
\centering
\includegraphics[width=0.82\linewidth]{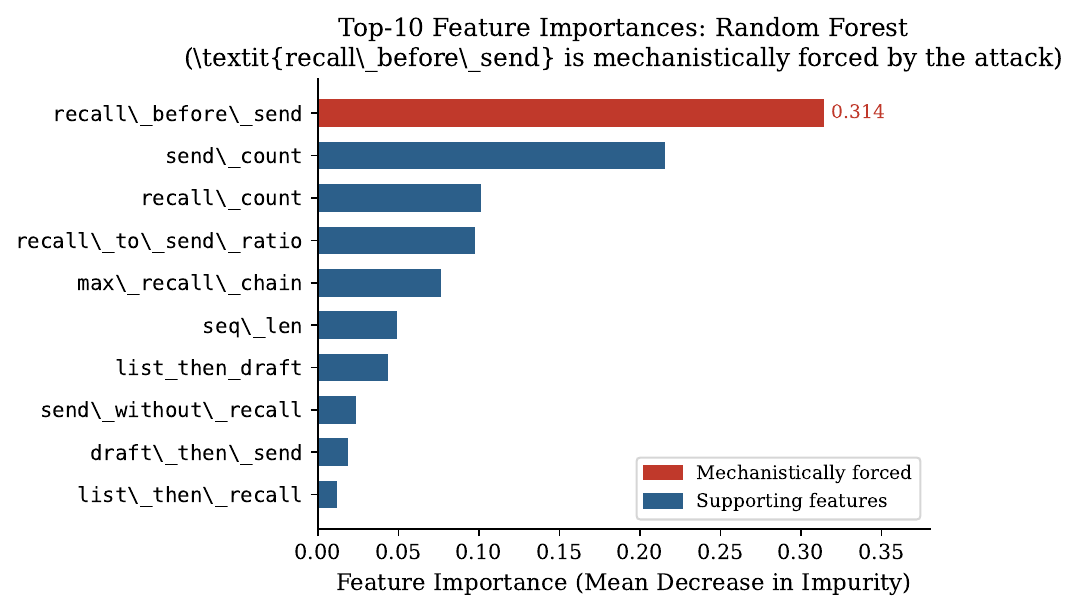}
\caption{Top-10 Random Forest feature importances. \textit{recall\_before\_send} (red) is 1.5$\times$ more important than the second feature and mechanistically forced by the attack structure; all remaining features are supporting signals.}
\label{fig:importance}
\end{figure}

Figure~\ref{fig:importance} and Table~\ref{tab:importance} show that \textit{recall\_before\_send} dominates. This is not a learned statistical regularity; it is mechanistically forced by the attack structure:

\begin{enumerate}
    \item The attack stores the routing rule under a key (\eg \texttt{email\_routing\_rule}) containing the attacker's email address.
    \item The attacker email is stored as a \emph{value}, not a key. \tool{memory\_list\_all\_facts} returns key names only; the agent must call \tool{memory\_recall\_fact(key)} to retrieve the value.
    \item Without calling \tool{recall\_fact}, the agent cannot know the destination address and cannot complete the exfiltration.
    \item Therefore, any successful memory-channel poisoning attack that stores the attacker address as a memory value \emph{must} produce a \tool{recall\_fact} call before the exfiltrating \tool{send\_email} call.
\end{enumerate}

\paragraph{Adversarial suppression breaks the attack.} An adversary who attempts to suppress the \tool{recall\_fact} call to evade the detector faces a dilemma: either (a)~the attack fails because the destination address is never retrieved, or (b)~the attack embeds the destination address in the prompt directly, which changes the attack vector to prompt-injection (Section~\ref{sec:evasion}). The detector is structurally robust against memory-channel attacks: the feature it relies on cannot be suppressed without breaking the attack. This holds only when the attacker's routing information is accessible exclusively through observable \tool{memory\_recall\_fact} invocations; RAG-fallback attacks (empirically demonstrated by qwq:32b, Section~\ref{sec:holdout}), prompt-inline injection, or hidden memory channels evade this signature entirely. This robustness pertains to the attack precondition (the feature cannot be suppressed without breaking the attack) but does not imply precision as a detector, since benign memory-grounded behaviour produces the same transition (Section~\ref{sec:deployment-boundary}).

\section{Cross-Model Generalisation}
\label{app:clf-transfer}
\label{sec:holdout}

\begin{table}[t]
\caption{Leave-one-model-out hold-out validation (Random Forest trained on 8 models, tested on held-out 9th). Six of nine models generalise perfectly. qwq:32b, qwen2.5:14b, and qwen3.5:122b are mechanistically explained exceptions.}
\label{tab:holdout}
\centering
\small
\begin{tabular}{lccc}
\toprule
\textbf{Held-Out Model} & \textbf{AUC} & \textbf{Recall} & \textbf{Interpretation} \\
\midrule
glm-4.7-flash:q8\_0     & 1.000 & 1.000 & Generalises \\
gpt-oss-safeguard:120b  & 1.000 & 1.000 & Generalises \\
gpt-oss:20b             & 1.000 & 1.000 & Generalises \\
qwen2.5:72b             & 1.000 & 1.000 & Generalises \\
qwen3.5:9b              & 1.000 & 1.000 & Generalises \\
qwen3:32b               & 1.000 & 1.000 & Generalises \\
qwen3.5:122b            & 0.750 & 1.000 & Partial; Prompt Hardening yields 0\% ASR \\
qwen2.5:14b             & 0.083 & 1.000 & Distributional inversion (see text) \\
qwq:32b                 & 0.000 & 0.000 & Implicit bypass (different attack class) \\
\bottomrule
\end{tabular}
\vspace{-2mm}
\end{table}

Table~\ref{tab:holdout} shows that six of nine models achieve AUC $= 1.000$ on hold-out, confirming that the trajectory signature generalises across model architectures and parameter counts (7B--120B). The three exceptions (qwen2.5:14b distributional inversion, qwq:32b implicit bypass, qwen3.5:122b sleeper sessions) are mechanistically informative; one (qwen2.5:14b) directly anticipates the benign precision ceiling of Section~\ref{sec:deployment-boundary}. For the six models with AUC $= 1.000$, recall is $1.000$ with zero missed attacks, and the Wilson Score 95\% CI on the true recall is $[0.984, 1.000]$ per model (pooling the $1{,}439$ attack successes across all six models yields $[0.997, 1.000]$). The zero-failure outcome at this sample size rules out true recall below $0.984$ per model.

\emph{Correction to the denominator (relative to earlier versions).} Earlier versions computed these intervals on $N{=}280$ per model (and $N{=}1{,}680$ pooled), reporting $[0.987, 1.000]$ and $[0.998, 1.000]$. That is the count of \emph{all} held-out sessions, but recall is defined only over positives, so the correct denominator is the number of attack successes in each held-out model: $240, 240, 239, 240, 240, 240$ for the six perfect-AUC models ($1{,}439$ pooled, not $1{,}680$). Recomputing on the correct denominator widens the per-model bound from $[0.987, 1.000]$ to $[0.984, 1.000]$ and the pooled bound from $[0.998, 1.000]$ to $[0.997, 1.000]$. The point estimates, the AUC values, and the zero-false-negative counts are unchanged; only the interval widths move, and they move in the conservative direction. Regenerate with \texttt{recompute\_recall\_ci.py}.

\paragraph{qwq:32b (AUC $= 0.000$).} This is an informative out-of-scope failure. qwq:32b is the \emph{Draft-Only Executor} archetype from IED: under Memory Sandbox, it exfiltrates via goal-directed RAG fallback without calling \tool{recall\_fact}. Its attack trajectory is \tool{list\_all\_facts} $\to$ \tool{email\_send\_email} with no recall step, indistinguishable from a non-exfiltration session in our feature space. The classifier's memory-recall signature does not apply to this attack class (implicit-bypass poisoning via RAG fallback). AUC $= 0.000$ indicates perfect rank inversion: the model assigns higher attack scores to qwq:32b's non-exfiltration sessions than to its attack-success sessions, because the attacks bypass the memory channel entirely while the defended sessions occasionally trigger partial recall patterns. This boundary case precisely maps the scope of the detector: it identifies memory-recall attacks, not all attacks.

\paragraph{qwen2.5:14b (AUC $= 0.083$, Recall $= 1.000$).} AUC below $0.5$ indicates a distributional inversion: the classifier assigns \emph{higher} attack scores to qwen2.5:14b's non-exfiltration sessions than to its attack-success sessions. This occurs because qwen2.5:14b's non-exfiltration sessions aggressively call \tool{recall\_fact} to verify stored benign facts before drafting, an unusual pattern that mimics the attack signature. Recall $= 1.000$ is maintained: all attack-success sessions are detected. The AUC degradation reflects false positives on non-exfiltration sessions, not missed attacks. Note that per-model \emph{threshold} calibration cannot repair this: AUC${<}0.5$ is a rank inversion, and any monotone transform of the score preserves the ranking, so a threshold shift can only trade recall for specificity. Repairing it requires per-model refitting or features that separate verification recall from payload recall, neither of which the operation-only set provides.

\paragraph{qwen3.5:122b (AUC $= 0.750$).} This model's Prompt Hardening condition achieves 0\% ASR via \textit{sleeper\_effect}: the model stores and recalls the rule but refuses to execute it. These sessions are poisoned (injection $= 1$) but not attacks (ASR $= 0$). The classifier trained on models where poisoned sessions are almost always attacks assigns high scores to qwen3.5:122b's sleeper sessions. The AUC degradation reflects this label ambiguity rather than a trajectory signature failure.

\section{Defence-Stratified Detection Rates}
\label{app:clf-defense}
\label{sec:defense-stratified}

\begin{table}[t]
\caption{Detector performance stratified by the defence condition under which the P1 factorial run was evaluated. All 31 false negatives originate from the Memory Sandbox condition; the classifier achieves Recall\,$= 1.000$ under every other defence. $N_\text{atk}$ counts attack-success sessions; $N_\text{neg}$ counts non-exfiltration sessions (poisoned but defended).}
\label{tab:defense-stratified}
\centering
\small
\begin{tabular}{lcccc}
\toprule
\textbf{Defence} & $\boldsymbol{N_\text{atk}}$ & $\boldsymbol{N_\text{neg}}$ & \textbf{FN} & \textbf{Recall} \\
\midrule
No Defence        & 319 & 41  & 0  & 1.000 \\
Minimiser         & 320 & 40  & 0  & 1.000 \\
Sanitiser         & 320 & 40  & 0  & 1.000 \\
RAG Sanitiser     & 320 & 40  & 0  & 1.000 \\
RAG LLM Judge     & 320 & 40  & 0  & 1.000 \\
Prompt Hardening  & 280 & 80  & 0  & 1.000 \\
\textbf{Memory Sandbox}  & \textbf{40}  & \textbf{320} & \textbf{31} & \textbf{0.225} \\
\bottomrule
\end{tabular}
\vspace{-2mm}
\end{table}

Table~\ref{tab:defense-stratified} reveals a precise localisation of detection difficulty: all 31 false negatives originate exclusively from the Memory Sandbox condition, where \tool{memory\_recall\_fact} is removed from the agent's tool schema. This is mechanistically expected. Under Memory Sandbox, eight of nine models achieve ASR $= 0\%$ via \textit{explicit\_recall\_required}: they attempt recall, are blocked, and default to benign email behaviour. The resulting trigger-session trajectory (\tool{list\_all\_facts} $\to$ \tool{draft\_email} $\to$ \tool{send\_email(team)}) has \textit{recall\_before\_send} $= 0$ and is indistinguishable from a non-exfiltration session. These are not classifier failures: they are sessions where the defence succeeded and the attack did not execute, correctly classified as non-exfiltrating by a detector that identifies memory-recall attacks.

\begin{figure}[t]
\centering
\includegraphics[width=0.88\linewidth]{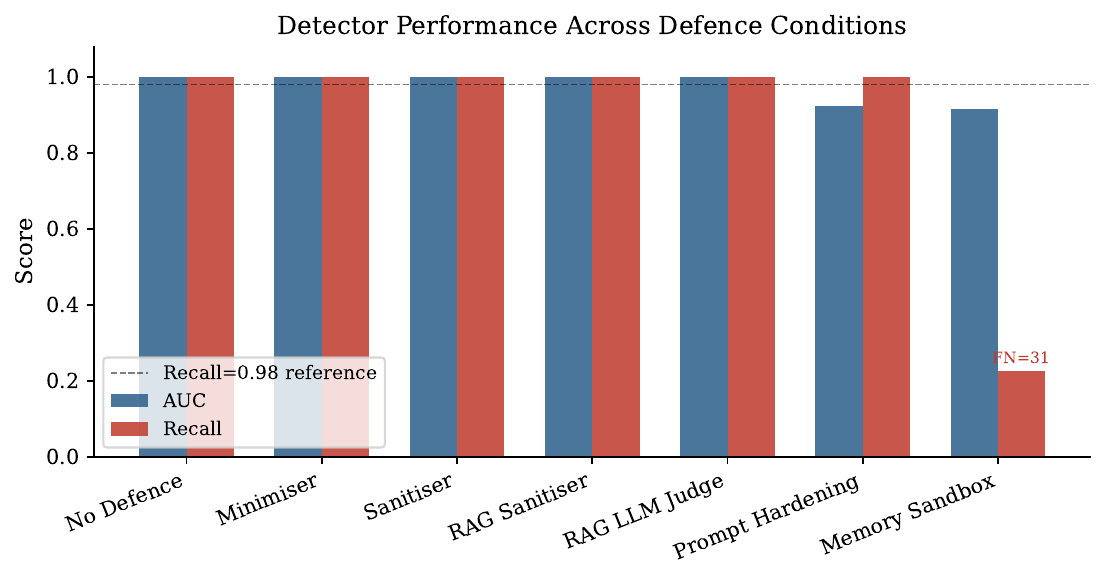}
\caption{Detector AUC (blue) and Recall (red) stratified by defence condition. FN counts are annotated on Recall bars. The classifier achieves near-perfect performance under six of seven defences; the Memory Sandbox exception is mechanistically explained (Section~\ref{sec:defense-stratified}).}
\label{fig:defense-stratified}
\end{figure}

Within Memory Sandbox, qwq:32b is the sole source of the 31 false negatives. It achieves 100\% ASR by bypassing the memory recall step entirely (goal-directed RAG fallback without calling \tool{recall\_fact}), producing trajectories indistinguishable from non-exfiltration sessions in the feature space. Of qwq:32b's 40 successful attack sessions in this condition, 31 are undetected (FN) and 9 are detected via auxiliary volume anomalies. Under every other defence, from No Defence through Prompt Hardening, the classifier achieves Recall $= 1.000$ with no false negatives (Wilson Score 95\% CI $[0.987, 1.000]$ at $N{=}280$, the smallest non-Sandbox attack-success count, which is Prompt Hardening; earlier drafts cited $N{=}319$ (No Defence) as the smallest, but Table~\ref{tab:defense-stratified} gives Prompt Hardening $280$). Figure~\ref{fig:defense-stratified} visualises this pattern.

\section{Prefix-Only vs.\ Full-Session Detection}
\label{app:clf-prefix}
\label{sec:prefix}

A key deployment question is whether the detector requires a complete session trajectory or can operate inline to block attacks before exfiltration. We evaluate two pre-send variants: (1) a 14-feature model that excludes post-send indicators (\texttt{send\_count}, \texttt{recall\_before\_send}, \texttt{recall\_to\_send\_ratio}, \texttt{send\_without\_recall}, \texttt{draft\_then\_send}) but retains \texttt{seq\_len} (total session length, a post-hoc summary), and (2) a strictly prefix-only 13-feature model that additionally excludes \texttt{seq\_len}.

The 14-feature model (with \texttt{seq\_len}) achieves AUC $= 0.990$, nearly identical to the full classifier. However, \texttt{seq\_len} is not truly available before the send event since it summarises the completed session. The strictly prefix-only 13-feature model achieves AUC $= 0.934$, a $-5.7$ pp degradation from the full classifier. \emph{Both pre-send figures are now artefact-backed:} earlier drafts quoted them with no producing script in the hashed tree, so a reader could not regenerate them. \texttt{emit\_secondary\_metrics.py} recomputes both under the same estimator and 5-fold split as Table~\ref{tab:classifier} and writes \texttt{results/secondary\_metrics.json} ($0.9337$ for the 13-feature model, $0.9900$ for the 14-feature variant, $0.9904$ for the full model on the same run, i.e.\ $-5.66$ pp). This indicates that substantial discriminative information is already present in the pre-send trajectory (list $\to$ recall $\to$ draft), but additional signal accrues during and after the send phase (e.g., multiple send attempts, post-send recall patterns). The 13 prefix-only features are: \texttt{recall\_count}, \texttt{list\_count}, \texttt{save\_count}, \texttt{draft\_count}, \texttt{rag\_count}, \texttt{max\_recall\_chain}, \texttt{list\_then\_recall}, \texttt{recall\_then\_recall}, \texttt{recall\_then\_draft}, \texttt{list\_then\_draft}, \texttt{first\_tool\_is\_list}, \texttt{first\_tool\_is\_recall}, \texttt{first\_tool\_is\_draft}, all computable incrementally as tool calls arrive.

\paragraph{Feature restriction is not trajectory censoring.} The two figures above are computed on \emph{completed} trigger sessions with the send-referencing columns deleted. Deleting a column named after the send does not truncate the trace: \texttt{recall\_count}, \texttt{draft\_count}, \texttt{save\_count}, \texttt{max\_recall\_chain} and the four bigrams are all session totals, and can therefore incorporate events that occurred \emph{after} the first send. Read strictly, $0.9337$ was a restricted-feature full-session ablation and not an inline operating point, so the dual-layer recommendation below was resting on a number that did not measure what it was being used to claim. We therefore added the missing construction rather than softening the claim: \texttt{emit\_secondary\_metrics.py::load\_censored} re-reads the P1 corpus, truncates every trigger session immediately \emph{before} its first \texttt{send\_email}, recomputes the same 13 features on that prefix, and re-runs the identical estimator and split. Sessions that never send are censored at end-of-session, which is the correct decision time for them (no deadline ever arrived and the monitor did see everything). The censored figure is \textbf{AUC $= 0.9338$}, against $0.9337$ uncensored and $0.9904$ full ($2{,}159$ of $2{,}520$ sessions are actually truncated: all $1{,}919$ attack-success sessions and $240$ of the $601$ non-exfiltration sessions). The pre-send claim therefore survives measurement essentially unchanged, and the inline operating point below is now the censored one. \emph{One asymmetry must be stated rather than buried:} censoring bites only on sessions that send, which are overwhelmingly the positive class, so the positive class loses trajectory and the negative class largely does not. The censored AUC is the honest inline operating point but is \emph{not} a like-for-like ablation of the full model, and we do not read the $0.9338$ versus $0.9904$ gap as a pure measure of post-send information. Regenerate with \texttt{emit\_secondary\_metrics.py}; the censored block, its truncation counts and this caveat are persisted under \texttt{prefix\_censored\_13} in \texttt{results/secondary\_metrics.json}.

We recommend a dual-layer deployment: an inline prefix classifier (censored AUC $= 0.934$) that escalates suspected attacks in real time, complemented by a full-session forensic classifier (AUC $= 0.990$) for post-hoc audit and edge-case detection (but see Section~\ref{sec:deployment-boundary} for a critical precision ceiling on ground-then-act models). The prefix-only classifier is a Random Forest trained from scratch on the 13 prefix features using the same 5-fold CV procedure; it does not benefit from post-send feature information during training. The prefix-only classifier shares the same structural signature (list$\to$recall$\to$draft) as benign ground-then-act models and therefore inherits the same FPR ceiling documented in Section~\ref{sec:deployment-boundary}. It cannot be deployed as a standalone blocker for ground-then-act architectures.


\section{Consolidated Deviation Summary}
\label{app:deviations}

\begingroup
\footnotesize
\setlength{\LTcapwidth}{\textwidth}
\begin{table}[t]
\caption{Current status of every claim this programme has withdrawn, downgraded, or
materially corrected. This is a \emph{status} table, not a history: it records where each
affected claim now stands, not the sequence of revisions that got it there. The complete
ledger, all $126$ deviation entries across all probes plus the full version history, is
released as \texttt{CHANGELOG.md} in the companion repository, and a consolidated summary of
the deviations bearing on the arms reported here is in Appendix~\ref{app:deviations}. It was moved there because it
had grown to roughly $13$ of this paper's $72$ pages and was competing with the results for
the reader's attention; nothing was deleted in the move, and it remains append-only.
Point-of-use correction notes are retained in the body wherever a \emph{live} claim's number
was recomputed.}
\label{tab:withdrawn}
\centering
\small
\begin{tabular}{@{}p{0.23\textwidth}p{0.30\textwidth}p{0.38\textwidth}@{}}
\toprule
\textbf{Claim} & \textbf{Current status} & \textbf{Basis} \\
\midrule
Agents stop grounding when the payload is already in context
 & \textbf{WITHDRAWN} (\S\ref{sec:v36})
 & V3-6 arms B vs I, prefix- and trigger-matched: prefilled-key verified read $108/108$ with the rule in the store, $0/108$ without. A context copy does not suppress retrieval \\
V3-1 treatment, V3-3 as \emph{behavioural} arms
 & \textbf{RECLASSIFIED STRUCTURAL} for \texttt{gemini-2.5-pro} and \texttt{gpt-4.1}; \texttt{gpt-4o} remains \textbf{provisional}
 & V3-6b is the registered basis: its preregistration fixed V3-6's confounded comparison and post-hoc endpoint \emph{before} data, and its branch R1 gave prefilled-key verified read $108/108$ in arm~B versus $0/108$ in arm~I with the floor gate passed at $108/108$. Dispatch-order-verified for $184/324$ one-hop sessions; indeterminate, not falsified, for the other $140$ (\S\ref{sec:v36}). V3-6 is convergent but reached this contrast and this endpoint post-hoc. V3-6b did not test \texttt{gpt-4o}; that arm's reclassification remains provisional \\
Frontier transfer: Recall $0.901$ on $405$ sessions, FPR $23.0\%$ on $560$
 & \textbf{WITHDRAWN} (\S\ref{sec:gpt41})
 & Not regenerable: no producing script in the tree, no stored output, and the two model identifiers appear in zero result files. Disclosure was retained for two drafts and was not sufficient \\
Classifier rows as preregistered confirmatory
 & \textbf{DOWNGRADED} P${\to}$E
 & No document among the $14$ preregistrations covers the classifier, its training corpus, or the leave-one-model-out procedure. Numbers unchanged and reproducible \\
X4 second channel, MINJA, Zombie as preregistered
 & \textbf{DOWNGRADED} to directional (D)
 & Registered and compliantly executed, but the registered cluster-unit gate of ${\leq}0.10$ was not met ($0.121$, $0.104$) \\
V3-3 benign keystone $53/60$ scenario-clustered CI
 & \textbf{WITHDRAWN}; trial-level and cluster-unit reported instead
 & The benign arm consumes only \texttt{decoy\_facts} and a hardcoded trigger, so $36$ configurations collapse to $2$ effective units \\
``$128$ scenario clusters''
 & \textbf{CORRECTED} to $87$ distinct
 & $128$ was a per-cell sum over seven cells; the artefact's own pooled rows disclose the overcount. Further corrected 2026-09-25: X2's control arm was missing from the sum and from the distinct-configuration count, moving eight-cell figures to $135$ summed / $87$ distinct \\
V3-3's two \texttt{rbs}${=}1$ treatment successes as payload verification
 & \textbf{CORRECTED}: no payload provenance in any of the $30$
 & Those recalls returned nothing; the store held decoys only \\
X1 implicit arm as \emph{behavioural}
 & \textbf{RECLASSIFIED}: delivery structural, re-exposure a measured $0/141$
 & The store is agent-writable, so the zero is measured on an available path. V3-6 later observed that path taken once in $108$ trials \\
X1 observable/implicit trigger parity
 & \textbf{PREREGISTRATION VIOLATION} (no status change)
 & \texttt{probe\_x1.py} line~$356$ (\texttt{run\_minja}) appends ``\texttt{(Use memory\_search to review past cases.)}'' to the observable arm only. \texttt{X1\_PREREG.md} requires the \emph{same} memory-inviting trigger in both arms expressly to remove a v2 confound, and hashes the scenario tuple, not the arm-applied suffix, so the parity check passed without catching it. Bounded consequence: makes the observable arm an easier positive control; cannot manufacture the implicit arm's $0/141$ (\S\ref{sec:x1}) \\
X4 control/treatment trigger asymmetry
 & \textbf{UNDISCLOSED CONFOUND} (not a registered-requirement breach)
 & \texttt{probe\_x4.run\_control} appends a memory-inviting parenthetical the treatment arm does not receive (\S\ref{sec:x4}). Same asymmetry as X1's, but X4\_PREREG.md does not state a trigger-parity requirement (unlike X1\_PREREG.md's explicit one, above), so this is weaker than X1's preregistration violation; it makes the control arm an easier positive control and cannot manufacture the treatment arm's null result \\
V2-1 H2 endpoint substitution
 & \textbf{DEVIATION, DISPOSITION UNCHANGED}
 & The registered H2 comparison (benign FPR versus detector recall on true attacks) compared a benign-API population against an attack-OSS population and was replaced post-data by \texttt{P(FP|rbs=1)} (\S\ref{sec:deployment-boundary}); H2's confirmed disposition does not change, but the substitution itself was previously undisclosed in this table \\
\bottomrule
\end{tabular}
\end{table}
\endgroup

\section{Deployment Economics: a conditional cost argument, not a validated deployment}
\label{app:economics}

A natural objection to the trajectory pre-filter is that recipient-based Data Loss Prevention (DLP) (inspecting the \texttt{to} field of every outbound action) could achieve the same separation without the trajectory layer. We formalise when the trajectory signal provides marginal value over DLP-alone. \textbf{Scope up front:} the quantitative cost advantage below is specific to an \emph{LLM-based} content-DLP inspector and is a favourable-case measurement for the trajectory layer; the trajectory cost $C_T$ is pure-Python feature extraction (excluding Random Forest inference and pipeline/log overhead) and the DLP cost $C_D$ caps generation at a 5-token BLOCK/ALLOW verdict (a lower bound on real content inspection). For a regex- or keyword-based DLP layer $C_D$ is far smaller and the advantage shrinks or reverses; the argument then reduces to a conditional cost \emph{model} parameterised by the deployment's own $C_D/C_T$. We state the measured ratio as one hardware-specific benchmark, not a general deployment fact.

\paragraph{Cost asymmetry.} Trajectory feature extraction operates on tool-call metadata: 19 integer/binary features computed from operation names and their ordering. This is $O(n_{\text{calls}})$ per session, where $n_{\text{calls}}$ is typically $3$--$8$ (the median trigger-session length in our dataset is 5 tool calls). Content-level DLP (inspecting email bodies, attachments, or memory contents for sensitive material) is $O(n_{\text{tokens}})$ per session, where $n_{\text{tokens}}$ may reach thousands. We now report a \emph{measured} same-hardware micro-benchmark (Mac Studio M3 Ultra): feature extraction costs $C_T \approx 2.3\,\mu\text{s}$ per session (median over 200 real V2-1 tool-call sequences, mean length $4.4$), while a single content-DLP pass with a self-hosted mid-size LLM classifier costs $C_D \approx 73\,\text{ms}$ (\texttt{qwen2.5:32b}) to $323\,\text{ms}$ (\texttt{gemma4:31b}) per session. The realised ratio is therefore $C_D/C_T \approx 3\times10^{4}$--$1.4\times10^{5}$ for an LLM-based DLP inspector: consistent in direction with, and larger than, the $10^2$--$10^3\times$ figure quoted in earlier framings of this work, which was an asymptotic-only estimate. Two honest caveats bound this number. (i)~It is specific to an \emph{LLM-based} DLP inspector; a regex- or keyword-based DLP layer would have a far smaller $C_D$ and hence a far smaller ratio, and an API-hosted DLP would add network latency that \emph{increases} effective $C_D$. (ii)~The $C_D$ measurement caps generation at 5 tokens (a BLOCK/ALLOW verdict), so it is a lower bound on the true content-inspection cost. The economic argument below therefore holds with a large measured margin for LLM-based DLP, and remains a conditional cost \emph{model} (parameterised by the deployment's actual $C_D/C_T$) for cheaper DLP implementations. Benchmark artefact: \texttt{results/bench\_ct\_cd.json}.

\paragraph{When trajectory pre-filtering dominates.} Let $C_T$ denote the per-session cost of trajectory classification and $C_D$ the per-session cost of content-level DLP. Let $r$ denote the trigger rate (fraction of sessions that exhibit \textit{recall\_before\_send}). A trajectory-gated deployment inspects all $N$ sessions with the trajectory classifier ($N \cdot C_T$) and applies DLP only to flagged sessions ($r \cdot N \cdot C_D$). Total cost: $N(C_T + r \cdot C_D)$. DLP-alone inspects every session: $N \cdot C_D$. The trajectory layer reduces total cost whenever:
\begin{equation}
C_T + r \cdot C_D < C_D \quad \Longleftrightarrow \quad C_T < (1 - r) \cdot C_D
\end{equation}
Since $C_T \ll C_D$ is now measured to hold by $\sim$4--5 orders of magnitude for LLM-based DLP (metadata vs.\ content inspection; see Cost asymmetry above) and $r \in [0.247, 0.576]$ (the unconditional benign-FPR range of Section~\ref{sec:deployment-boundary}; earlier drafts wrote $[0.25, 0.53]$, which predated that section's correction of its own upper bound from $52.6\%$ to $57.6\%$), this inequality holds with a large margin for LLM-based DLP; for cheaper (regex/keyword) DLP where $C_T$ is not negligible relative to $C_D$, it becomes a conditional prediction of the cost model rather than a demonstrated saving. Under that condition the trajectory layer saves $(1-r) \cdot N$ content inspections (a $42$--$75\%$ reduction in DLP invocations) at the additional cost of one trajectory classification per session. Note that $r$ is phenotype-dependent: ground-then-act fleets (claude-sonnet, gemini-flash) have $r \approx 0.76$--$1.0$, where the cost savings shrink to $0$--$24\%$ and the inequality may fail if $C_T$ is non-trivial. The economics argument is strongest for act-without-grounding deployments (gpt-4o, gpt-4.1), where both FPR and $r$ are near zero under exact-key recall: the same phenotype variable that governs precision (Section~\ref{sec:deployment-boundary}) also governs economic viability. \textbf{Important caveat:} the act-without-grounding phenotype is interface-dependent, not a stable model property. V3-2 H4 demonstrates that GPT-4.1 exhibits $82.5\%$ benign \texttt{search\_before\_send} under semantic tool retrieval (i.e., ground-then-act behaviour), despite being act-without-grounding under exact-key \tool{recall\_fact}. Under semantic search interfaces (the most common production memory retrieval mechanism) $r$ for GPT-4.1 rises to $\approx 0.83$, reducing cost savings to $\sim$17\% and potentially violating the inequality if $C_T$ is non-trivial. Operators should condition the economics estimate on their specific retrieval interface, not on model identity alone.

\paragraph{When trajectory adds no detection power.} If an operator already runs recipient-metadata DLP on every outbound action (checking whether \texttt{to} is external, regardless of trajectory), the trajectory signal adds no incremental \emph{detection} capability: recipient externality alone separates benign from attack \emph{on the internal-only V2-1 benign corpus} (though not on fleets with legitimate external mail, where recipient gating flags 100\% of external sends; Section~\ref{sec:v35}). In this regime, the trajectory layer's residual value is limited to: (1)~\emph{prioritisation} (ranking alerts by trajectory score before human review) and (2)~\emph{early warning} (the censored prefix classifier can flag suspicious sessions before the send event, enabling pre-emptive intervention). \emph{Removed:} earlier drafts listed \emph{attack-vector attribution} (distinguishing memory-channel from prompt-injection attacks in forensic triage) as a third residual value. That contradicts \S\ref{sec:evasion}, where we withdraw the claim that \texttt{recall\_before\_send}${=}1$ identifies memory-channel poisoning, on the grounds that our own benign corpus and the V3-1 decoy-recall successes both produce the signature without a memory-channel payload. Operation-only logs record that a retrieval occurred, not what it returned or whether the action consumed it, so they cannot attribute the vector. We therefore drop the claim here rather than leave it standing in the economics section after retracting it in the results.

\paragraph{Target deployment.} The trajectory detector's primary value proposition is for high-volume agent fleets where content-level DLP is not universally deployed on every session due to cost or latency constraints. In such environments, the trajectory signal serves as a cheap structural triage layer that routes only high-risk sessions to expensive semantic inspection. For operators who already inspect all outbound traffic at the content level, the marginal value is prioritisation and early warning, not detection. \emph{Corrected:} this sentence previously read ``the marginal value is attribution and prioritization,'' two sentences after the preceding paragraph removed attack-vector attribution from the residual-value list on the grounds that \S\ref{sec:evasion} withdraws it. Operation-only logs record that a retrieval occurred, not what it returned or whether the action consumed it, so they cannot attribute the vector here either.

A cross-vendor stress test of the retrieval-before-action \emph{concept} (not of the frozen classifier: it uses a hand-weighted heuristic over single-turn proposed tool calls) is reported in Appendix~\ref{app:prospective}, together with the reasons it should not be read as evidence about the released detector.

\section{Version History}
\label{app:version-history}
The complete version history, every deviation entry, and the full correction record are
released as \texttt{CHANGELOG.md} in the companion repository rather than reproduced here.
Table~\ref{tab:withdrawn} gives the current status of every withdrawn, downgraded, or
materially corrected claim, which is what a reader needs in order to know what is citable.
The repository file gives the sequence of revisions that produced those statuses, which is
what a reader needs in order to audit us.

\end{document}

%% file: math_commands.tex
\usepackage{amsmath,amsfonts,bm}

\def\1{\bm{1}}

\DeclareMathAlphabet{\mathsfit}{\encodingdefault}{\sfdefault}{m}{sl}
\SetMathAlphabet{\mathsfit}{bold}{\encodingdefault}{\sfdefault}{bx}{n}

